\documentclass[preprint,12pt]{elsarticle}

\usepackage{hyperref}

\usepackage{amsmath, amsfonts, amssymb,  theorem}
\usepackage{subfiles}
\usepackage{subfig}
\usepackage{epstopdf}
\usepackage{booktabs}
\usepackage{cleveref}
\crefname{equation}{Eq.}{Eqs.}
\Crefname{equation}{Equation}{Equations}
\crefname{figure}{Fig.}{Figs.}
\Crefname{figure}{Figure}{Figures}
\usepackage[usenames,dvipsnames]{color} 
\usepackage{nomencl}
\usepackage{framed} 

\journal{Engineering Fracture Mechanics}

\begin{document}

\begin{frontmatter}



\title{Experimental characterization of cohesive laws for mode-II interlaminar fracture in geometrically scaled composites using through-thickness deformation analysis}


\author[author1]{Han-Gyu Kim\corref{cor1}} 
 \ead{hkim@ae.msstate.edu}
 \cortext[cor1]{Corresponding author.}
\author[author2]{Ryan Howe} 
\author[author3]{Richard Wiebe} 
\author[author4]{S. Michael Spottswood} 
\author[author4]{Patrick J. O'Hara} 
\author[author5]{Marco Salviato} 

\affiliation[author1]{organization={Department of Aerospace Engineering, Mississippi State University},
            city={Mississippi State},
            postcode={39762}, 
            state={MS},
            country={USA}}

\affiliation[author2]{organization={56th Fighter Wing, 310th Fighter Squadron, Luke AFB, U.S. Air Force},
            city={Glendale},
            postcode={85309}, 
            state={AZ},
            country={USA}}

\affiliation[author3]{organization={Department of Civil and Environmental Engineering, University of Washington},
            city={Seattle},
            postcode={98195}, 
            state={WA},
            country={USA}}
            
\affiliation[author4]{organization={Structural Sciences Center, Air Force Research Laboratory, AFRL/RQHF},
            city={Wright-Patterson AFB},
            postcode={45433}, 
            state={OH},
            country={USA}}   
            
\affiliation[author5]{organization={Department of Aeronautics and Astronautics, University of Washington},
            city={Seattle},
            postcode={98195}, 
            state={WA},
            country={USA}}

\begin{abstract}
This work proposes an experimental framework to characterize a cohesive law for mode-II interlaminar fracture and demonstrates its implementation. 
For a size effect study, geometrically scaled end-notched flexure specimens were tested using microscopic and macroscopic digital image correlation (DIC) systems. 
The fracture energy was characterized using a compliance calibration method and Bažant's type-II size effect law for comparison.
In the proposed experimental framework, the DIC data were post-processed using three steps: coordinate transformation, curve fitting, and through-thickness deformation analysis. 
Different magnitudes of separation values were measured from different sizes at fracture loads, implying size effect and partial development of cohesive laws. 
Modeling and simulations were intended to validate the proposed method and demonstrate the utilization of the experimental data. 
Additionally, challenges related to finding a single cohesive law for geometrically scaled specimens of a single material were exposed.
A single cohesive law for the scaled specimens was developed and proposed as a material property of the specimen material. 
The fracture energy of the single law was smaller than the energy obtained from the size effect analysis, while the sizes of fracture process zones at fracture loads were smaller than the experimental measurements.
However, the global fracture behaviors of the models showed good agreement with the experimental data of the mid-size specimen while showing reasonable agreement with the other sizes.
Furthermore, the single law successfully captured local fracture behaviors by showing partial cohesive zone development at the fracture loads and matching the microscopic measurement of the separation values.
\end{abstract}






\end{frontmatter}

\begin{framed}
\begin{thenomenclature} 
\nomgroup{A}
  \item [{\(\alpha\)}]\begingroup Normalized effective crack length\nomeqref {0}\nompageref{4}
  \item [{\(\bar{E}_\text{1f}\)}]\begingroup Effective flexural modulus\nomeqref {0}\nompageref{4}
  \item [{\(\Delta u_0\)}]\begingroup Separation at softening transition in a cohesive law\nomeqref {0}\nompageref{4}
  \item [{\(\Delta u_\text{1}\)}]\begingroup Separation in a cohesive law\nomeqref {0}\nompageref{4}
  \item [{\(\Delta u_\text{f}\)}]\begingroup Final (or complete) separation in a cohesive law\nomeqref {0}\nompageref{4}
  \item [{\(\Delta u_\text{i}\)}]\begingroup Separation at damage initiation in a cohesive law\nomeqref {0}\nompageref{4}
  \item [{\(\Delta u_\text{max}\)}]\begingroup Experimental measurement of the maximum separation\nomeqref {0}\nompageref{4}
  \item [{\(\Delta u_\text{min}\)}]\begingroup Experimental measurement of the minimum separation\nomeqref {0}\nompageref{4}
  \item [{\(\delta\)}]\begingroup Displacement applied by a loading roller\nomeqref {0}\nompageref{4}
  \item [{\(\nu_\text{12}\)}]\begingroup Poisson's ratio\nomeqref {0}\nompageref{4}
  \item [{\(\sigma_0\)}]\begingroup Pseudo-plastic limit\nomeqref {0}\nompageref{4}
  \item [{\(\sigma_\text{Nc}\)}]\begingroup Nominal strength\nomeqref {0}\nompageref{4}
  \item [{\(\tau_0\)}]\begingroup Traction at softening transition in a cohesive law\nomeqref {0}\nompageref{4}
  \item [{\(\tau_\text{13}\)}]\begingroup Traction in a cohesive law\nomeqref {0}\nompageref{4}
  \item [{\(\tau_\text{f}\)}]\begingroup Maximum (or critical) traction in a cohesive law\nomeqref {0}\nompageref{4}
  \item [{\(\varepsilon_\text{13}\)}]\begingroup Shear strain\nomeqref {0}\nompageref{4}
  \item [{\(a_\text{0}\)}]\begingroup Initial crack length\nomeqref {0}\nompageref{4}
  \item [{\(b\)}]\begingroup Nominal width\nomeqref {0}\nompageref{4}
  \item [{\(c_\text{f}\)}]\begingroup Effective size of a fracture process zone\nomeqref {0}\nompageref{4}
  \item [{\(D\)}]\begingroup Damage variable\nomeqref {0}\nompageref{4}
  \item [{\(E^*\)}]\begingroup Equivalent elastic modulus\nomeqref {0}\nompageref{4}
  \item [{\(E_\text{1t}\)}]\begingroup $0^{\circ}$ Tensile modulus\nomeqref {0}\nompageref{4}
  \item [{\(E_\text{2t}\)}]\begingroup $90^{\circ}$ Tensile modulus\nomeqref {0}\nompageref{4}
  \item [{\(g\)}]\begingroup Dimensionless energy release rate\nomeqref {0}\nompageref{4}
  \item [{\(G_\text{12}\)}]\begingroup In-plane shear modulus\nomeqref {0}\nompageref{4}
  \item [{\(G_\text{f,LEFM}\)}]\begingroup Mode-II fracture energy based on linear elastic fracture mechanics\nomeqref {0}\nompageref{4}
  \item [{\(G_\text{f}\)}]\begingroup Mode-II fracture energy\nomeqref {0}\nompageref{4}
  \item [{\(h\)}]\begingroup Nominal thickness\nomeqref {0}\nompageref{4}
  \item [{\(h_0\)}]\begingroup Transitional characteristic length\nomeqref {0}\nompageref{4}
  \item [{\(K\)}]\begingroup Initial interface stiffness in a cohesive law\nomeqref {0}\nompageref{4}
  \item [{\(L\)}]\begingroup Gauge length\nomeqref {0}\nompageref{4}
  \item [{\(L_\text{m}\)}]\begingroup Field length of a microscope\nomeqref {0}\nompageref{4}
  \item [{\(P\)}]\begingroup Load applied by a loading roller\nomeqref {0}\nompageref{4}
  \item [{\(P_\text{max}\)}]\begingroup Peak load (or fracture load)\nomeqref {0}\nompageref{4}
  \item [{\(u_1\)}]\begingroup Displacement along the principal material axis $1$\nomeqref {0}\nompageref{4}
  \item [{\(u_3\)}]\begingroup Displacement along the principal material axis $3$\nomeqref {0}\nompageref{4}
  \item [{\(u_\text{x}\)}]\begingroup Displacement along the geometric axis $x$\nomeqref {0}\nompageref{4}
\end{thenomenclature}
\end{framed}



\section{Introduction}
\label{sec:intro}

\subsection{Motivations}\label{sec:intro_motivations}

\begin{figure}[t!]
\centering
\includegraphics[width=1.0\textwidth]{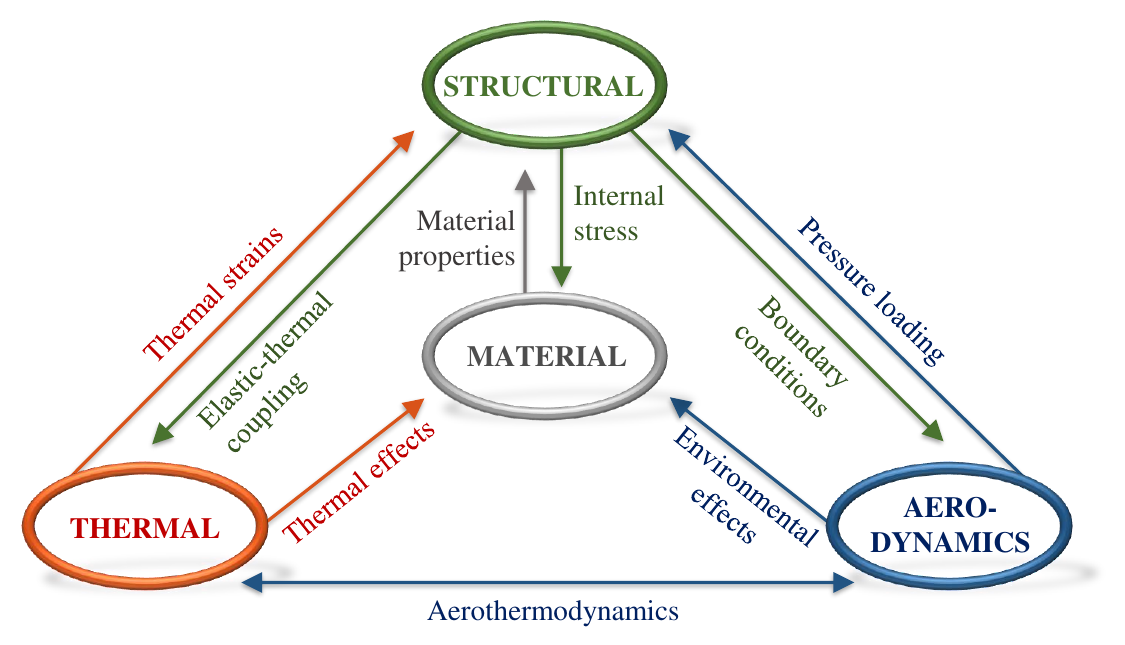}
\caption{Schematics of the multi-physics elements and their couplings for high-speed aircraft under aerothermodynamic loading.}
\label{fig_aerodynamics}
\end{figure}

Modern aircraft designs have extensively adopted composites for primary structures to achieve lightweight structure designs. 
The carbon-fiber composite fuselage structures of the Boeing 787 Dreamliner (a subsonic commercial airliner) and the Lockheed F-35 Joint Strike Fighter (a supersonic fighter) are representative examples \cite{baker2014boeing,DOTE2014F35}. 
Composite structures in high-speed aircraft, however, would be subjected to extreme environments induced by aerothermodynamic couplings (see \cref{fig_aerodynamics}), which could lead to internal stress amplification and mode-II interlaminar failure \cite{kim2020}.
To address this issue, this paper focuses on developing an experimental framework that could contribute to developing high-fidelity damage models to predict the structural life of composite structures in high-speed aircraft.

\subsection{Quasibrittle fracture process and size effect in composites}\label{sec:intro_quasibrittle}

The anisotropy and heterogeneity of composites could lead to complex damage initiation and progression compared to fracture and fatigue in isotropic and homogeneous materials \cite{nairn1994}.
Bažant et al. \cite{bazant2017} described that the degree of brittleness and ductility of materials would be dictated by the size of fracture process zones (FPZs) in front of crack tips. 
It was noted by Bažant et al. that the size of FPZs in brittle materials would be small enough for linear elastic fracture mechanics (LEFM) to be applicable, while a large nonlinear plastic (or yielding) zone could be observed along the FPZs of ductile or elastoplastic materials (mainly metals). 
Bažant et al. categorized composites as quasibrittle materials that would not demonstrate notable plastic behaviors due to their negligible hardening areas but can deliver larger crack opening displacements or interlaminar separation with the formation of extensive softening areas in FPZs compared to brittle materials.

For experimental characterization of the interlaminar fracture toughness (or energy) of quasibrittle materials, Williams \cite{williams1989} and Davidson et al. \cite{davidson2006-1,davidson2006-2} proposed a compliance calibration method based on LEFM for modes I and II, respectively.
These methods have been adopted in the ASTM standards \cite{ASTM5528,ASTM7905}.
Salviato et al. \cite{salviato2019}, however, experimentally showed that these LEFM-based methods estimated multiple fracture energies for geometrically scaled composites of a single material: that is, higher fracture energies with higher geometric scaling factors.
This analysis result contradicts the fact that fracture energies are unique material properties like elastic moduli. 
To address this issue, Salviato et al. applied Bažant's type-II size effect law \cite{bazant1984} to the experimental load-displacement data in conjunction with numerical J-integral analysis \cite{rice1968} and obtained the fracture energy of the specimen material as a single value.
Also, Salviato et al. pointed out that the damage mechanisms such as matrix microcracking, crack deflection, and plastic yielding occurring under mode-II loading could lead to a significantly larger FPZ compared to mode I.
Therefore, for sufficiently small specimens, the size of the highly nonlinear FPZ under mode-II interlaminar fracture would not be negligible compared with the specimen characteristic size (more specifically, the distance between the initial crack tip and the loading roller) and would result in a strong size effect (i.e., a significant deviation from LEFM).

\subsection{Damage modeling for composites}\label{sec:intro_CZM}

The virtual crack closure technique (VCCT) and cohesive zone modeling have been widely adopted in finite element models (FEMs) to simulate damage propagation in composites. 
Rybicki and Kanninen \cite{rybicki1977} proposed VCCT based on LEFM principles to evaluate stress intensity factors in homogeneous isotropic materials. 
A thorough review of this technique can be found in Ref. \cite{krueger2004}.
The LEFM-based technique reportedly has several drawbacks for composite damage modeling.
To be specific, VCCT requires a precrack and an extremely fine mesh near the crack tip; additionally, complex moving mesh techniques are needed to advance the crack front \cite{abrate2015}.
On the other hand, cohesive zone models (CZMs) can be applied to composites without an existing crack, and an FEM mesh around CZMs does not need to be modified to simulate damage progression.
The early concepts of CZMs were proposed by Barenblatt \cite{barenblatt1962}, Dugdale \cite{dugdale1960}, and Hillerborg \cite{hillerborg1976}.
Each composite material would have a single traction-separation law as its material property and thus the fracture mechanisms of CZMs for the material should follow the law. 
Detailed discussions on the characteristics of CZMs can be found in Ref. \cite{bazant1998-1}, while Refs. \cite{abrate2015,elices2002,park2011} presented thorough reviews of CZMs. 

For CZMs, de Moura et al. \cite{moura2012,fernandes2013,silva2014} experimentally characterized traction-separation laws (or cohesive laws) for mode-I and mode-II fracture in adhesive-bonded joints using both the direct and inverse methods. 
de Moura et al. adopted a digital image correlation (DIC) technique to measure separation at crack tips from two DIC data points above and below the crack tips.
Traction values were estimated based on Timoshenko beam theory. 
The simulation results of de Moura et al. showed good agreement with the experimental data. 
Compared to composites, adhesive-bonded joints show more stable crack growth (dependent on the sizes of the joints) with larger separation values, which are beneficial to the experimental characterization of their cohesive law parameters. 
Composites, on the other hand, show unstable crack propagation in mode II with a sudden drop (or a snapdown) of applied loads induced by snapback instability \cite{bazant1987-1,bazant1987-2} under small separation between layers. 
These characteristics make CZM characterization for composites more challenging. 
Khaled et al. \cite{khaled2019} applied a similar approach to T800S/F3900 carbon/epoxy unidirectional composites for mode-I and mode-II CZM characterization. 
Precracking was performed to propagate $5$ mm-long precracks from Teflon inserts in the specimens.
Khaled et al. measured separations by manually selecting two DIC data points above and below the potential crack tip locations while estimating traction values using a closed-form solution for the J-integral \cite{leffler2007} based on a compliance calibration method. 
Khaled et al. reported that the CZMs showed good performance overall; however, with underestimated maximum loads from the simulations, they reported difficulties in locating the crack tips exactly on DIC images and accurately capturing the separations with their DIC resolution.
It should be noted that all studies introduced in this paragraph used single-size specimens and thus size effect was not considered in these studies. 

\subsection{Objectives}\label{sec:intro_objectives}

The objective of the work herein is to propose and implement an experimental framework to characterize a cohesive law for mode-II interlaminar fracture in geometrically scaled specimens.  
The proposed framework aims to find cohesive law parameters that can incorporate size effect in quasibrittle fracture.
Characterizing the parameters directly from experimental data without employing complex analytical equations is of primary interest. 
This paper also focuses on exposing issues related to the experimental characterization of quasibrittle fracture process from a single-size data set and discussing difficulties in finding a single cohesive law to meet the experimental data from different sizes of specimens at the same time.
It needs to be noted that extensive modeling and simulations were done for this work; however, only some of them are introduced in this paper to validate the experimental framework and demonstrate its implementation for modeling and simulations. 
The other data will be presented in a separate paper with more detailed discussions on modeling and simulation aspects.

\section{Experimental work}\label{sec:exp_work}

\subsection{Specimen details}\label{sec:exp_work_specimen}

The composite specimens were laid up using the Toray T700G/2510 prepreg system and were cured in an autoclave. 
The prepreg system is made of unidirectional carbon fibers, and its elastic properties are given in \Cref{tab_material}.
The specimens had initial delamination along their midplanes at the ends induced by embedded inserts. 
Some researchers \cite{perez2007,cahain2015,kuppusamy2016} studied other methods of making precracks by applying mechanical or fatigue loads to specimens. 
These methods, however, could induce unintended initial damage to FPZs in front of designed crack-tip locations in the form of microcracks. 
In this work, keeping the FPZs intact prior to fracture tests was crucial to the experimental characterization of cohesive law parameters.
Thus, this work employed the method of embedding non-adhesive inserts before curing to induce precracks in the specimens.
A Teflon\textsuperscript{\textregistered} FEP film ($12.7$ \textmu m-thick) was selected as an insert to meet the ASTM D7905/D7905M-19e1 specification of non-adhesive film insert type and thickness \cite{ASTM7905}.

\begin{table}[t!]\caption{Elastic properties of the Toray T700G/2510 prepreg system \cite{Toray_P707AG}}\label{tab_material}
\centering
\begin{tabular}{lcc}
 \toprule
 \textbf{Property} && \textbf{Value}\\
 \midrule
 $0^{\circ}$ tensile modulus, ${E_\text{1t}}$ && $125$ GPa\\
 $90^{\circ}$ tensile modulus, ${E_\text{2t}}$ && $8.41$ GPa\\
 In-plane shear modulus, ${G_\text{12}}$ && $4.23$ GPa\\
 Poisson's ratio, ${\nu_\text{12}}$ && $0.31$ \\
 \bottomrule
\end{tabular}
\end{table}

\begin{figure}[t!]
\centering
\includegraphics[width=1.0\textwidth]{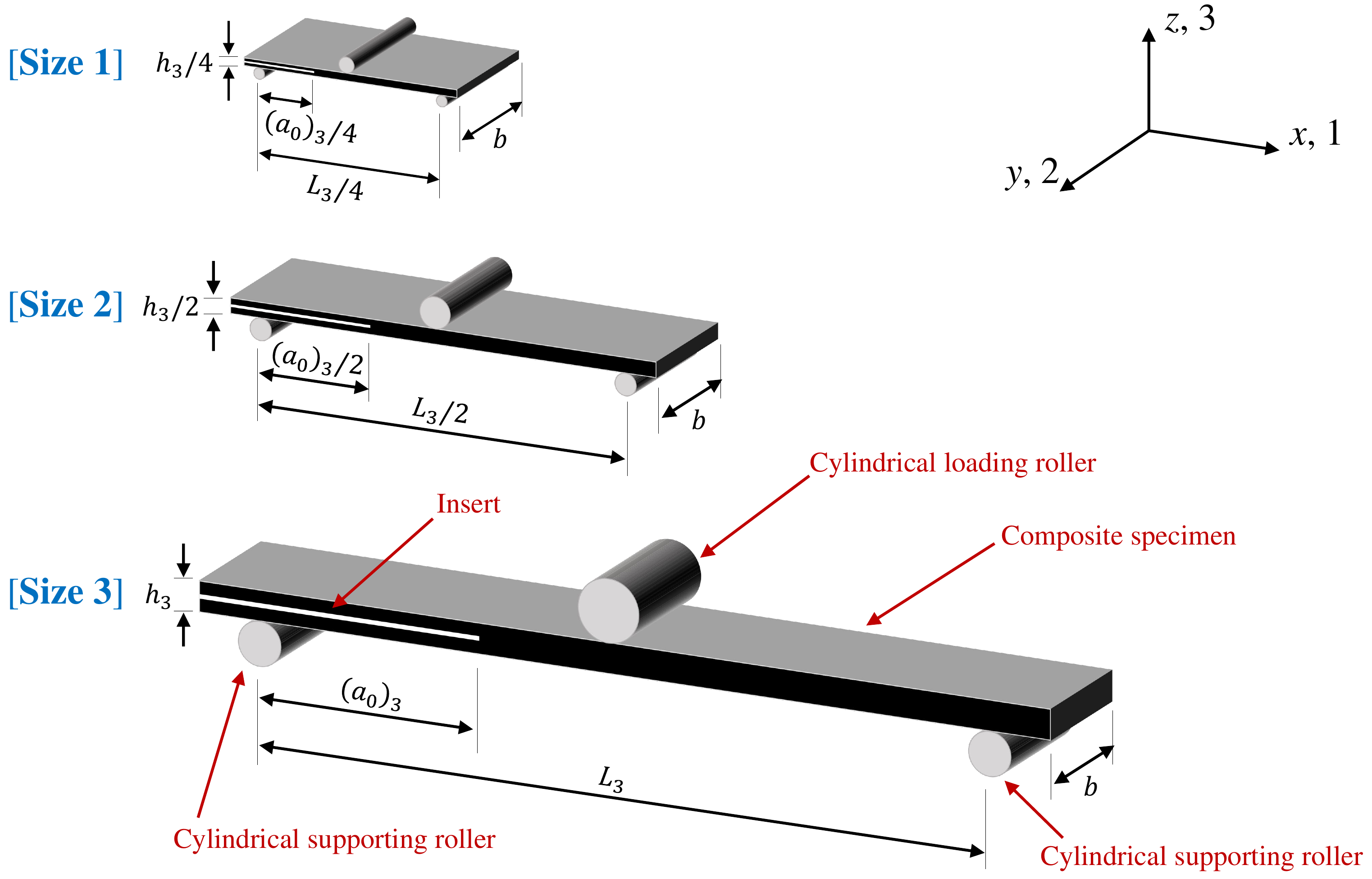}
\caption{
	Schematics of the scaled ENF specimens. 
	The thickness, gauge length, width, and initial crack length of the Size-3 specimen are	represented by $h_3$, $L_3$, $b$, and $(a_0)_3$, respectively.
}
\label{fig_scaling}
\end{figure}

For a size effect study, the end-notched flexure (ENF) specimens were geometrically scaled in three levels as shown in \cref{fig_scaling}.
In this paper, the principal material axes \cite{tuttle2013} are represented by the axes $1$, $2$, and $3$, while the axes $x$, $y$, and $z$ indicate the geometric axes.
The direction of the unidirectional fibers is represented by the $1$ axis.
Before beam bending, the axes $1$, $2$, and $3$ were aligned with the axes $x$, $y$, and $z$, respectively.
The dimensions of the Size-3 specimens were determined based on the ASTM D7905/D7905M-19e1 specification and were used as a baseline for scaling.
Given the beam characteristics of these specimens, 2D scaling was made to the geometric dimensions on the $x$--$z$ plane.
To be specific, the gauge length, thickness, and initial crack length of the Size-3 set were scaled down, while the width was kept constant. 
The target scaling factors were $1/2$ and $1/4$ for the Size-2 and Size-1 sets, respectively.
Additionally, symmetric layups were required for all sizes to embed the Teflon films in the midplanes as inserts.
To meet these requirements, the ASTM dimensions had to be slightly modified for the Size-3 specimens.
As a result, 8, 16, and 32 plies of Toray T700G/2510 prepreg were laid up for Sizes 1, 2, and 3, respectively.
The geometric dimensions of the scaled specimens are summarized in \Cref{tab_dimension}.
Four specimens per size were manufactured, and their actual dimensions are tabulated in \Cref{tab_exp}.


\begin{table}[t!]\caption{Geometric dimensions of the geometrically scaled ENF specimens}\label{tab_dimension}
\centering
\begin{tabular}{lrrrrrr}
 \toprule
\textbf{Dimension} && \textbf{Size 1} && \textbf{Size 2} && \textbf{Size 3}\\
 \midrule
Gauge length, ${L}$ (mm) && $36$ && $72$ && $144$\\ 
Nominal thickness, ${h}$ (mm) && $1.25$ && $2.5$ && $5.0$\\
Nominal width, ${b}$ (mm) && $25$ && $25$ && $25$\\
Initial crack length, ${a_\text{0}}$ (mm) && $10$ && $20$ && $40$\\
 \bottomrule
\end{tabular}
\end{table}

\begin{table}[t!]\caption{Actual specimen dimensions and experimental data}\label{tab_exp}
\centering
\begin{tabular}{cccccccc}
 \toprule
\textbf{Size} & \textbf{Specimen} & \textbf{Label} & $\boldsymbol{b}$ & $\boldsymbol{h}$ & $\boldsymbol{P_\text{max}}$ & $\boldsymbol{\bar{E}_\text{1f}}$ & $\boldsymbol{G_\text{f,LEFM}}$ \\
& & & (mm)& (mm) & (N) & (GPa) & (N/mm) \\
\midrule
1 & 1 & S1SP1 & 25.7 & 1.30 & 439 & 110.1 & 0.63\\
 & 2 & S1SP2 & 25.4 & 1.32 & 443 & 102.6 & 0.40\\
 & 3 & S1SP3 & 25.4 & 1.29 & 462 & 112.0 & 0.54\\
 & 4 & S1SP4 & 23.1 & 1.31 & 431 & 107.7 & 0.52\\
\midrule
2 & 1 & S2SP1 & 25.0 & 2.46 & 656 & 112.0 & 0.79\\
 & 2 & S2SP2 & 24.0 & 2.50 & 731 & 119.8 & 1.07\\
 & 3 & S2SP3 & 27.0 & 2.54 & 748 & 114.1 & 0.79\\
 & 4 & S2SP4 & 25.8 & 2.54 & 719 & 115.8 & 0.85\\
\midrule
3 & 1 & S3SP1 & 25.8 & 4.98 & 1141 & 108.4 & 0.70\\
 & 2 & S3SP2 & 26.4 & 4.95 & 1233 & 117.0 & 1.06\\
 & 3 & S3SP3 & 24.8 & 5.01 & 1154 & 121.2 & 1.01\\
 & 4 & S3SP4 & 26.7 & 5.04 & 1151 & 118.7 & 0.91\\
\bottomrule
\end{tabular}
\end{table}

\subsection{Experimental setup}\label{sec:exp_work_setup}

\begin{figure}[t!]
\centering
    \subfloat[\label{fig_expsetup_a}]{%
      \includegraphics[height=0.6\textwidth]{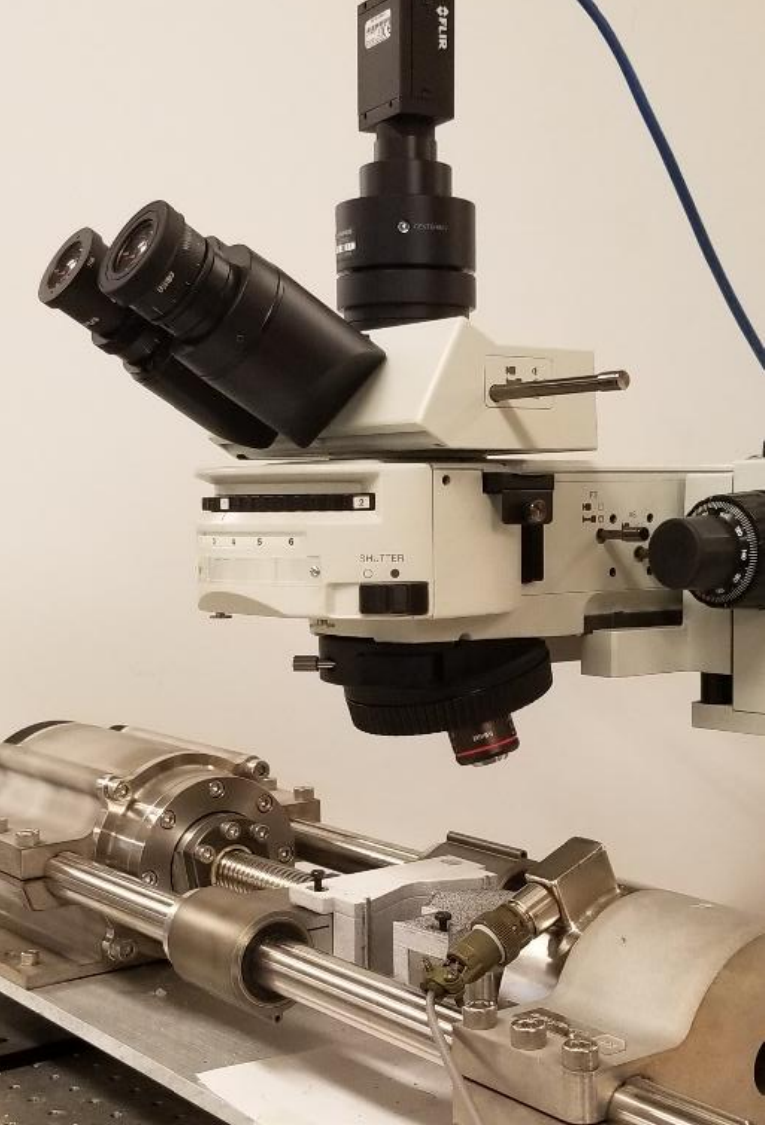}      
    } 
    \subfloat[\label{fig_expsetup_b}]{%
      \includegraphics[height=0.6\textwidth]{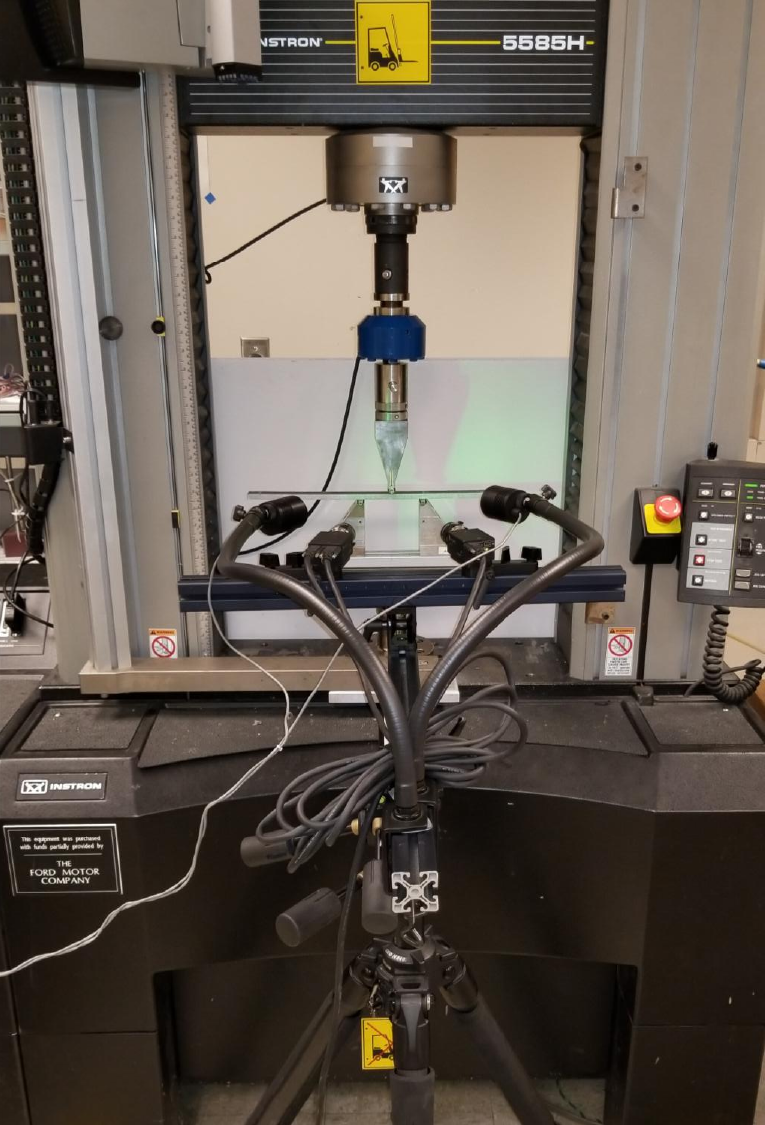}      
    }     
    \caption{Experimental setups for two different scales of DIC tests. (a) 2D microscopic DIC setup. (b) 3D macroscopic DIC setup.}
    \label{fig_expsetup}
  \end{figure}

For the experimental characterization of mode-II interlaminar fracture in these ENF specimens, three-point bending tests were done using displacement control.
Two different types of DIC systems were employed as shown in \cref{fig_expsetup}.
The microscopic-scale tests (see \cref{fig_expsetup_a}) were intended to capture detailed information on damage progression in the vicinity of the initial crack tips.
Therefore, a small part of the specimen surface on the $x$--$z$ plane including the initial crack tip was captured in the digital images through the microscope. 
The Size-1 and Size-2 specimens were tested using this microscopic DIC system; however, the Size-3 specimens could not fit into the system due to their large gauge lengths.
The 3D macroscopic DIC system (see \cref{fig_expsetup_b}) was instead used for the Size-3 tests capturing the entire $x$--$z$ surface. 
The details on the DIC systems are described in the following sections based on the reporting guidelines outlined in Ref. \cite{jones2018}.

\subsubsection{2D microscopic DIC system}
For the 2D microscopic DIC tests, a Psylotech \textmu TS testing system and an Olympus BXFM microscope were used as shown in \cref{fig_expsetup_a}.
The microscope had an Olympus MPLFLN $1.25\times$ objective lens having a $180$-mm focal length and a $17.6$-mm field of view.
For digital image acquisition, a FLIR Grasshopper 3 GS3-U3-51S5M-C camera having a resolution of $2448 \times 2024$ pixels was attached to the microscope, and a Correlated Solutions VIC-Snap package \cite{vic_snap} was used.
The details on the image scale are described with illustrations in \Cref{sec:exp_results_local}.
The stand-off distance between the front of the objective lens and the specimen surface was approximately $4$ mm. 
The image acquisition rate was $1$ Hz, and the loading speed was designed to take nearly six hundred images up to fracture.
For microscopic DIC, fine black speckle patterns were generated on white backgrounds using an airbrush, and the dot size was between 22 and 81 $\mu$m.

\subsubsection{3D macroscopic DIC system}
For the 3D macroscopic DIC tests, an Instron 5585H testing system and a Correlated Solutions VIC-3D HR system were employed as shown in \cref{fig_expsetup_b}.
For digital image acquisition, two Point Grey GRAS-50S5M cameras having a resolution of $2448 \times 2024$ pixels were used with a Correlated Solutions VIC-Snap package \cite{vic_snap}. 
The cameras had Schneider Xenoplan 1.9/35-0901 lenses having a $35$-mm focal length and an $18^\text{o}$ field of view.
The image scale of the region of interest of the image was $154.7$ mm, and the stereo angle was $4.6^\text{o}$.
The stand-off distance was approximately $700$ mm, and the distance between the two cameras was $56.8$ mm.
The image acquisition rate and loading speed were set to match those of the 2D microscopic testing setup.
Black speckle patterns were generated on white backgrounds using spray paint, and the dot size was between 80 and 161 $\mu$m.

\section{Experimental results}\label{sec:exp_results}

\subsection{Experimental characterization of global fracture behaviors}\label{sec:exp_results_global}

\subsubsection{Load-displacement analysis}\label{sec:exp_results_global_load_disp}

\begin{figure}[t!]
\centering
    \subfloat[\label{fig_loading_curves_a}]{%
      \includegraphics[width=0.48\textwidth]{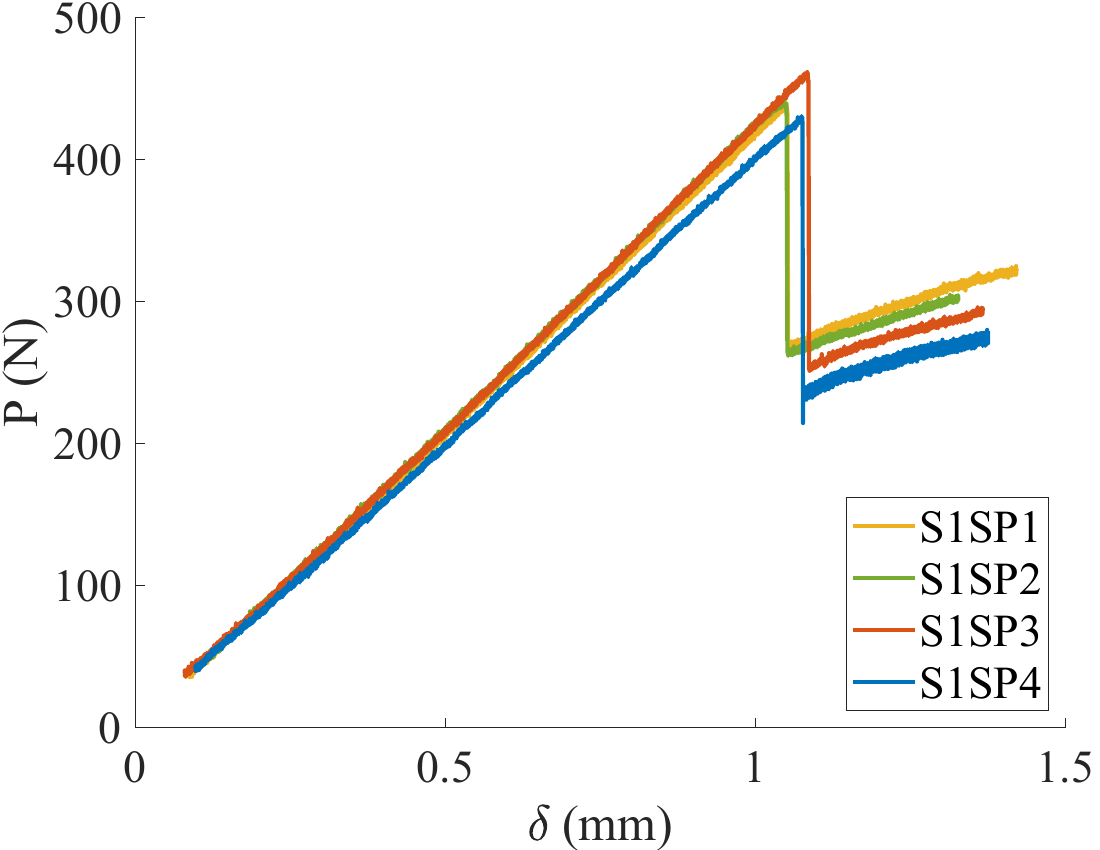}      
    } 
    \hfill
    \subfloat[\label{fig_loading_curves_b}]{%
      \includegraphics[width=0.48\textwidth]{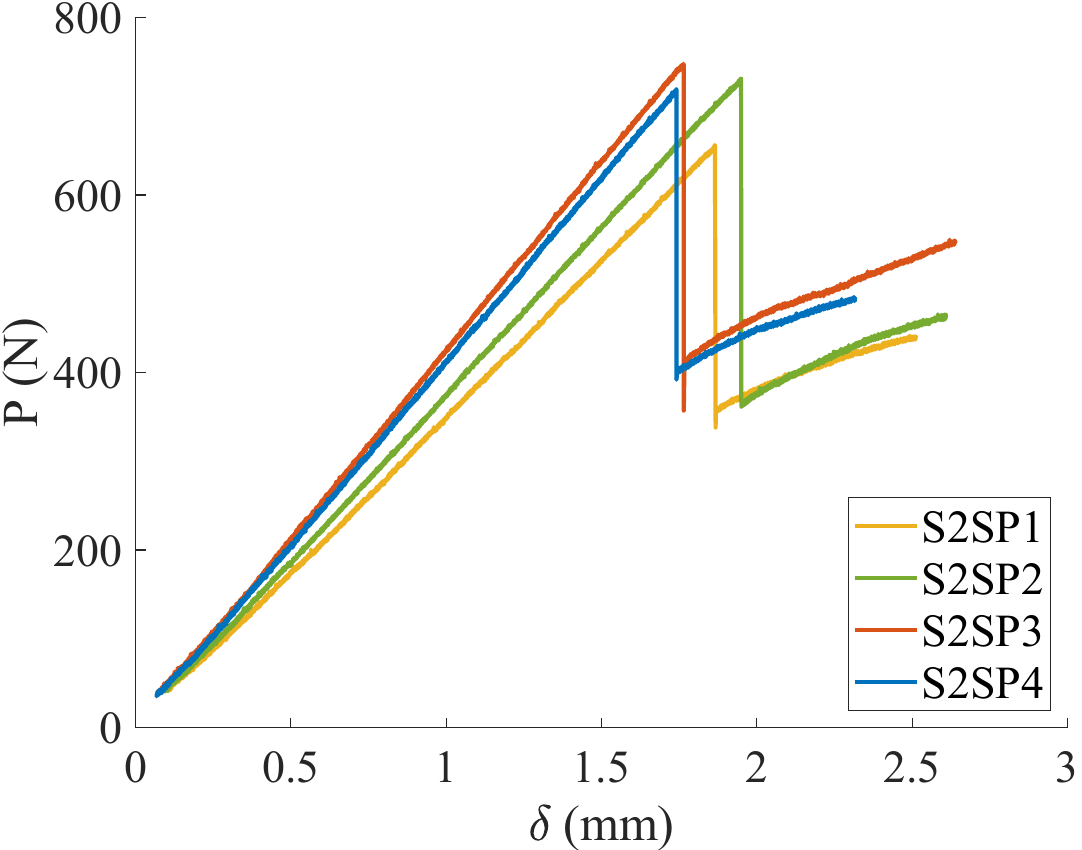}      
    }     
    \hfill
    \subfloat[\label{fig_loading_curves_c}]{%
      \includegraphics[width=0.48\textwidth]{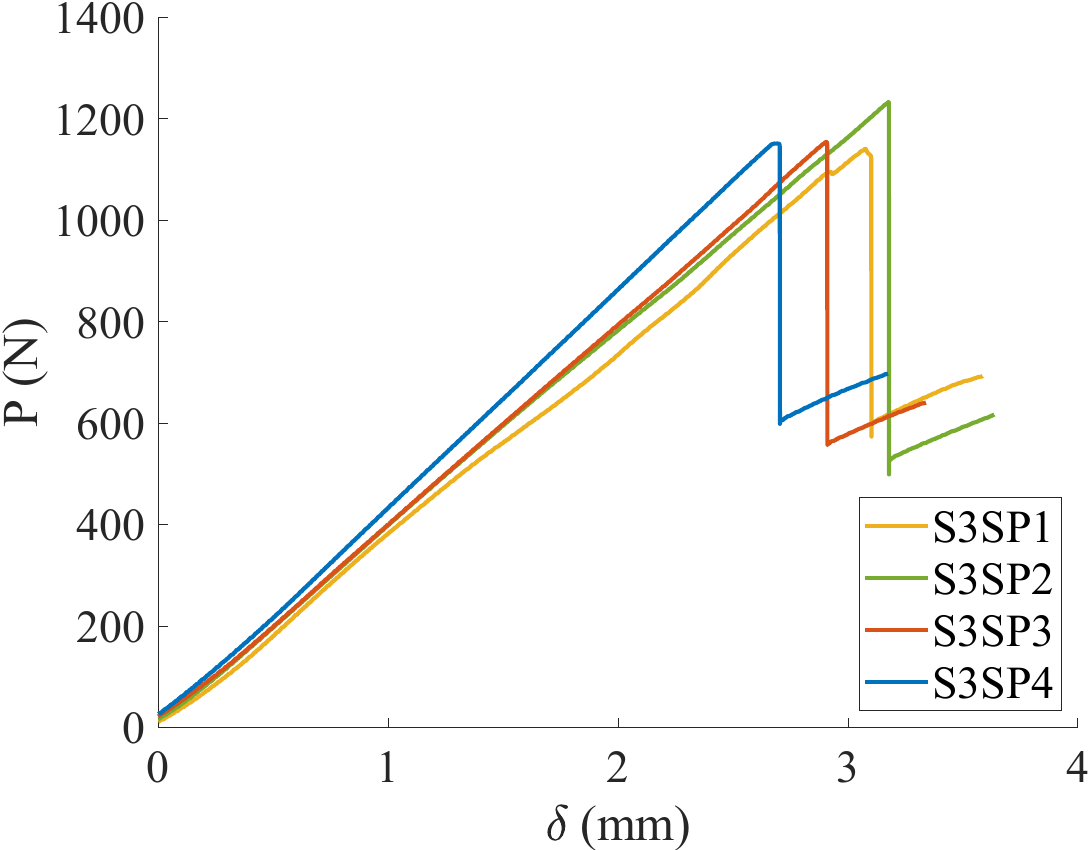}      
    }     
    \hfill
    \subfloat[\label{fig_loading_curves_d}]{%
      \includegraphics[width=0.48\textwidth]{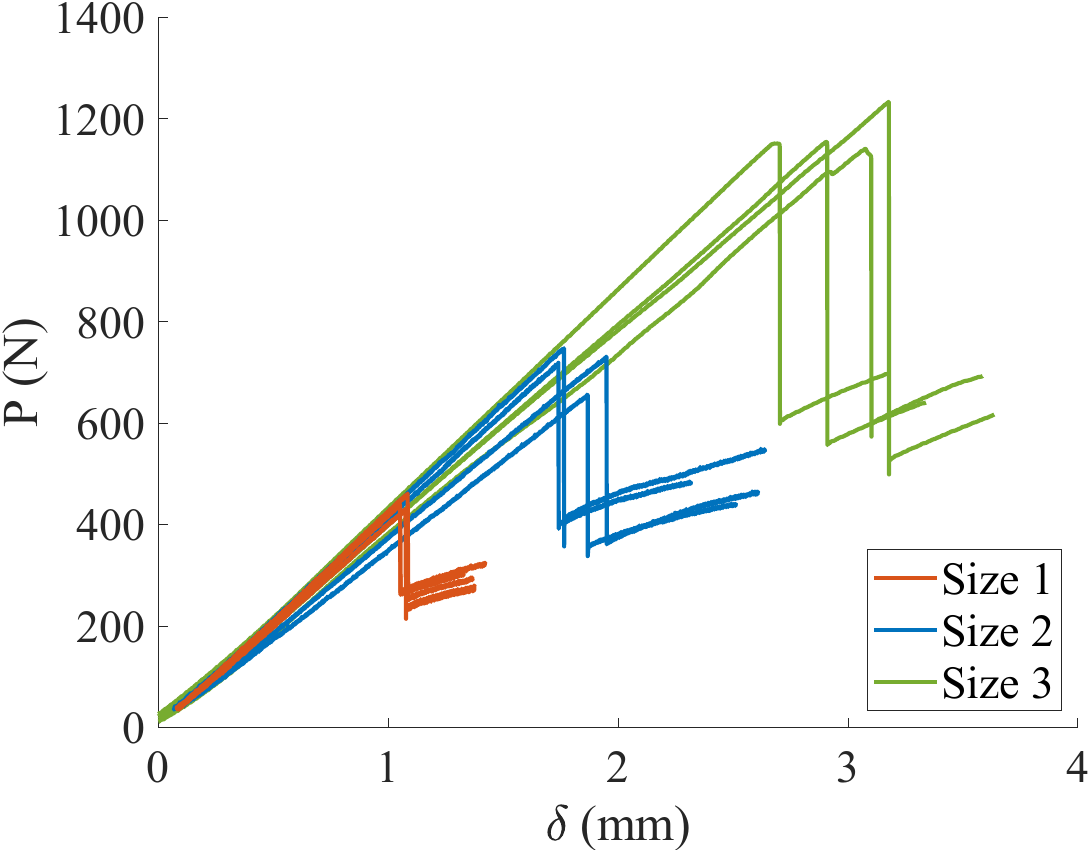}      
    }          
    \caption{Load-displacement analysis of the ENF tests. 
    (a) Size-1 data. (b) Size-2 data. (c) Size-3 data. (d) Collection of all test data.}
    \label{fig_loading_curves}
  \end{figure}

The test results are illustrated in the form of load-displacement ($P$-$\delta$) curves in \cref{fig_loading_curves}. 
The load-displacement data were intended to analyze the fracture energy of the specimen material and validate the global fracture behaviors simulated by the numerical models.
It needs to be noted that the horizontal testing configuration of the 2D microscopic DIC system (see \cref{fig_expsetup_a}) required small preloads (around $40$ N) to keep the specimens in the fixtures prior to the tests. 
Thus, the load-displacement curves of the Size-1 and Size-2 specimens (see \cref{fig_loading_curves_a,fig_loading_curves_b}) begin at the preloads.
In \cref{fig_loading_curves_d}, all the experimental data are plotted together to demonstrate the impact of geometric scaling on global fracture behaviors. 

The analysis results are tabulated in \Cref{tab_exp} for the peak load (or the fracture load) $P_\text{max}$, the effective flexural modulus $\bar{E}_\text{1f}$, and the LEFM mode-II interlaminar fracture energy $G_\text{f,LEFM}$.
The compliance calibration method described in the ASTM D7905/D7905M-19e1 testing manual was applied to characterize $G_\text{f,LEFM}$ and $\bar{E}_\text{1f}$.
The manual specifies that the maximum load $P_j$ ($j=1,2$) for compliance calibration should be set to achieve $15\%$ to $35\%$ of $G_\text{f,LEFM}$.
In this work, it was attempted to set $P_j$ to satisfy the minimum requirement (i.e., $15\%$ of $G_\text{f,LEFM}$) to keep the FPZs intact prior to actual fracture tests.
It needs to be noted that $\bar{E}_\text{1f}$ of the specimens were broadly consistent, and thus the inconsistency in the load-displacement curve slopes of the same-size specimens could be caused by some variances in the actual dimensions of the specimens.
The larger scaling factors led to higher $P_\text{max}$ and $G_\text{f,LEFM}$ than the smaller ones, while $P_\text{max}$ and $G_\text{f,LEFM}$ did not proportionally increase with an increase in the scaling factor. 

\subsubsection{Size effect analysis}\label{sec:exp_results_global_size_effect}

As Salviato et al. \cite{salviato2019} pointed out, $G_\text{f,LEFM}$ increased with the specimen size increase, and thus the LEFM-based compliance calibration method would not be suitable to characterize the fracture energy of the specimen material as a unique material property. 
To address this issue, Bažant's type-II size effect law \cite{bazant1984} was applied to the experimental data for fracture energy analysis.
Detailed descriptions of the application of the size effect law can be found in Refs. \cite{salviato2019,salviato2016}.
For completeness, some parts of these works are briefly introduced in this section. 
The mode-II interlaminar fracture energy $G^\text{(II)}$ and effective FPZ size $c_\text{f}$ of the material can be obtained by \cite{bazant1990,bazant1998-2}:
\begin{equation}\label{e_Gf}
G_\text{f} = \frac{g(\alpha_0)}{E^*A},~c_\text{f} = \frac{C}{A}\frac{g(\alpha_0)}{g'(\alpha_0)},
\end{equation}
where
$g$ is the dimensionless energy release rate, $\alpha=a/h$ is the normalized effective crack length, and $E^*$ is the equivalent elastic modulus.
Taking into account the unidirectional fiber directions of the specimens, the equivalent elastic modulus was established as $E^*=E_\text{1t}$. 
The effective flexural modulus values $\bar{E}_\text{1f}$ in \Cref{tab_exp} could also be used for $E^*$; however, the $\bar{E}_\text{1f}$ values overall showed good agreement with $E_\text{1t}$, and thus $E^*=E_\text{1t}$ was adopted for simplicity.
The constants $A$ and $C$ in \cref{e_Gf} can be obtained from a linear regression analysis:
$Y = A X + C$,
where $Y=\sigma_\text{Nc}^{-2}$ and $X=h$.
The nominal strength $\sigma_\text{Nc}$ is defined as $\sigma_\text{Nc}={P_\text{max}}/{bh}$.
Finally, the size effect plot can be given by
\begin{equation}\label{e_sigma}
\sigma_\text{Nc}=\frac{\sigma_0}{\sqrt{1+h/h_0}},
\end{equation}
where the pseudo-plastic limit $\sigma_0$ and the transitional characteristic length $h_0$ are given by
\begin{equation}\label{e_sigma0}
\sigma_0=\sqrt{\frac{G_\text{f}~ E^*}{c_\text{f}~ g'(\alpha_0)}},~
h_0=\frac{c_\text{f} ~ g'(\alpha_0)}{g(\alpha_0)}.
\end{equation}

\begin{figure}[t!]
\centering
    \subfloat[\label{fig_size_effect_a}]{%
      \includegraphics[width=0.48\textwidth]{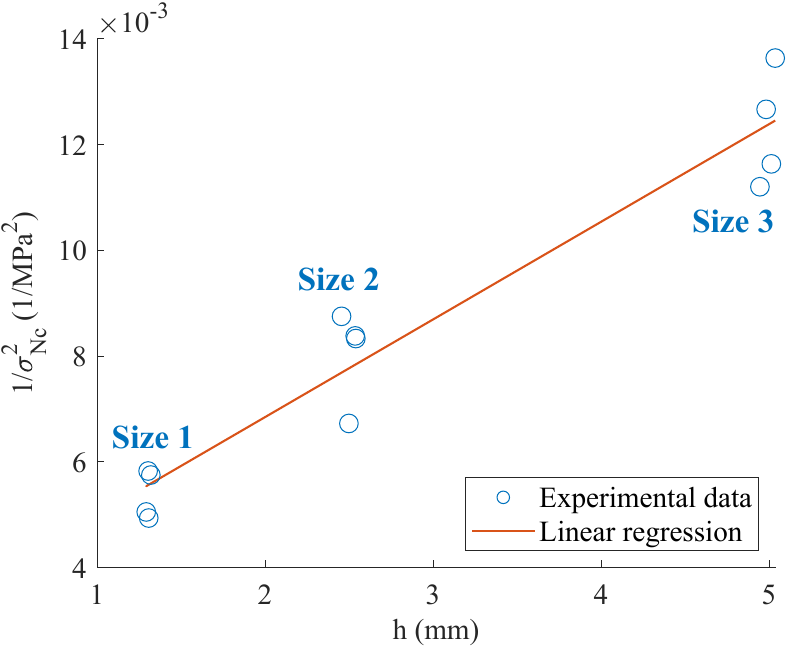}      
    } 
    \hfill
    \subfloat[\label{fig_size_effect_b}]{%
      \includegraphics[width=0.48\textwidth]{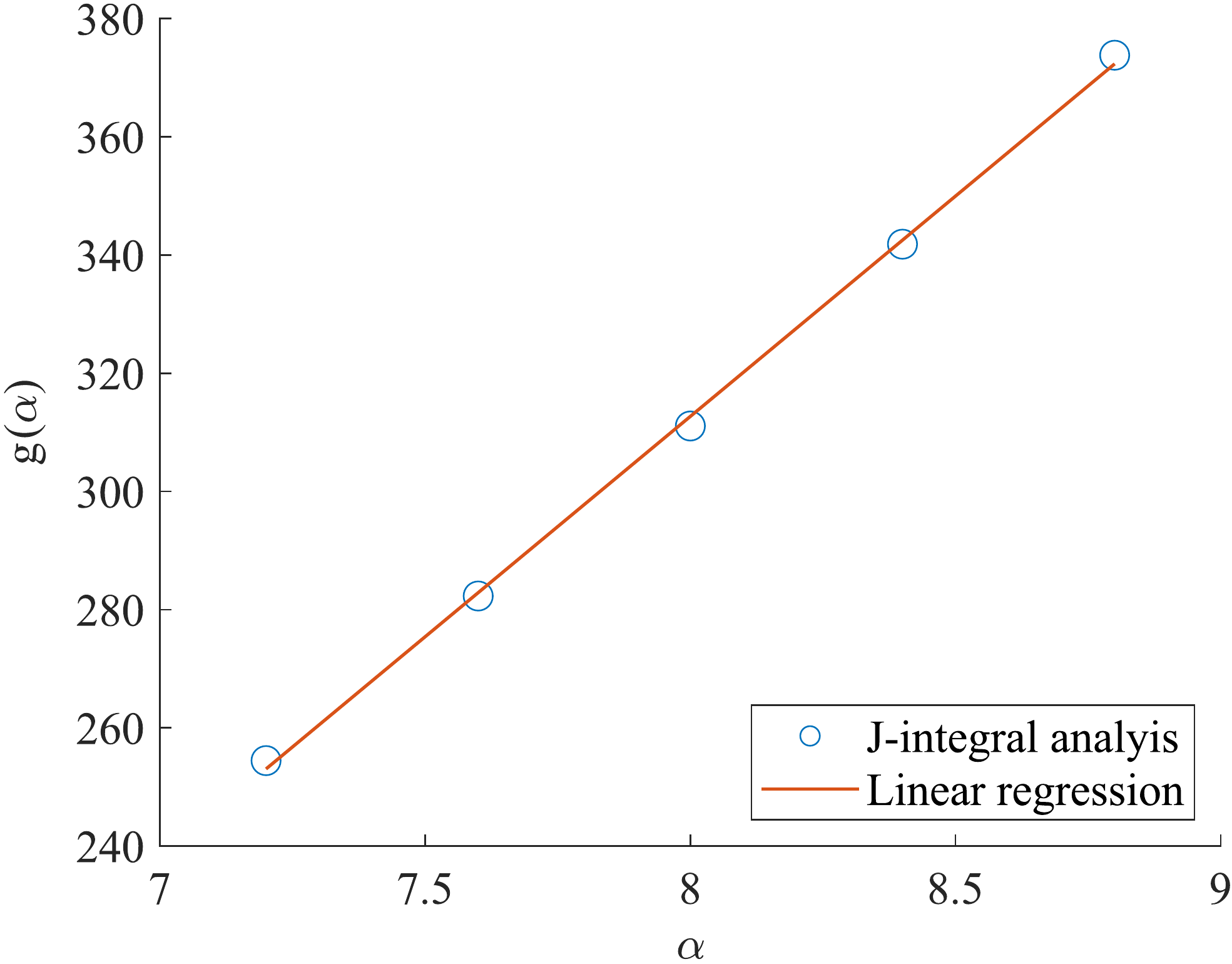}      
    }     
    \hfill
    \subfloat[\label{fig_size_effect_c}]{%
      \includegraphics[width=0.8\textwidth]{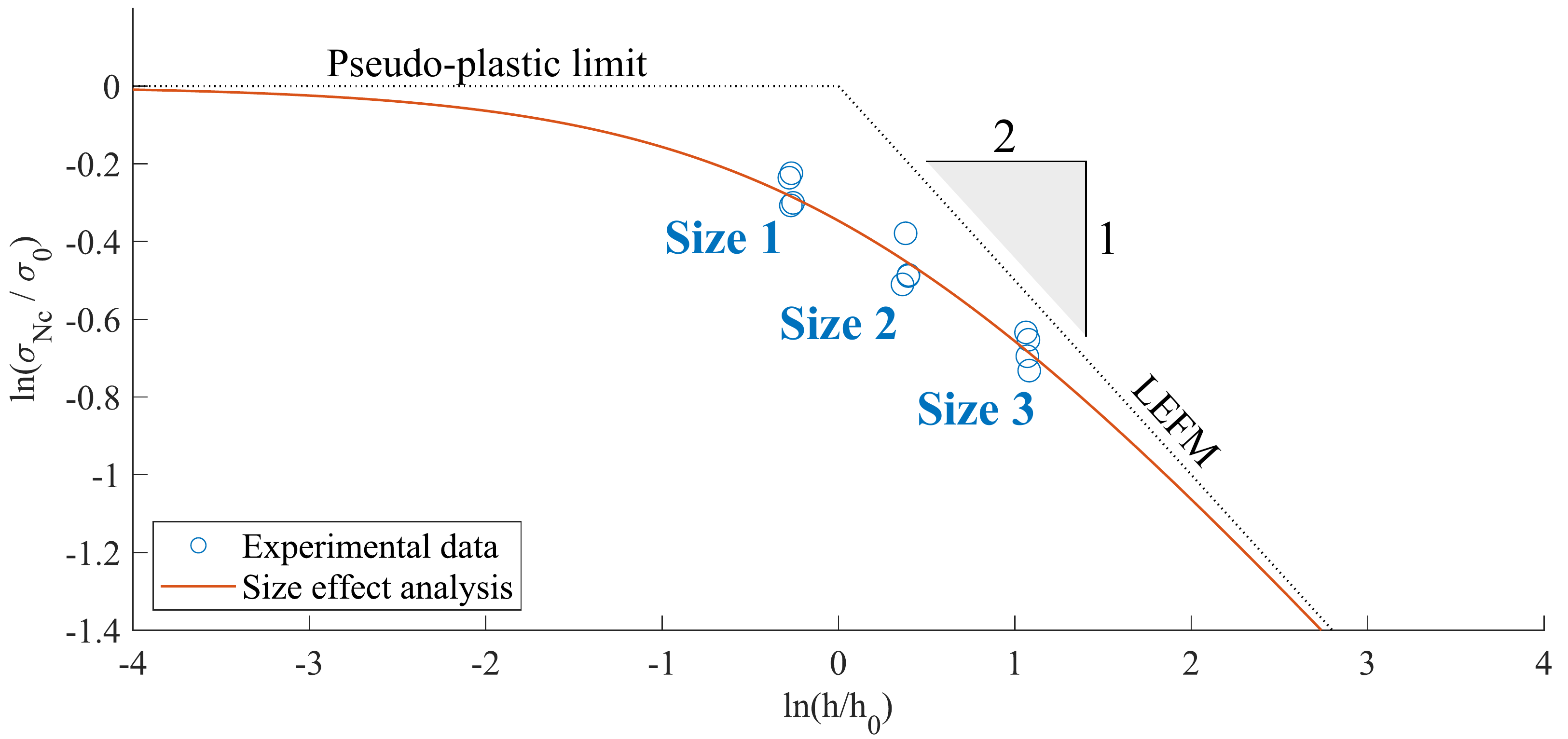}      
    }        
    \caption{Size effect analysis of the experimental data. (a) Linear regression analysis of the experimental data. (b) J-integral simulation results for $g$ and $g'$. (c) Size effect plot.}
    \label{fig_size_effect}
  \end{figure}

The size effect analysis result is presented in \cref{fig_size_effect}.
The linear regression result in \cref{fig_size_effect_a} showed $A=1.847\times 10^{-3}~ \text{mm}^3/\text{N}^2$ and $C=3.155\times 10^{-3}~ \text{mm}^4/\text{N}^2$.
The dimensionless values $g$ and $g'$ were obtained from the J-integral analysis using the Abaqus 2024/Standard solver \cite{abaqus2024}. 
The nominal dimensions of the Size-2 specimen (see \Cref{tab_dimension}) were used for modeling, while additional simulation data were obtained by changing the initial crack length $a_0$ by $\pm1$ and $\pm2$ mm. 
The modeling results are illustrated in \cref{fig_size_effect_b} with $g(\alpha_0)=311.1$ and $\alpha_0=8$.
The dimensionless value $g'(\alpha_0)$ was obtained as $74.56$ from the linear regression analysis of the modeling data.
Substituting these values into \cref{e_Gf}, the fracture energy and effective FPZ size of the specimen material were obtained as $G_\text{f}=1.347$ N/mm and $c_\text{f}=7.126$ mm, respectively.
Additionally, the pseudo-plastic limit and the transitional characteristic length were obtained as $\sigma_0=17.80 \text{ MPa}$ and $h_0=1.708 \text{ mm}$, respectively.
Finally, the size effect plot was drawn based on \cref{e_sigma} as shown in \cref{fig_size_effect_c}.
The plot shows that the geometrically scaled specimens well captured both the pseudo-plastic limit and the LEFM regime.
To be specific, the Size-3 specimens were large enough to be in close proximity to the LEFM regime, while the thickness dimensions of the Size-1 set were scaled down enough to be smaller than $h_0$.

\subsection{Experimental characterization of local fracture behaviors}\label{sec:exp_results_local}

The DIC analysis of the experimental data was performed to capture the local fracture process along the FPZs in front of the initial crack tips. 
The DIC data were collected from all specimens; however, a single case was selected from each size set and was focused on for local fracture analysis and modeling in this paper. 
To be specific, the S1SP4, S2SP4, and S3SP3 data sets (see \Cref{tab_exp} for the specimen details) were chosen. 
These selections were made by evaluating discrepancies between the nominal and actual dimensions and consistencies in the compliance calibration data and DIC analysis results. 
As described in \Cref{sec:exp_work_setup}, the S1SP4 and S2SP4 specimens were tested using the 2D microscopic DIC system, while the 3D macroscopic DIC system was used for the S3SP3 specimen. 

\begin{figure}[t!]
\centering
    \subfloat[\label{fig_field_of_view_a}]{%
      \includegraphics[width=0.9\textwidth]{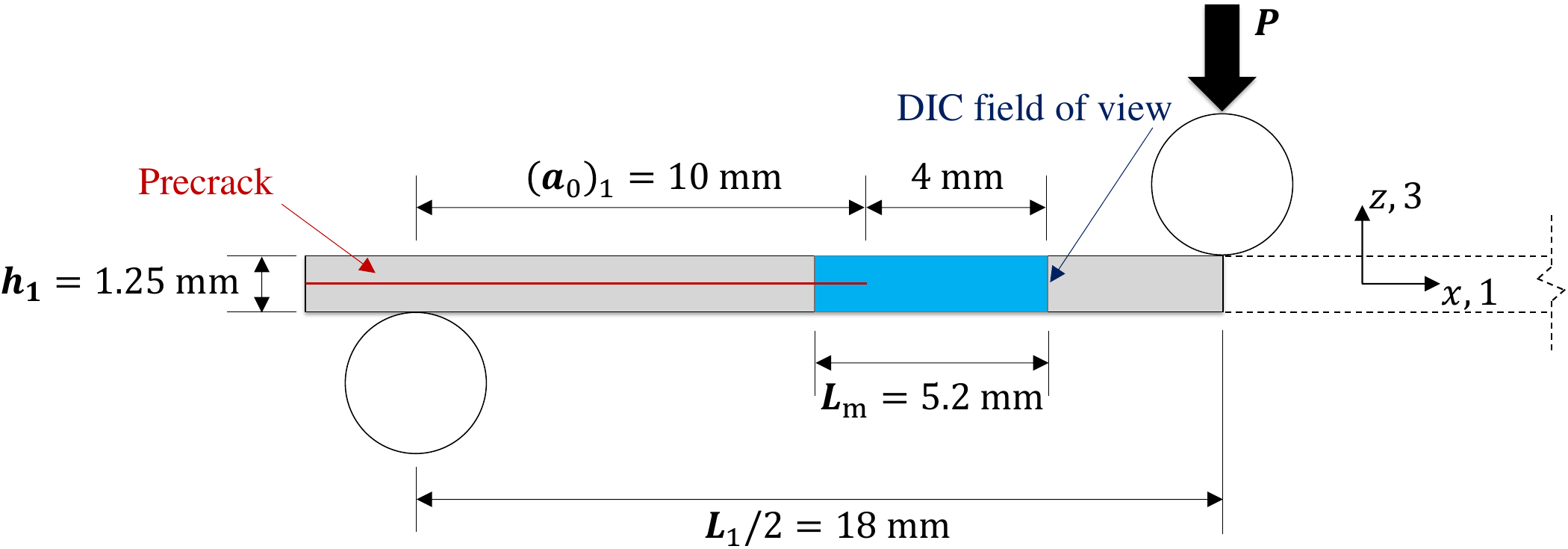}      
    } 
    \hfill
    \subfloat[\label{fig_field_of_view_b}]{%
      \includegraphics[width=0.9\textwidth]{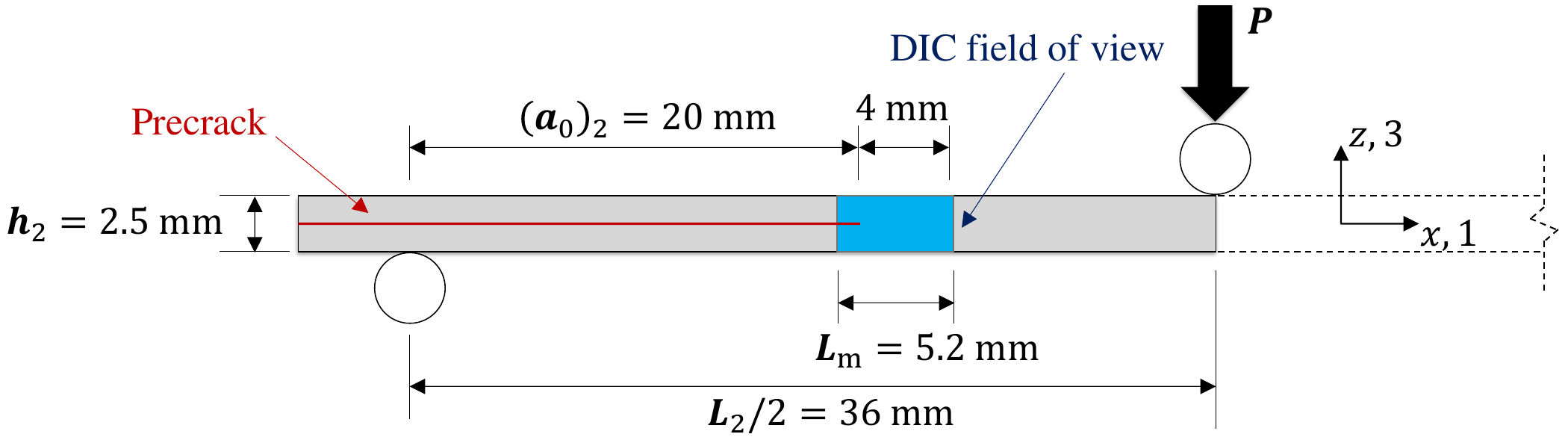}      
    }              
    \caption{Schematic drawings for the field of view of the microscopic DIC system. The field length is represented by $L_\text{m}$.
    (a) A half of the nominal Size-1 specimen. 
    (b) A half of the nominal Size-2 specimen.}
    \label{fig_field_of_view}
  \end{figure}

The 2D microscopic DIC system was focused on the initial crack tips and predicted crack path (i.e., the potential FPZ).
The field of view of the microscopic DIC system is illustrated in \cref{fig_field_of_view}.
The field length $L_\text{m}$ includes $1.2$ mm of the precrack and $4$ mm of the potential FPZ. 
The resolution of the digital images captured from the microscope was 2448$\times$2048 pixels.
For 2D DIC, these images were processed using a Correlated Solutions VIC-2D 2009 package \cite{vic2d2009} with a subset size \cite{sutton2009} of 51$\times$51 pixels and a step size of 5$\times$5 pixels.
The subset shape function used was an optimized 6-tap spline with normalized squared differences and Gaussian subset weights.
The calibration scale was between $367$ and $383$ pixels/mm.
For strain computation, the Lagrange strain formulation was used with a strain window of 15 data points.  
The 3D macroscopic DIC system, on the other hand, covered the entire surface of the S3SP3 specimen on the $x$--$z$ plane. 
The resolution of the digital images captured from the two cameras was 2448$\times$2048 pixels.
For 3D DIC, these images were processed using a Correlated Solutions VIC-3D 7 package \cite{vic3d7} with a subset size of 35$\times$35 pixels and a step size of 7$\times $7 pixels.
The calibration scale was between $14.3$ and $15.0$ pixels/mm.
The subset shape function and strain computation method adopted for 3D DIC were identical to the 2D DIC analysis. 

The local fracture analysis was made by processing the DIC data with three steps: coordinate transformation, curve fitting, and through-thickness deformation analysis.
The details of the three steps are described in the following sections. 

\subsubsection{Step 1: Coordinate-transformation process}\label{sec:exp_coord_transform}

\begin{figure}[t!]
\centering
    \subfloat[\label{fig_axis_rotation_a}]{%
      \includegraphics[width=0.48\textwidth]{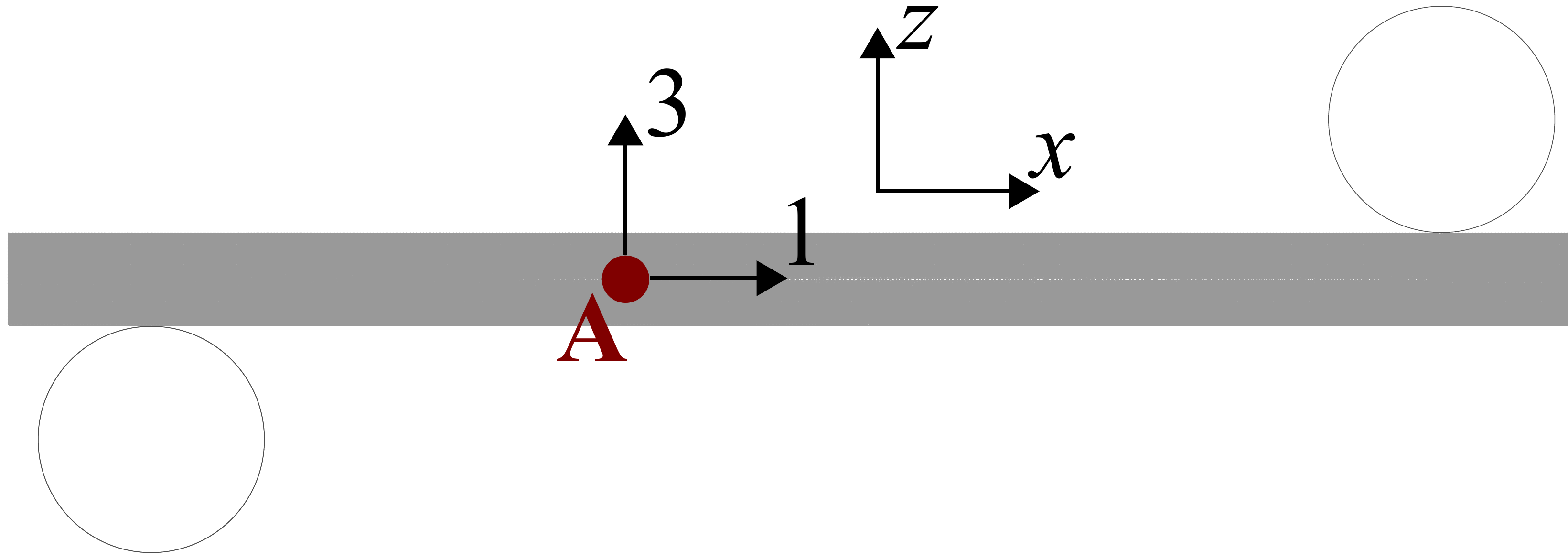}      
    } 
    \hfill
    \subfloat[\label{fig_axis_rotation_b}]{%
      \includegraphics[width=0.48\textwidth]{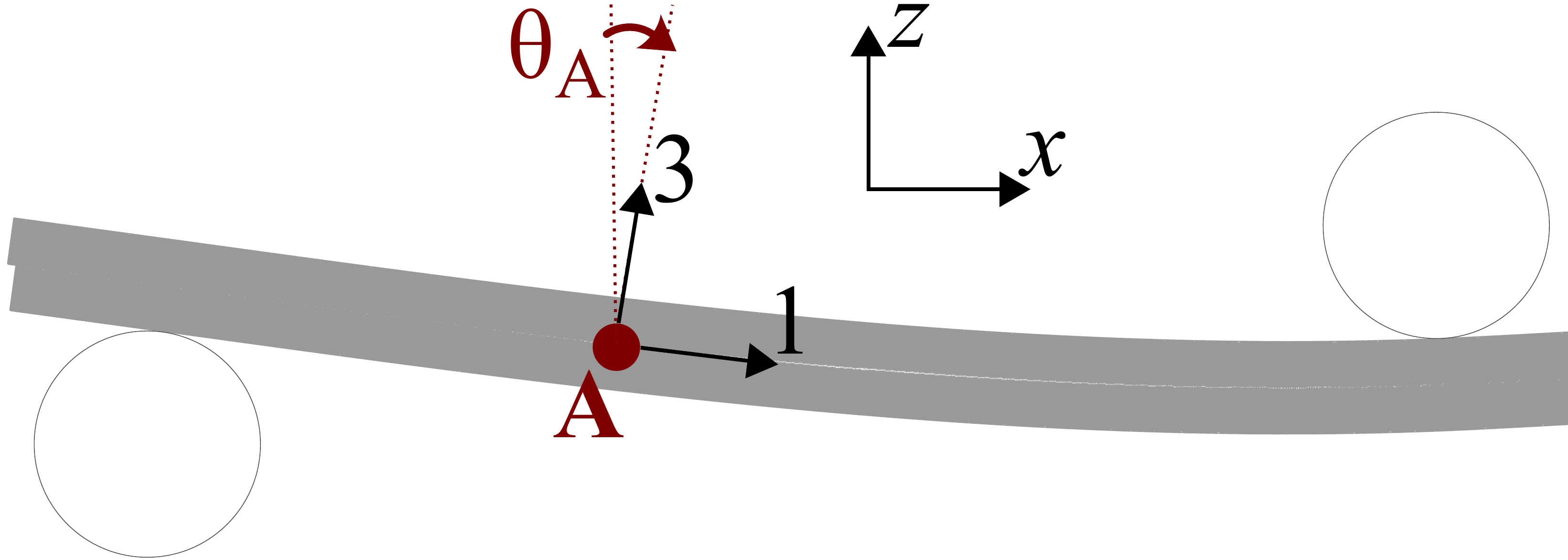}      
    }              
    \caption{Schematic drawings for beam bending and axis rotation.
    For a better visualization of axis rotation, only the left half of an ENF specimen is shown here. 
    (a) The undeformed beam prior to loading.  
    (b) The deformed beam under three-point bending.
    The $1$-$3$ axes at point $A$ were rotated by angle $\theta_A$.
    }
    \label{fig_axis_rotation}
  \end{figure}

Axis rotation from beam bending is illustrated in \cref{fig_axis_rotation}.
The principal material axes (or the local coordinate system) were aligned with the geometric axes (or the global coordinate system) prior to loading (See \cref{fig_axis_rotation_a}).
The principal material axes $1$ and $3$ rotated about the $2$ axis with specimen bending, while the geometric axes were fixed (See \cref{fig_axis_rotation_b}).
The raw VIC-2D data were obtained along the geometric axes; for example, displacements were measured in the axes $x$ and $z$ and were named $u_\text{x}$ and $u_\text{z}$, respectively.
Rotation angles varied with the locations of the DIC data points.

\begin{figure}[t!]
\centering
    \subfloat[\label{fig_coord_transform_a}]{%
      \includegraphics[width=0.92\textwidth]{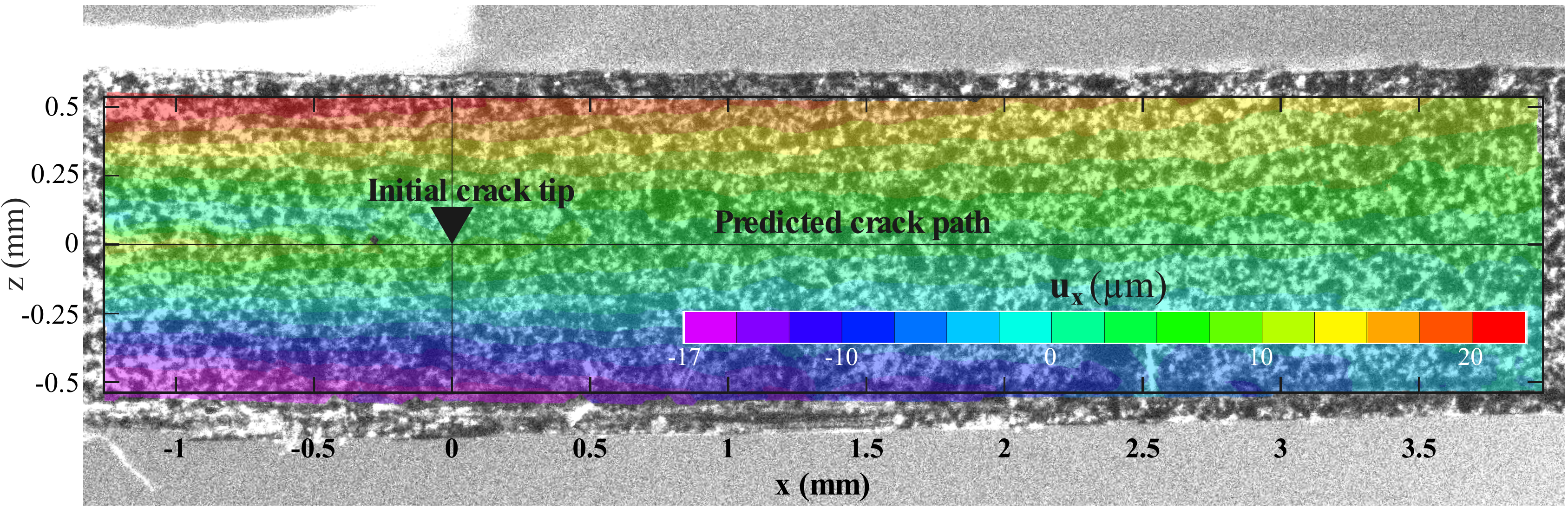}      
    } 
    \hfill
    \subfloat[\label{fig_coord_transform_b}]{%
      \includegraphics[width=0.9\textwidth]{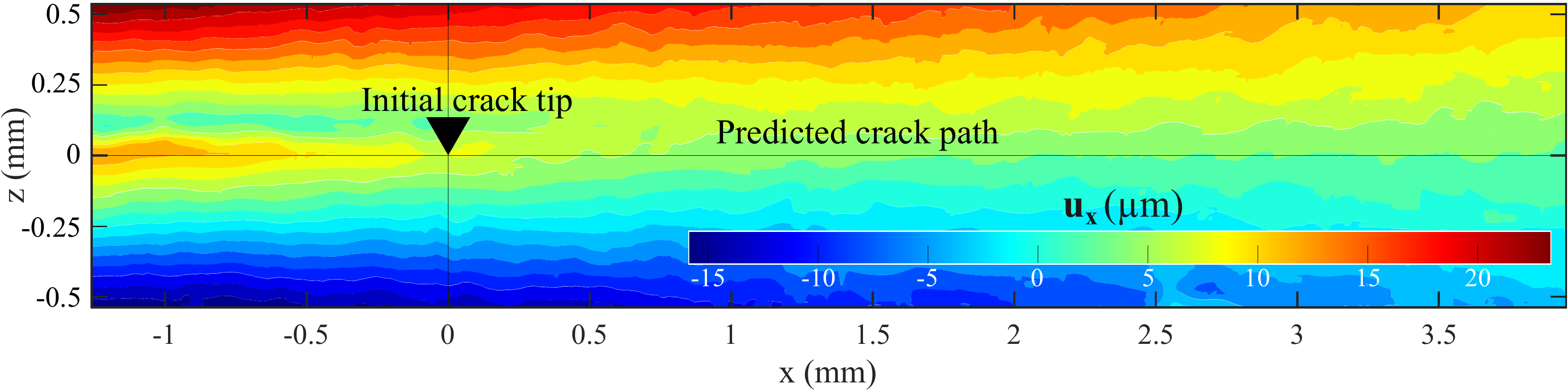}      
    }     
    \hfill
    \subfloat[\label{fig_coord_transform_c}]{%
      \includegraphics[width=0.9\textwidth]{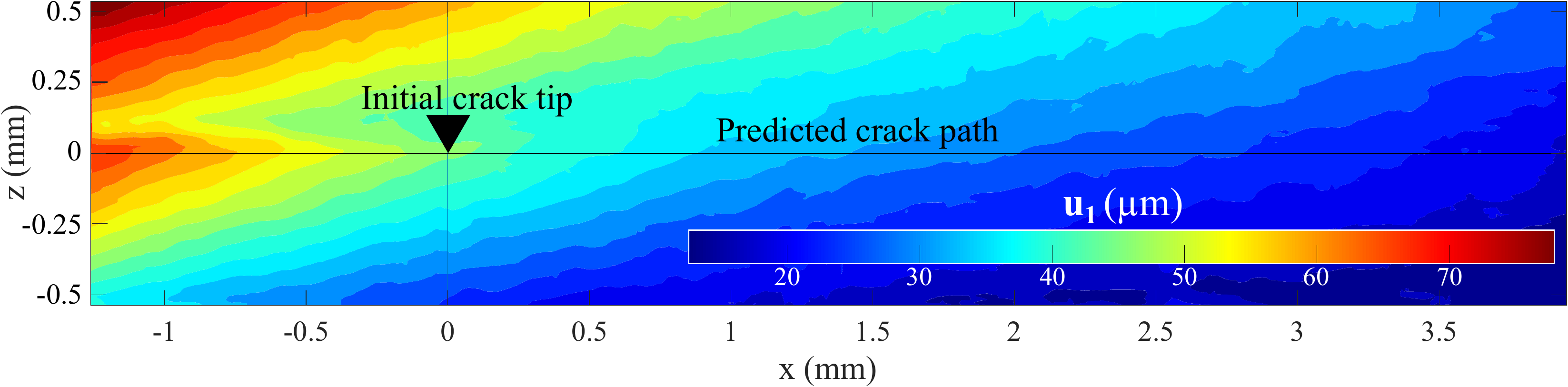}      
    }                 
    \caption{The contours of the raw and post-processed DIC data from the S1SP4 specimen at $P_\text{max}$. 
    (a) The raw DIC data of $u_x$.
    (b) The interpolated data of $u_x$.
    (c) The transformed data of $u_1$.}
    \label{fig_coord_transform}
  \end{figure}

Given that mode-II interlaminar fracture is expected to occur between the fibers (i.e., along the $1$ axis), a coordinate transformation of the raw data from the $x$-$z$ axes to the $1$-$3$ axes was required.
The coordinate transformation steps are illustrated in \cref{fig_coord_transform} to obtain the in-plane displacement $u_1$ along the $1$ axis of the S1SP4 specimen at $P_\text{max}$.
The contours in the figures were drawn on the undeformed specimen geometry.
In \cref{fig_coord_transform_a}, the contours of the raw data $u_x$ are superimposed on the corresponding digital image of the speckled S1SP4 specimen.
The raw data were initially interpolated to have additional data points between the raw data using the \textit{griddata} function of Matlab R2021a \cite{matlab} (see \cref{fig_coord_transform_b}).           
The rotation angle of each data point was calculated based on a beam theory.
Finally, the contours of $u_1$ were obtained by transforming the $u_x$ data from the $x$--$z$ coordinate system to the $1$--$3$ coordinate system (see \cref{fig_coord_transform_c}).
The raw DIC data of $u_z$, $\varepsilon_{x}$, $\varepsilon_{z}$, and $\varepsilon_{xz}$ were also transformed to $u_3$, $\varepsilon_{1}$, $\varepsilon_{3}$, and $\varepsilon_{13}$, respectively, using the same steps. 

\subsubsection{Step 2: Curve-fitting process}\label{sec:exp_curve_fitting}

\begin{figure}[t!]
\centering
    \subfloat[\label{fig_curve_fitting_a}]{%
      \includegraphics[width=0.47\textwidth]{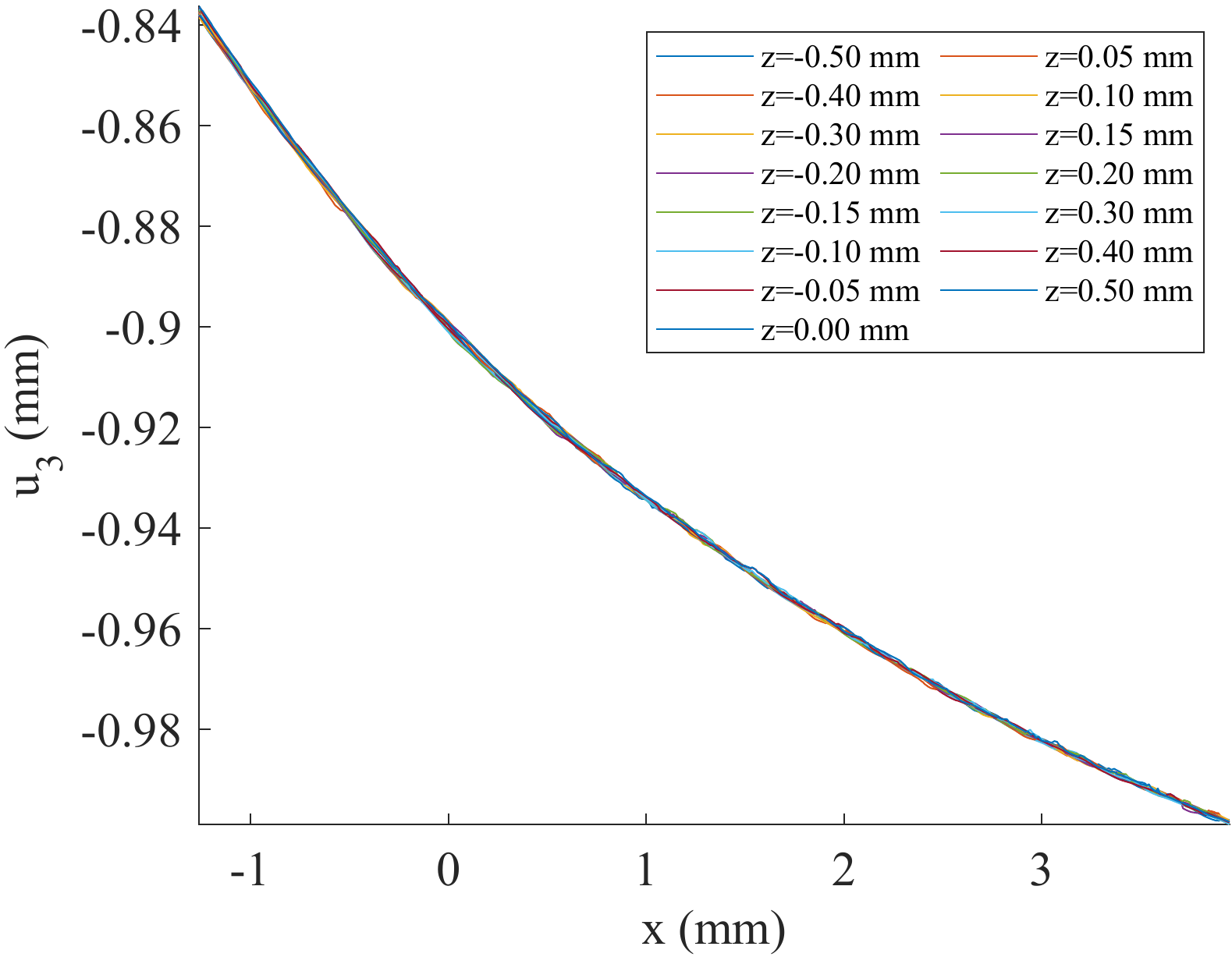}      
    }    
        \hfill
    \subfloat[\label{fig_curve_fitting_b}]{%
      \includegraphics[width=0.47\textwidth]{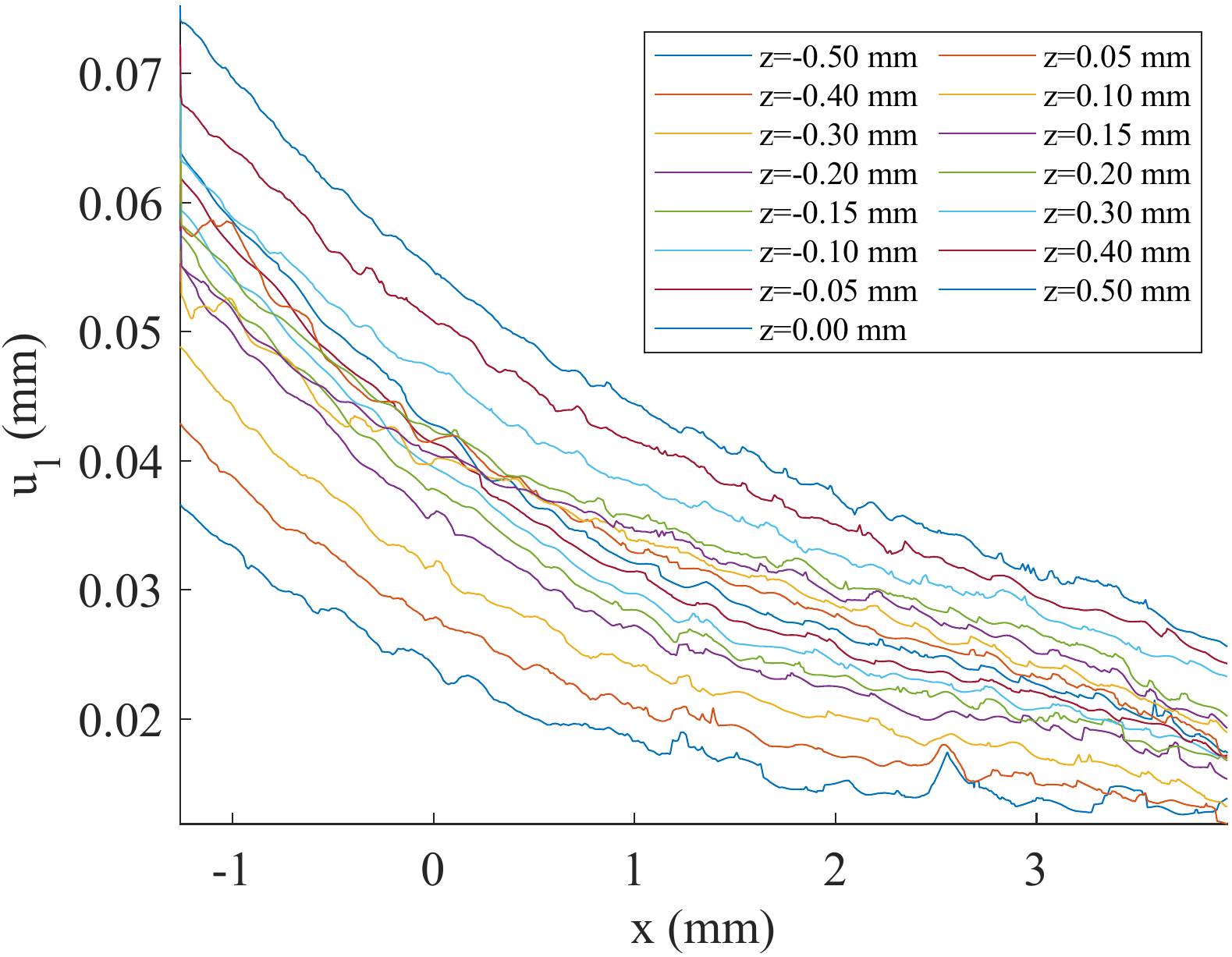}      
    }    
        \hfill
    \subfloat[\label{fig_curve_fitting_c}]{%
      \includegraphics[width=0.47\textwidth]{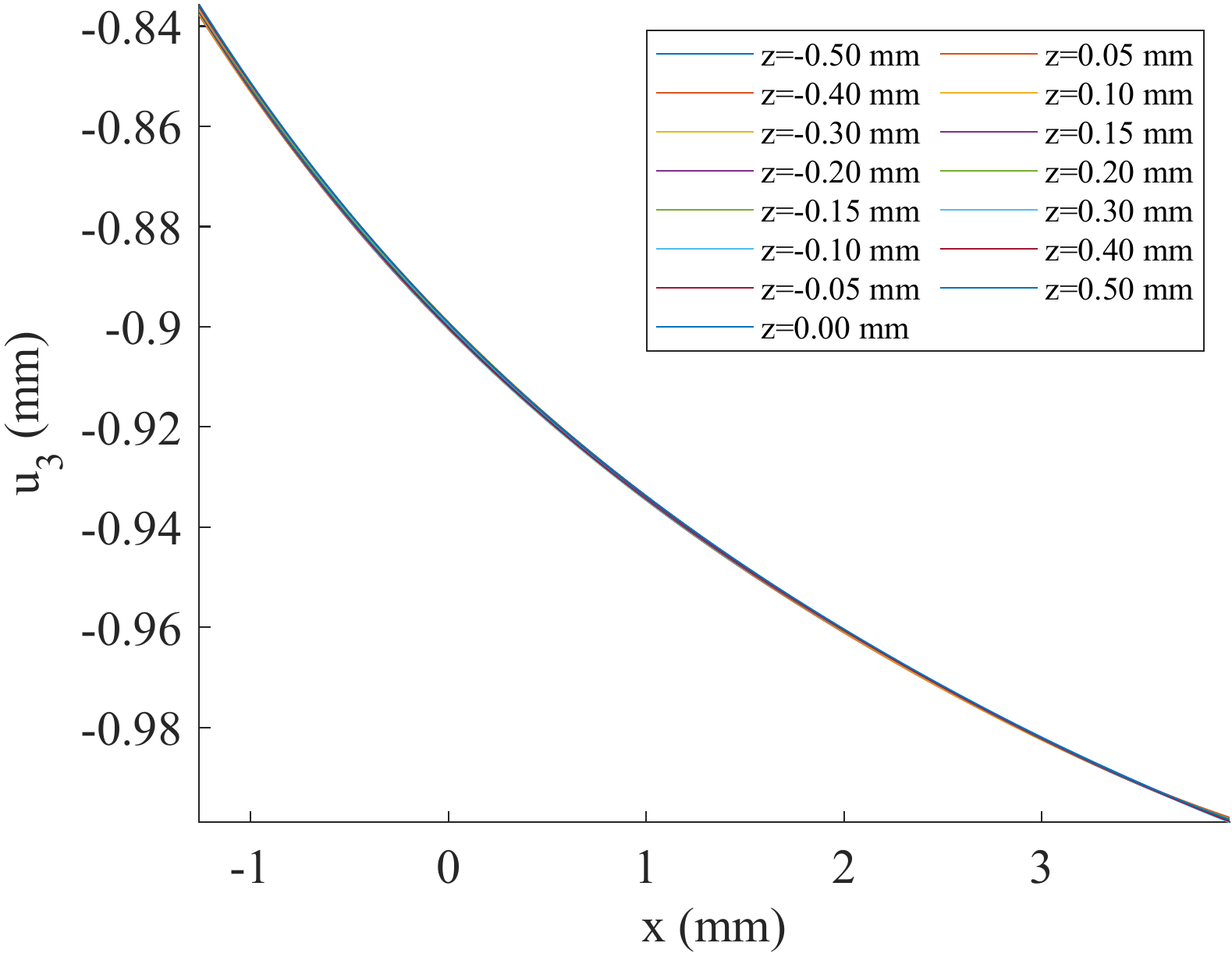}      
    }      
        \hfill
    \subfloat[\label{fig_curve_fitting_d}]{%
      \includegraphics[width=0.47\textwidth]{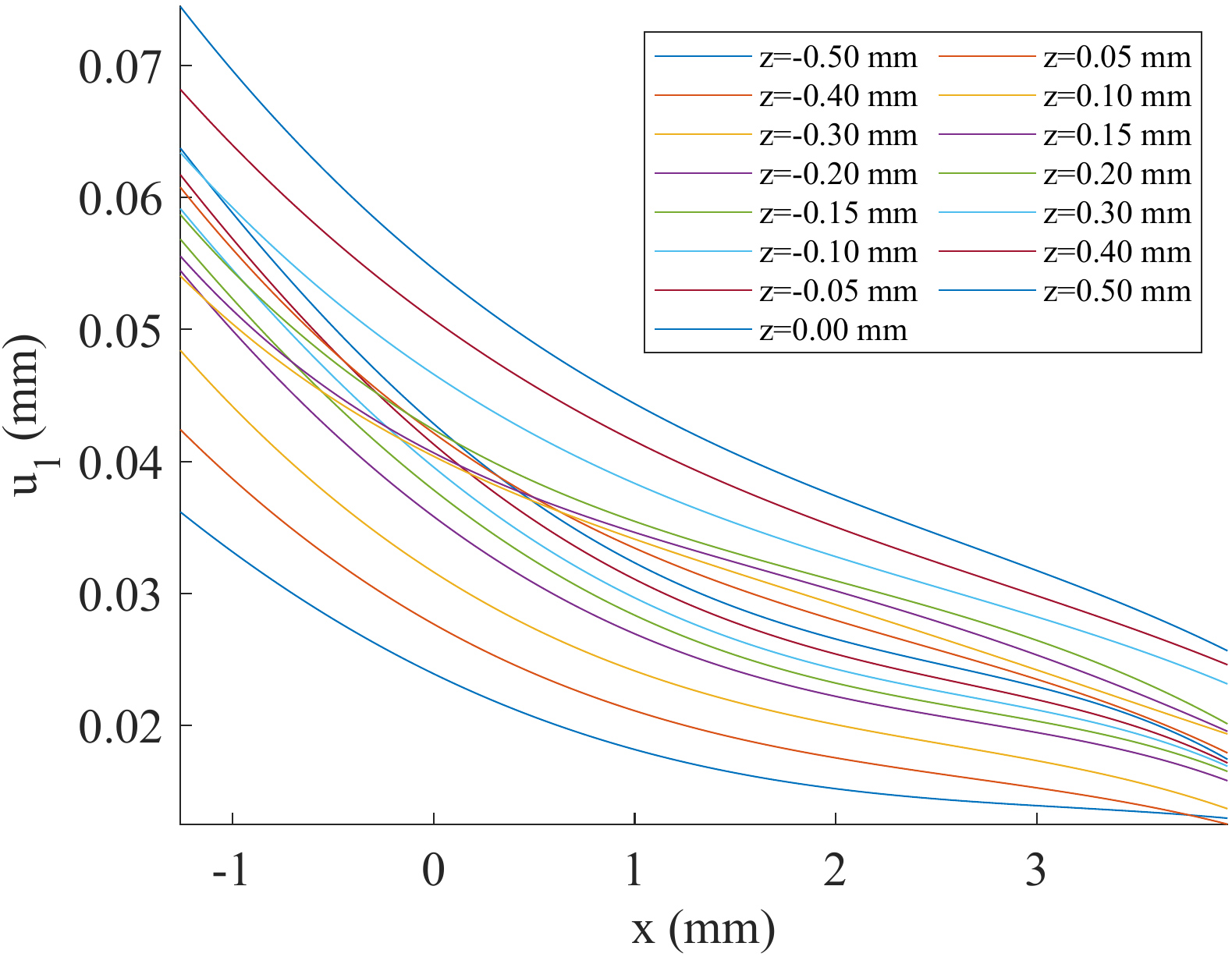}      
    }       
    \caption{Curve-fitting process of the transformed DIC data from the S1SP4 specimen at $P_\text{max}$. 
    (a) The transformed DIC data of $u_3$.
    (b) The transformed DIC data of $u_1$.
    (c) The curve-fitted DIC data of $u_3$.
    (d) The curve-fitted DIC data of $u_1$.}
    \label{fig_curve_fitting}
  \end{figure}

The transformed DIC data sets were post-processed one more time through a curve-fitting process. 
The curve-fitting process for the $u_1$ and $u_3$ data from the S1SP4 specimen at $P_\text{max}$ is presented in \cref{fig_curve_fitting}.
The data sets were taken between $z=-0.5$ and $0.5$ mm, and a smaller interval was made near the potential FPZs at $z=0$.
The $u_3$ curves are all well aligned and thus appear almost as a single curve in the figures.
The transformed data $u_1$ (see \cref{fig_curve_fitting_a}) showed significantly larger noises in the curves than $u_3$ (see \cref{fig_curve_fitting_b}) due to the smaller magnitude of $u_1$.
The curve-fitting process was made using the \textit{polyfit} function of Matlab R2021a.
Fourth-order polynomials were employed based on a beam theory. 
As shown in \cref{fig_curve_fitting_c,fig_curve_fitting_d}, the curve-fitting process was more beneficial to the $u_1$ data than $u_3$.
The curve-fitting process was also made to the data of $\varepsilon_{1}$, $\varepsilon_{3}$, and $\varepsilon_{13}$.

\begin{figure}[t!]
\centering
    \subfloat[\label{fig_contour_size1_a}]{%
      \includegraphics[width=0.9\textwidth]{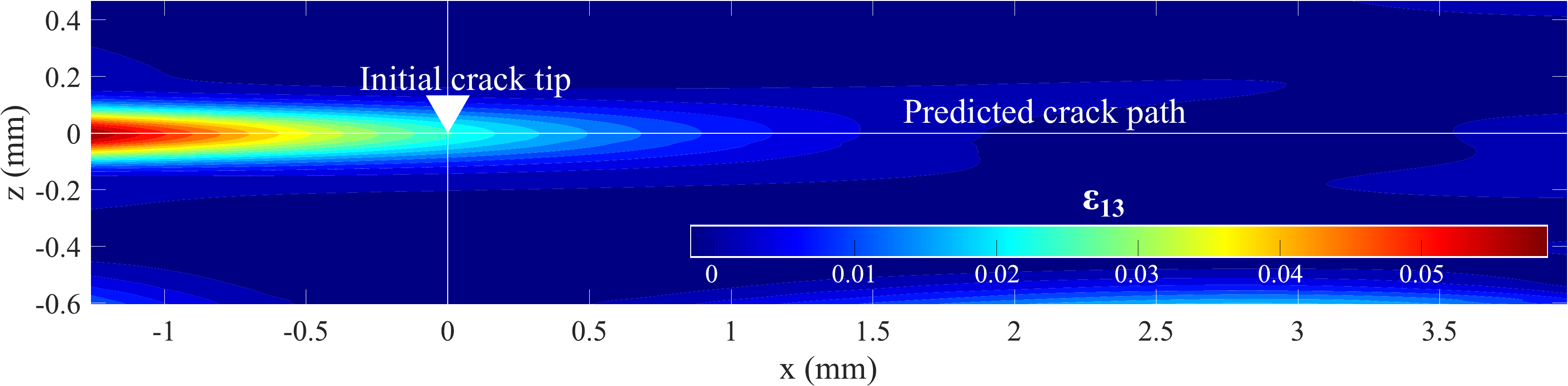}      
    }     
    \hfill
    \subfloat[\label{fig_contour_size1_b}]{%
      \includegraphics[width=0.9\textwidth]{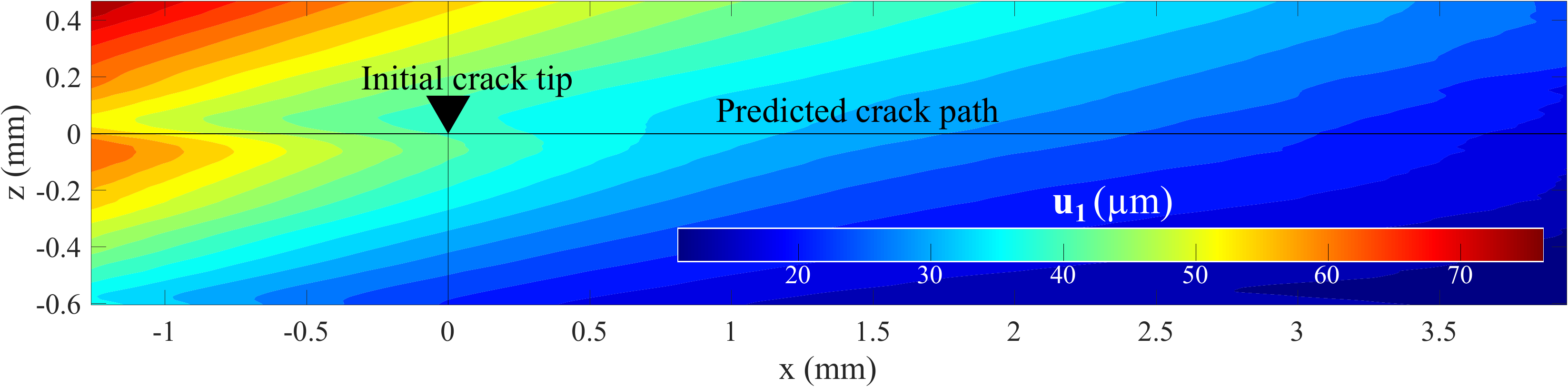}      
    }              
    \caption{The contours of the curve-fitted microscopic DIC data from the S1SP4 specimen at $P_\text{max}$. 
    The contours are presented in the vicinity of the initial crack tip on the undeformed geometry.
    (a) Shear strains $\varepsilon_\text{13}$. 
    (b) In-plane displacements $u_1$.}
    \label{fig_contour_size1}
  \end{figure}

\begin{figure}[t!]
\centering
      \includegraphics[width=0.9\textwidth]{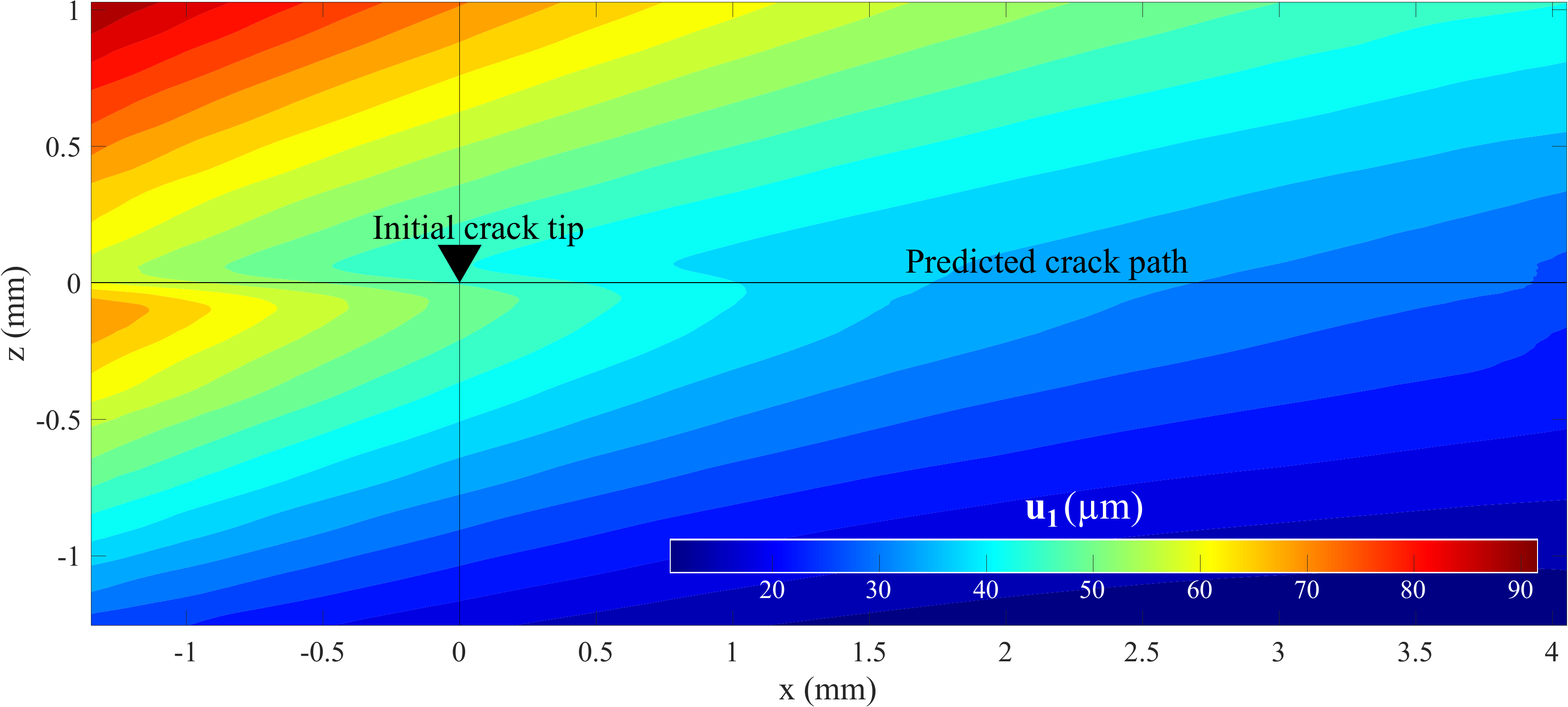}      
    \caption{The contours of the curve-fitted microscopic DIC data of $u_1$ from the S2SP4 specimen at $P_\text{max}$. 
    The contours are presented in the vicinity of the initial crack tip on the undeformed geometry.}
    \label{fig_contour_size2}
  \end{figure}

The curve-fitted microscopic DIC data from the S1SP4 and S2SP4 specimens at $P_\text{max}$ are presented in the form of contour plots in \cref{fig_contour_size1,fig_contour_size2}, respectively.
The contour plots of the S2SP4 specimen have the same length $L_\text{m}$ as the S1SP4 plots but cover larger surface areas due to the scaled-up thickness (See \cref{fig_field_of_view}).
It needs to be noted that the top and bottom boundary areas of the specimens were unevenly cut for the DIC analysis; as a result, the precracks and the predicted crack paths were not positioned in the middle of the contour drawings. 
Monitoring the development of shear strain $\varepsilon_\text{13}$ in the vicinity of the initial crack tip (see \cref{fig_contour_size1_a}) was helpful in validating the position of the predicted crack path at $z=0$.
Large shear strains were observed in the precrack region at $x\leq0$ and propagated along the predicted crack path. 
At the same time, the strongly zigzagging contours of in-plane displacement $u_1$ were formed along the precracks of the S1SP4 and S2SP4 specimens (see \cref{fig_contour_size1_b} and \cref{fig_contour_size2}, respectively) implying separation across the precracks. 
Weaker zigzags of the $u_1$ contours were observed near the initial crack tip and along the predicted crack path, while nearly linear contours (i.e., no separations) were formed away from the initial crack tip.  

\begin{figure}[t!]
\centering
      \includegraphics[width=.9\textwidth]{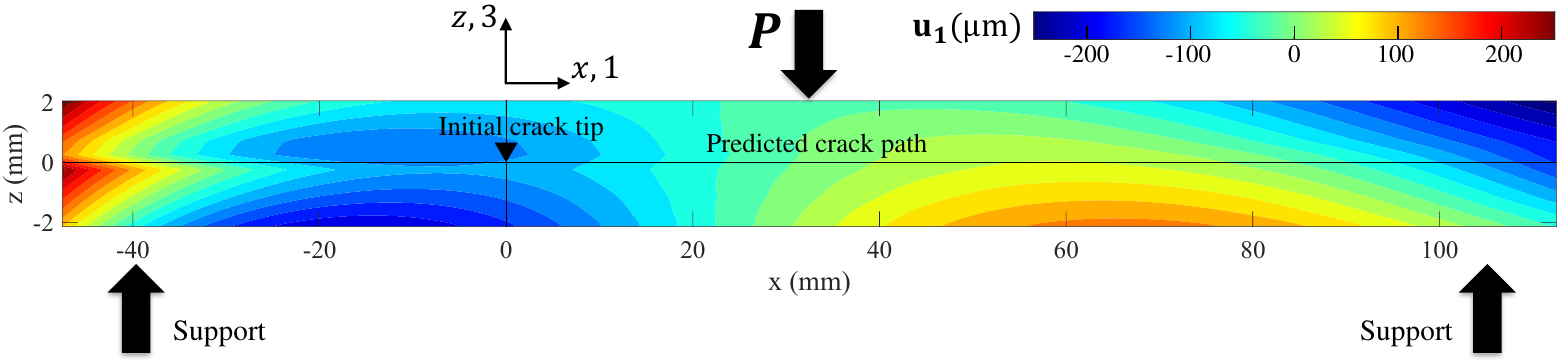}    
    \caption{
    The contours of the curve-fitted macroscopic DIC data of $u_1$ from the S3SP3 specimen at $P_\text{max}$.
    The contours are presented on the undeformed geometry.
    For better visualization, the $z$ axis is exaggerated.
    }
    \label{fig_contour_size3}
  \end{figure}

The macroscopic DIC data were also post-processed through the coordinate-transformation and curve-fitting steps as shown in \cref{fig_contour_size3}.
In contrast to the microscopic DIC data, the macroscopic DIC data covered the entire specimen surface.
As a result, the level of detail shown in the macroscopic DIC data was significantly lower than the microscopic data as expected. 
Similar to the microscopic data, the strongly zigzagging contours of $u_1$ were observed along the precrack and near the initial crack tip.

\subsubsection{Step 3: Through-thickness deformation analysis}\label{sec:exp_through_thickness}

For a detailed analysis of local fracture behaviors, this work proposes a through-thickness analysis of in-plane displacements as the final step. 
In this method, the DIC data sets were further analyzed in the form of $u_1$ variation through the thickness (i.e., the $z$ axis).
This analysis was intended to characterize the development of separation during the mode-II interlaminar fracture process. 
The implementation of the proposed method is introduced in this section.

\begin{figure}[t!]
\centering
    \subfloat[\label{fig_du_size1_a}]{%
      \includegraphics[width=0.48\textwidth]{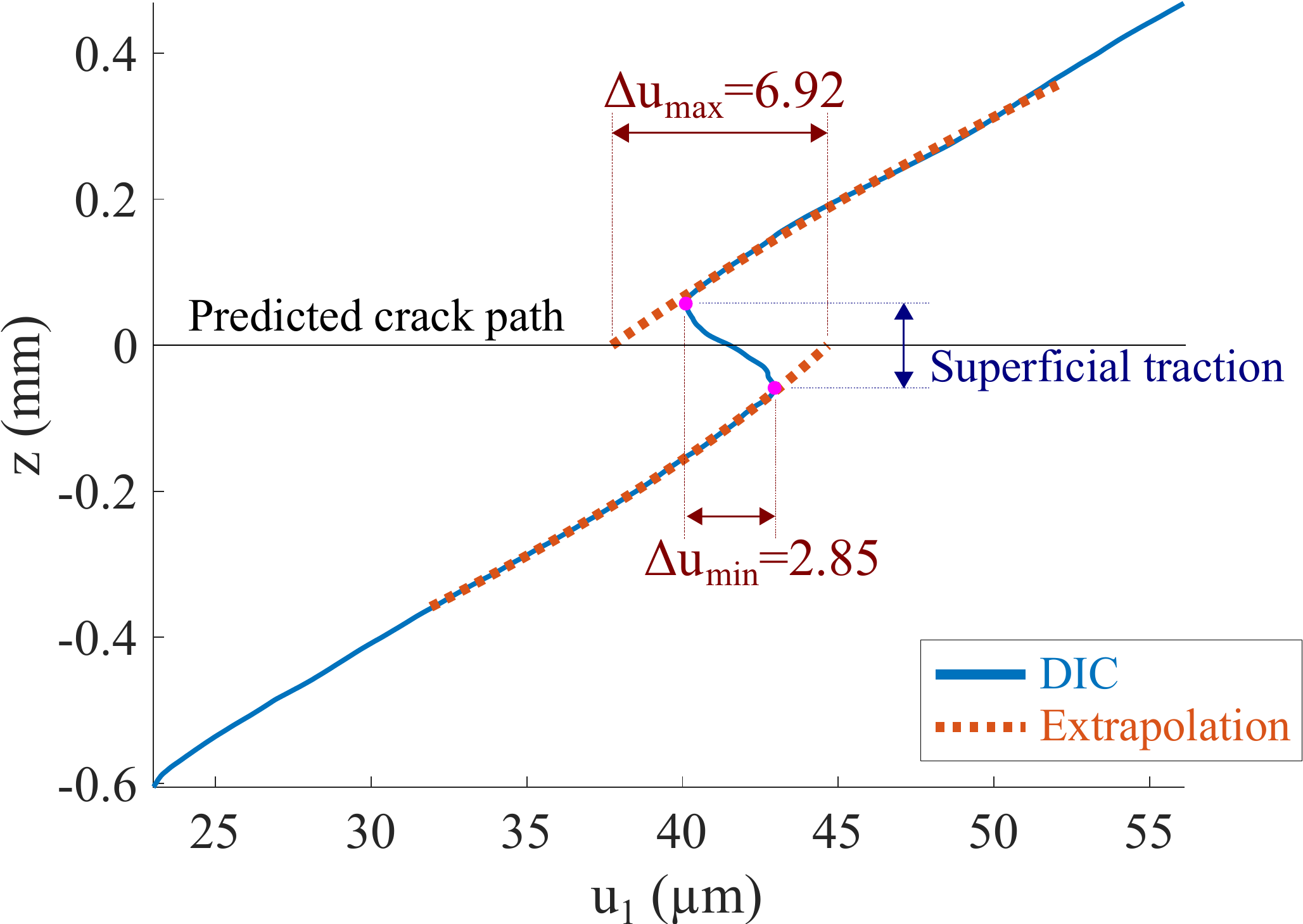}      
    } 
    \hfill
    \subfloat[\label{fig_du_size1_b}]{%
      \includegraphics[width=0.48\textwidth]{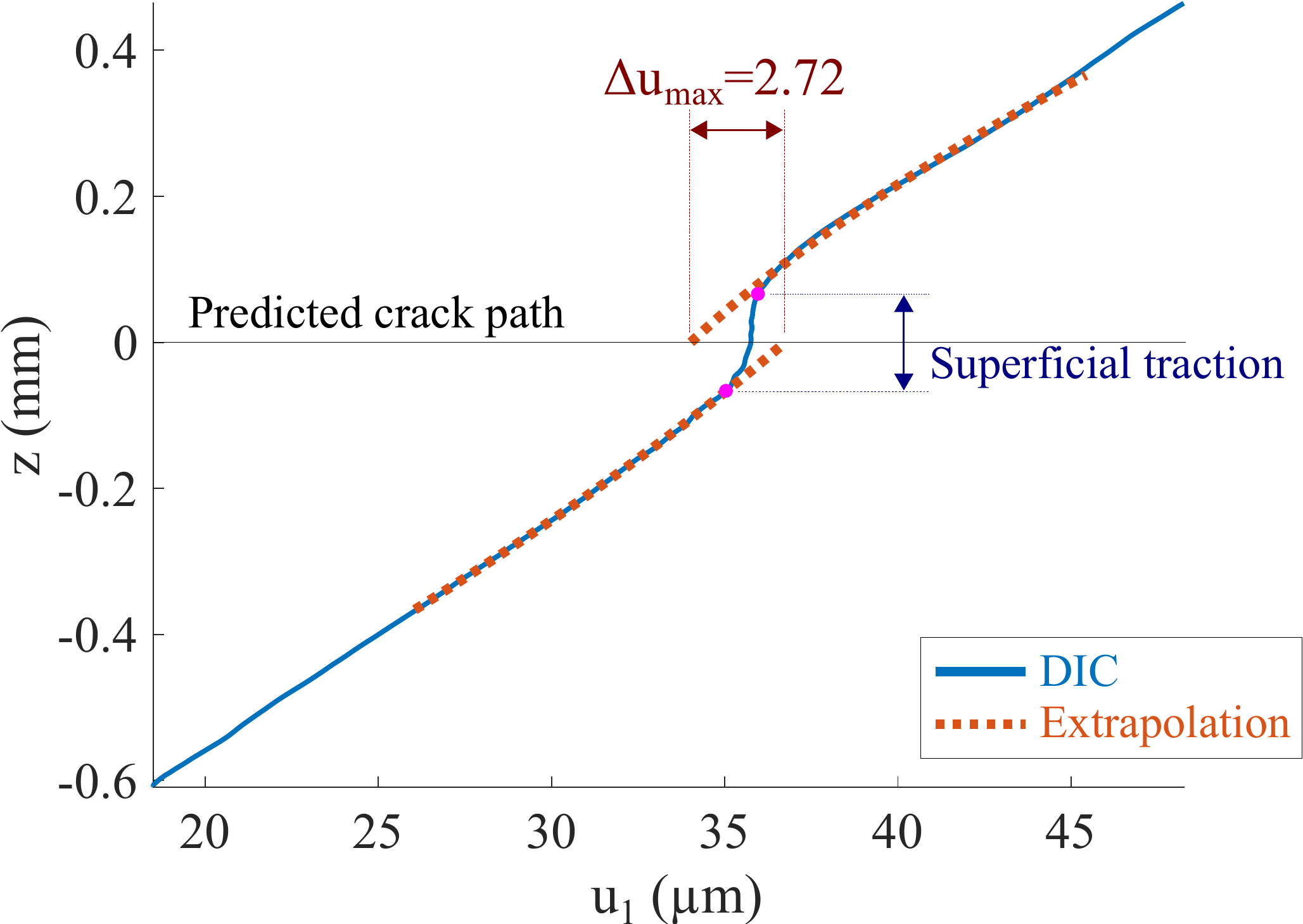}      
    }                 
    \caption{
    In-plane displacement $u_1$ variation through the thickness in the S1SP4 specimen at $P_\text{max}$. 
    (a) At $x=0$. (b) At $x=0.7$ mm.}
    \label{fig_du_size1}
  \end{figure}

For the S1SP4 specimen at $P_\text{max}$, in-plane displacement $u_1$ variation through the thickness is presented at two different locations in \cref{fig_du_size1}.
To be specific, \cref{fig_du_size1_a} shows the $u_1$--$z$ plots at the initial crack tip (i.e., $x=0$), while \cref{fig_du_size1_b} was captured $0.7$ mm away from the initial crack tip along the predicted crack path (i.e., $x=0.7$ mm).
A strongly zigzagging curve was obtained at the initial crack tip.
It should be noted that paints were sprayed on the specimen surface (i.e., the $x$--$z$ plane) to form black and white speckles for DIC (see \cref{fig_coord_transform_a}).
The strong zigzag implies a partial or complete separation between the layers; however, the paint coat on the surface might not have been separated yet and formed effective superficial traction.
The superficial traction could have led to the continuous zigzagging curve instead of discontinuity in the curve across the predicted crack path (i.e., the FPZ) induced by separation.
To address this issue, separation values were estimated in this work by extrapolating the plots except the superficial traction parts, which predicted the maximum separation values $\Delta u_\text{max}$.
The minimums $\Delta u_\text{min}$ were obtained using the $u_1$ values at the beginning and end of the superficial traction parts.
The actual separation magnitudes would exist between $\Delta u_\text{max}$ and $\Delta u_\text{min}$ and could be closer to $\Delta u_\text{max}$ than $\Delta u_\text{min}$.
This topic will be further discussed in the following section by comparing these plots with numerical simulation results.
Weaker zigzagging curves (i.e., smaller separation magnitudes) were observed from $x=0.7$ mm as shown in \cref{fig_du_size1_b} in which $\Delta u_\text{min}$ could not be obtained.
Nearly linear plots were observed at $x\geq2.0$ mm, implying that the size of the FPZ formed in the S1SP4 specimen at $P_\text{max}$ could be between $0.7$ and $2.0$ mm.

\begin{figure}[t!]
\centering
    \subfloat[\label{fig_du_size2_a}]{%
      \includegraphics[width=0.48\textwidth]{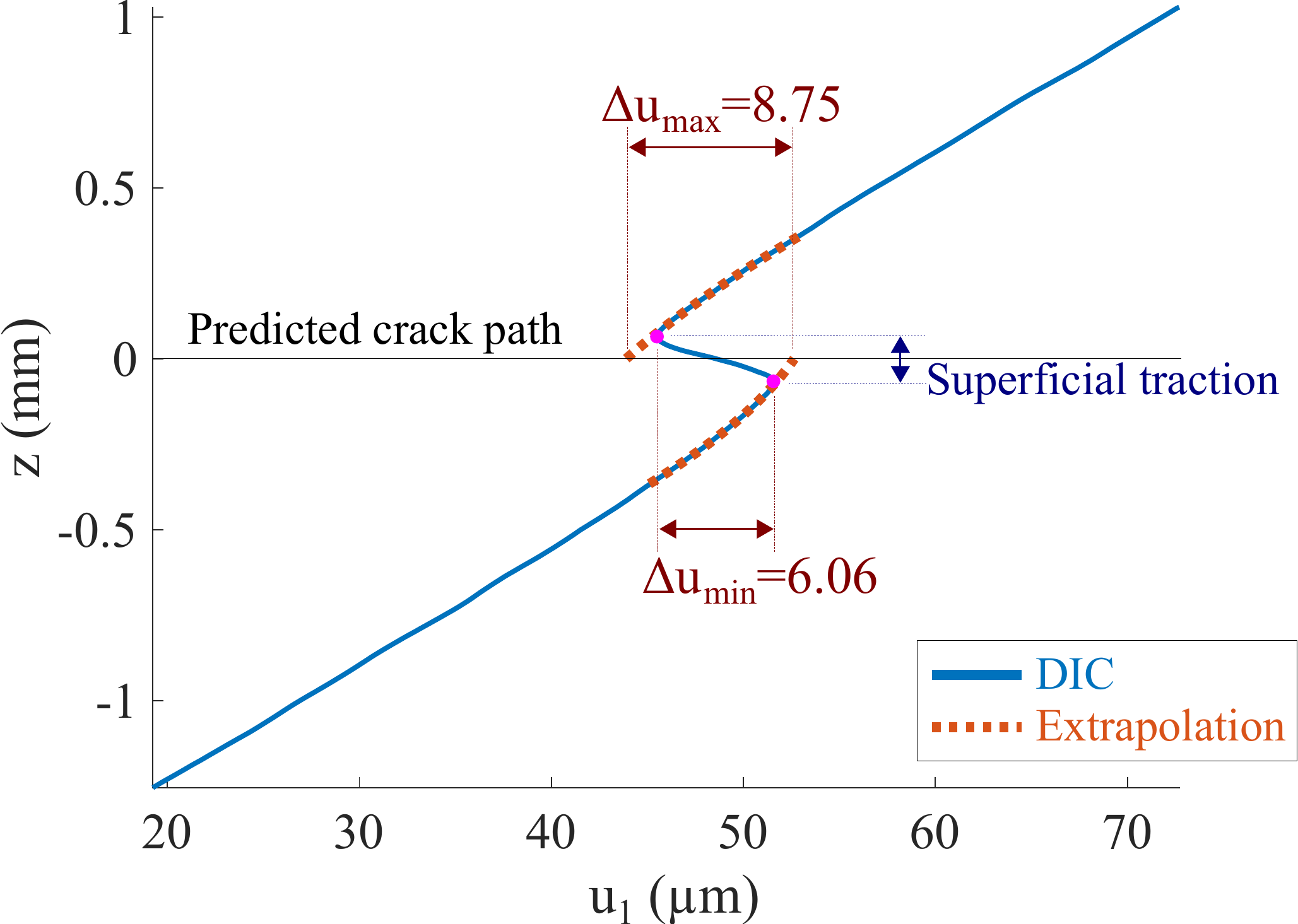}      
    } 
    \hfill
    \subfloat[\label{fig_du_size2_b}]{%
      \includegraphics[width=0.48\textwidth]{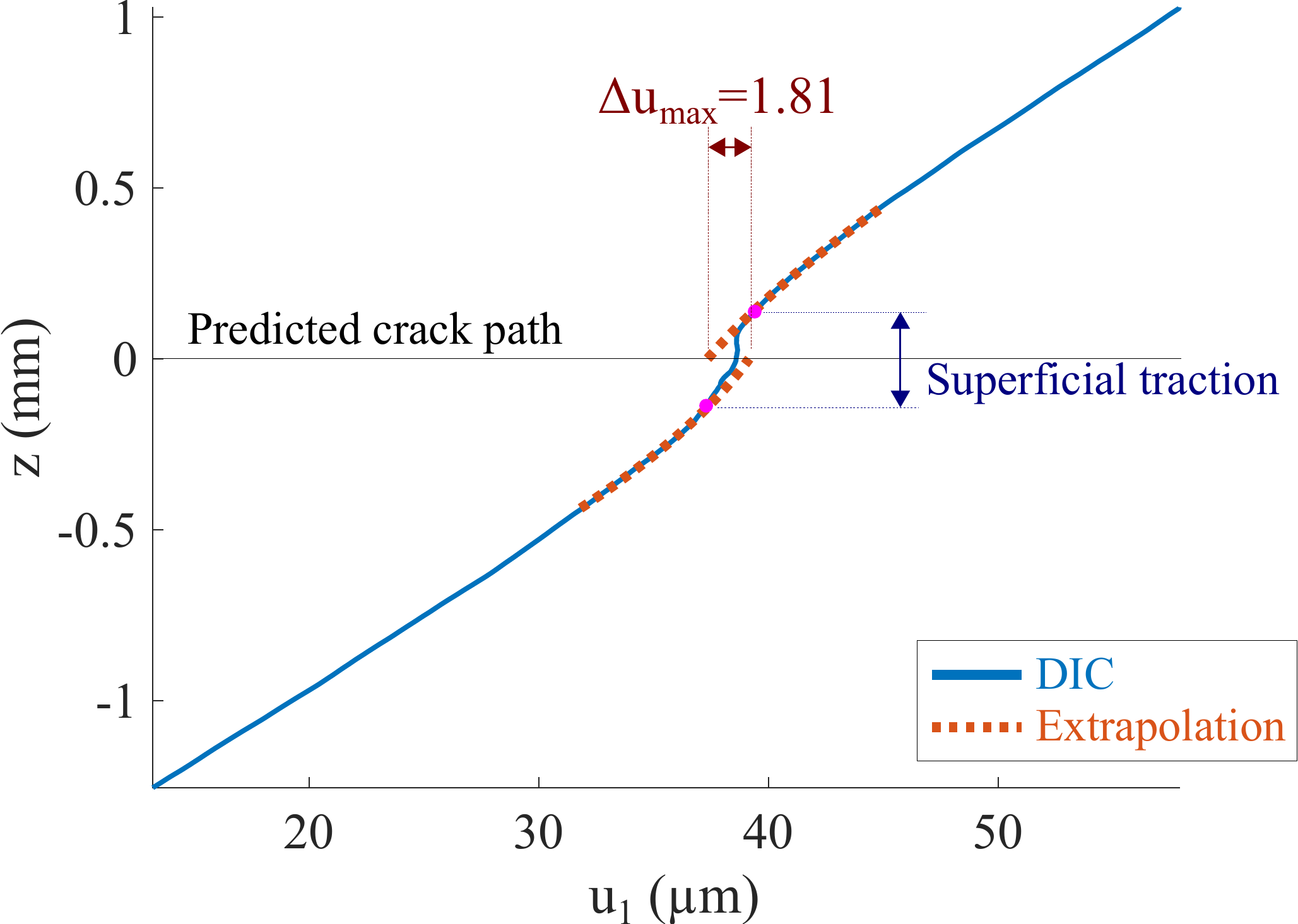}
    }               
    \caption{In-plane displacement $u_1$ variation through the thickness in the S2SP4 specimen at $P_\text{max}$. 
    (a) At $x=0$. (b) At $x=1.5$ mm..}
    \label{fig_du_size2}
  \end{figure}

For the S2SP4 specimen at $P_\text{max}$, the variation of $u_1$ through the thickness is presented in \cref{fig_du_size2}.
The separation values $\Delta u_\text{max}$ and $\Delta u_\text{min}$ at the initial crack tip (see \cref{fig_du_size2_a}) were larger than the corresponding values of the S1SP4 specimen (see \cref{fig_du_size1_a}). 
The minimum separation $\Delta u_\text{min}$ was no longer observed from $x=1.5$ mm (see \cref{fig_du_size2_b}), while nearly linear plots were observed at $x \geq 3.0$ mm. 
This implies that the size of the FPZ formed at the S2SP4 specimen under $P_\text{max}$ could be between $1.5$ and $3.0$ mm, which is larger than the predicted FPZ size of the S1SP4 specimen.

\begin{figure}[t!]
\centering
    \subfloat[\label{fig_du_size3_a}]{%
      \includegraphics[width=0.48\textwidth]{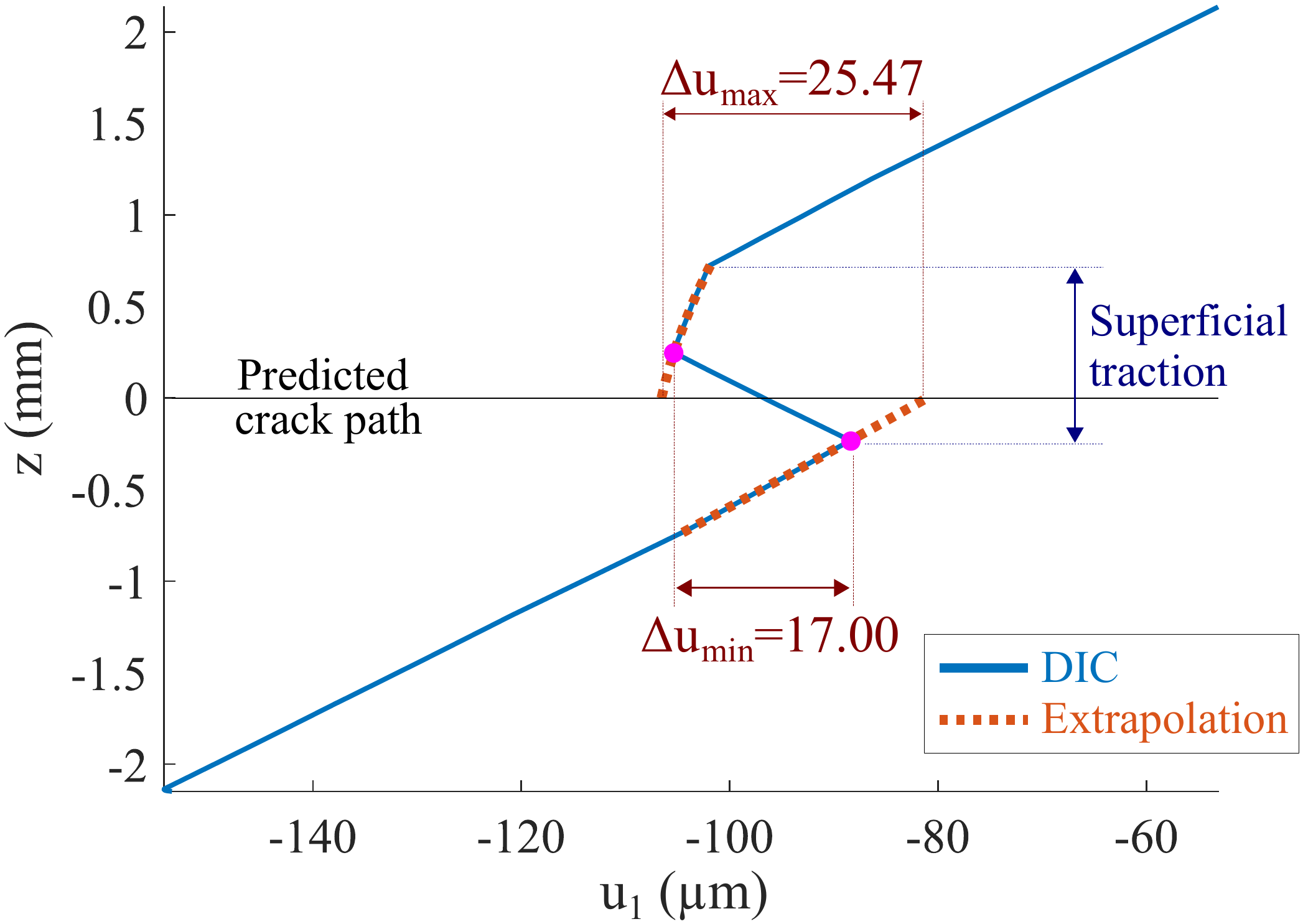}      
    }   
    \hfill
    \subfloat[\label{fig_du_size3_b}]{%
      \includegraphics[width=0.48\textwidth]{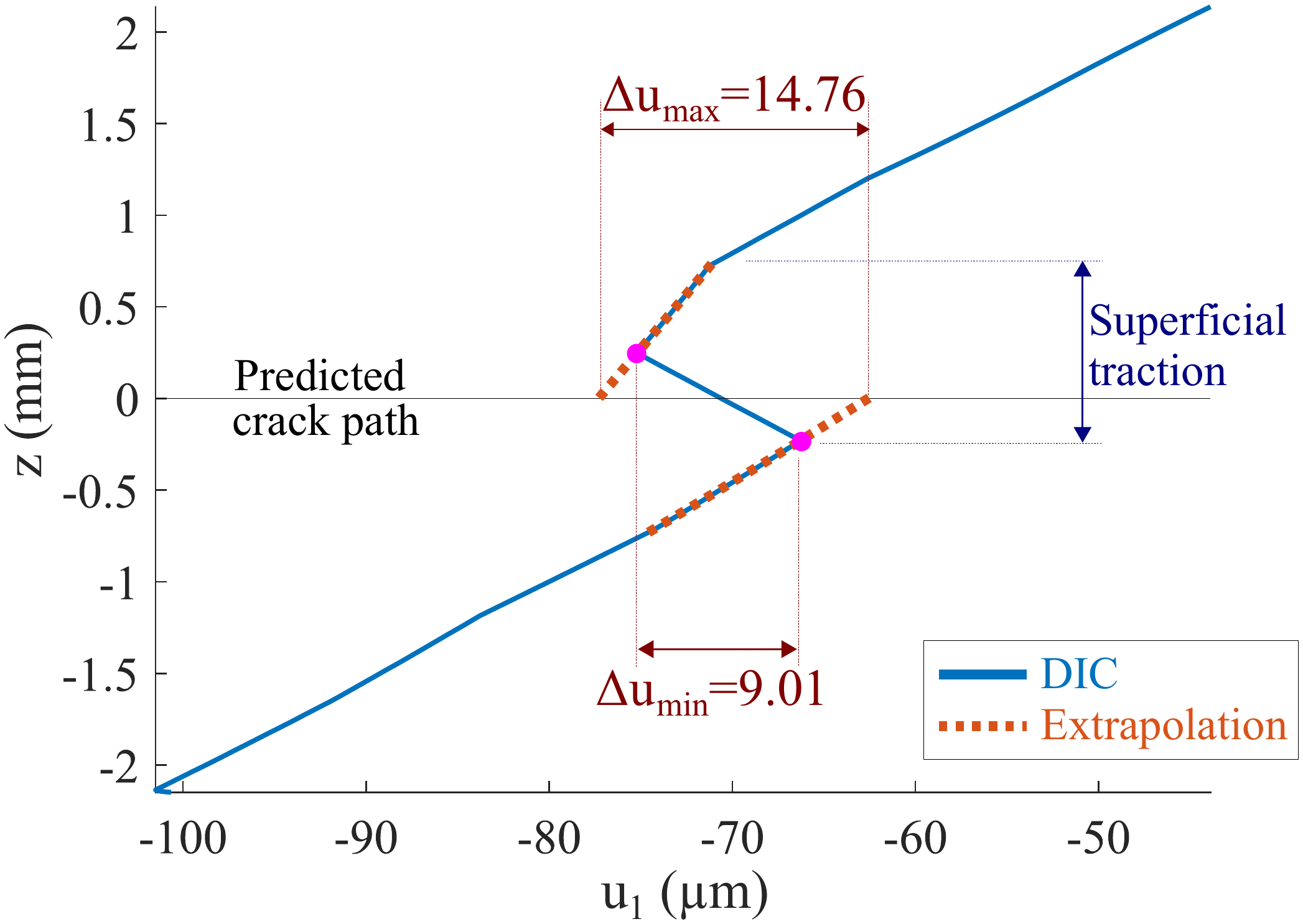}      
    }            
    \caption{In-plane displacement $u_1$ variation through the thickness in the S3SP3 specimen at $P_\text{max}$. 
    (a) At $x=0$. (b) At $x=10.0$ mm.}
    \label{fig_du_size3}
  \end{figure}

Lastly, the variation of $u_1$ through the thickness in the S3SP3 specimen at $P_\text{max}$ is shown in \cref{fig_du_size3}.
In contrast to the microscopic DIC data, the resolution of the macroscopic DIC data was not sufficient for the through-thickness deformation analysis. 
For example, the extrapolation of the upper parts of the $u_1$ curves was done based on the size of the superficial traction region observed from the S1SP4 and S2SP4 specimens; however, a sudden change in the slope around $z=0.9$ mm made the process difficult and the extrapolated values $\Delta u_\text{max}$ less reliable. 
Significantly larger separation values $\Delta u_\text{max}$ and $\Delta u_\text{min}$ were observed at the initial crack tip of the S3SP3 specimen (see \cref{fig_du_size3_a}) compared to the S1SP4 and S2SP4 specimens. 
Large separation values were also observed at $x = 10.0$ mm (see \cref{fig_du_size3_b}), while no linear variation was found due to the relatively low resolution of the macroscopic DIC data. 
Given a change in the sign of the $u_1$--$z$ plot slopes across $x=22.0$ mm (it can also be observed in the $u_1$ contour plots in \cref{fig_contour_size3}), the FPZ size of S3SP3 could possibly be around $22.0$ mm.
The separation values and FPZ sizes of these three specimens under $P_\text{max}$ are summarized in \Cref{tab_DIC}.
   
\begin{table}[t!]\caption{Separation values $\Delta u$ and FPZ sizes characterized from the DIC data}\label{tab_DIC}
\centering
\begin{tabular}{ccccc}
 \toprule
\textbf{Label} & $\Delta\boldsymbol{u}_\text{max}$ (\textmu m)& $\Delta\boldsymbol{u}_\text{min}$ (\textmu m)& \textbf{FPZ size} (mm)\\
\midrule
S1SP4 & 6.92 & 2.85 & 0.7--2.0 \\
S2SP4 & 8.75 & 6.06 & 1.5--3.0 \\
S3SP3 & 25.47 & 17.00 & 22.0 \\
\bottomrule
\end{tabular}
\end{table}

The proposed experimental framework was effective in characterizing separation values and FPZ sizes at $P_\text{max}$ from the microscopic DIC data.
However, finding the cohesive law of the specimen material still remained challenging due to difficulties in the experimental characterization of the corresponding tractions. 
Additionally, the separation values increased with the increase in the specimen sizes, which could imply different levels of traction-separation development at $P_\text{max}$ in different specimen sizes due to the size effect. 
This issue will be further discussed in the following section.

\section{Modeling and simulations}

This section is intended to introduce the utilization of the experimental data obtained from \Cref{sec:exp_results} for modeling and simulations.
It needs to be noted that extensive modeling and simulations were done for this work.
The main focus of this paper, however, is on proposing the experimental framework and demonstrating its implementation. 
Therefore, the modeling and simulation aspects will be simply introduced and discussed with some of the results in this paper. 
More detailed discussions on modeling and simulations will be made with the other data in a separate paper.

\subsection{Modeling approach}

\begin{figure}[t!]
\centering
    \subfloat[\label{fig_Abaqus_S1SP4_a}]{%
      \includegraphics[width=0.95\textwidth]{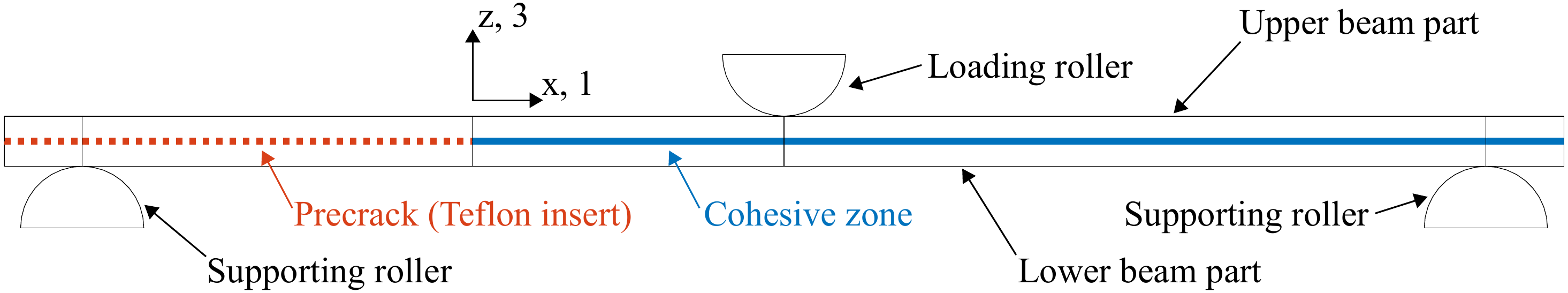}      
    } 
    \hfill
    \subfloat[\label{fig_Abaqus_S1SP4_b}]{%
      \includegraphics[width=0.9\textwidth]{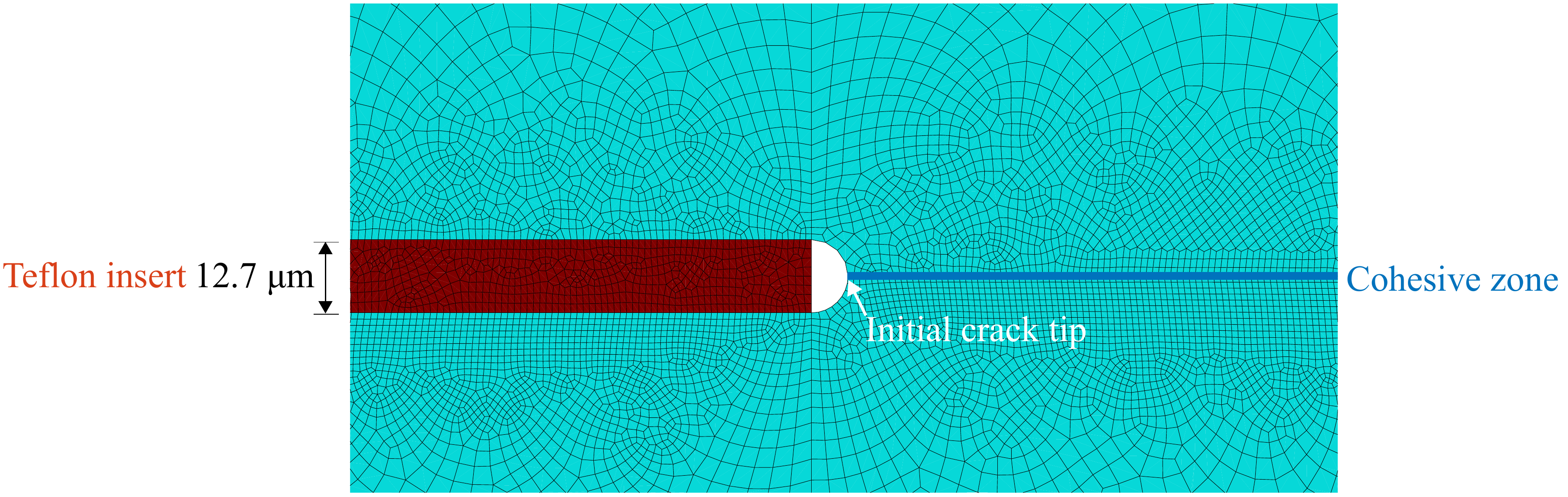}      
    }          
    \caption{Screenshots of the Abaqus model of the S1SP4 specimen with descriptions. 
    (a) The assembly of the following parts: the upper and lower beam parts and loading and supporting rollers. 
    (b) The zoomed-in mesh of the elements in the vicinity of the initial crack tip.}
    \label{fig_Abaqus_S1SP4}
  \end{figure}

Three models were built for the S1SP4, S2SP4, and S3SP3 specimens and were named the S1SP4, S2SP4, and S3SP3 models, respectively.
Abaqus/CAE 2024 was employed for modeling, and an identical modeling methodology was applied to the three models. 
Screenshots of the Abaqus S1SP4 model are presented in \cref{fig_Abaqus_S1SP4} to describe the modeling methodology.
The upper and lower beam sections were modeled using 2D-planar (on the $x$--$z$ plane) deformable parts, while the loading and supporting rollers were generated using 2D-planar analytical rigid parts.
These separate parts were assembled using contact interactions as shown in \cref{fig_Abaqus_S1SP4_a}. 
The cohesive zone included cohesive and damage parameters based on cohesive laws and was modeled along the midplane of the beam.
The high magnification of the microscopic images revealed a small, half-circular resin pocket in front of the Teflon insert. 
To incorporate this phenomenon, the Teflon insert was modeled with the elastic property of a Teflon\textsuperscript{\textregistered} FEP film \cite{Teflon}, $E=480$ MPa, as shown in \cref{fig_Abaqus_S1SP4_b}.
For simplicity, the resin pocket was modeled as a half-circular gap between the initial crack tip and the Teflon end.
 
Four-node bilinear plane strain quadrilateral, incompatible (CPE4I) elements \cite{wilson1973,taylor1976} were employed as shown in \cref{fig_Abaqus_S1SP4_b}.
The three models had different sizes and numbers of elements based on the FPZ sizes of the corresponding specimens. 
The mesh on the S1SP4 model was composed of 235,434 CPE4I elements and the size of the elements along the cohesive zone between the initial crack tip and loading point were 5$\times$5 \textmu m.
This element size was chosen to be small enough to capture the formation of FPZ during damage progression along the cohesive zone given that the FPZ size formed in the S1SP4 specimen at $P_\text{max}$ was observed to be approximately $0.7$--$2.0$ mm from the experimental data.
Given that the FPZ size of the S2SP4 specimen at $P_\text{max}$ was $1.5$--$3.0$ mm, the element size of the S2SP4 model along the cohesive zone between the initial crack tip and the loading point was chosen to be 10$\times$10 \textmu m.
As a result, the element mesh on the model was made of 312,772 CPE4I elements.
Lastly, the element size of the S3SP3 model along the cohesive zone between the initial crack tip and loading point was chosen to be $20$ \textmu m $\times 20$ \textmu m given the potentially large FPZ size of the S3SP3 specimen at $P_\text{max}$, $22.0$ mm.
The total number of CPE4I elements in the model was 310,525.
The fracture processes in these models were simulated using the Abaqus 2024/Dynamic, Implicit solver with a quasi-static option.

\subsection{Formulation of cohesive laws}\label{sec:CZM}

\begin{figure}[t!]
\centering
    \subfloat[\label{fig_CZM_types_a}]{%
      \includegraphics[width=0.32\textwidth]{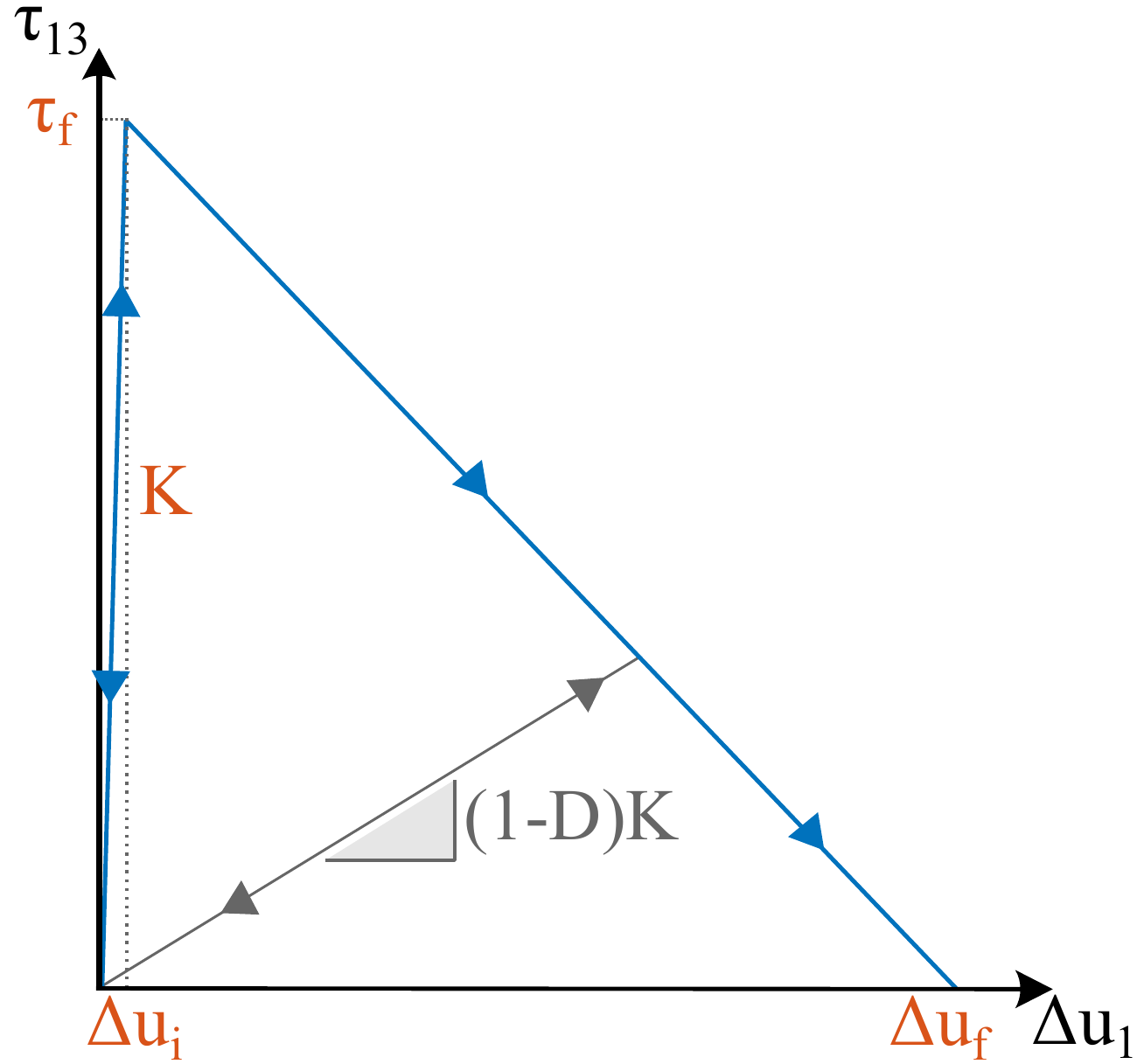}      
    } 
    \hfill
    \subfloat[\label{fig_CZM_types_b}]{%
      \includegraphics[width=0.32\textwidth]{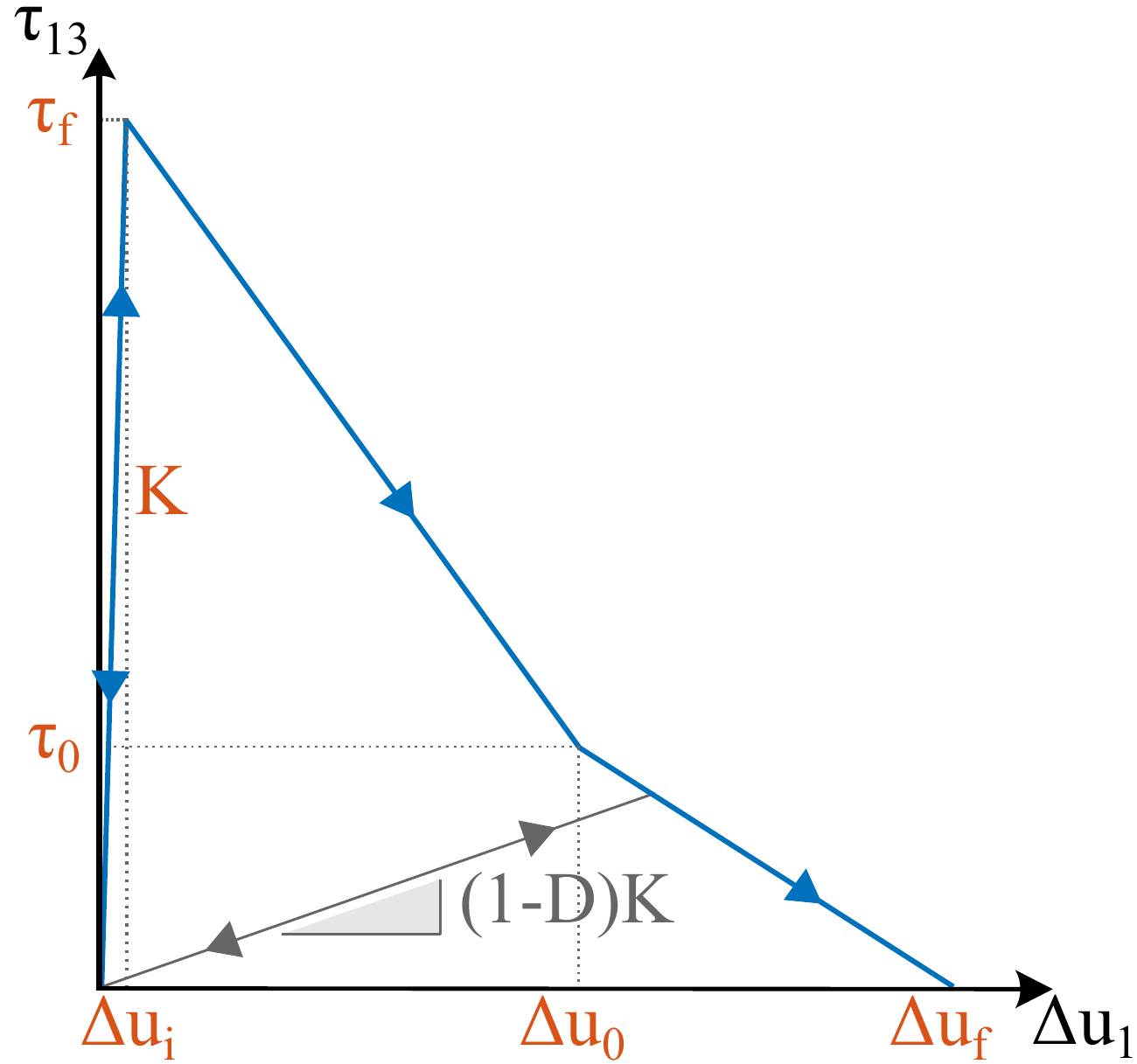}      
    }     
    \hfill
    \subfloat[\label{fig_CZM_types_c}]{%
      \includegraphics[width=0.32\textwidth]{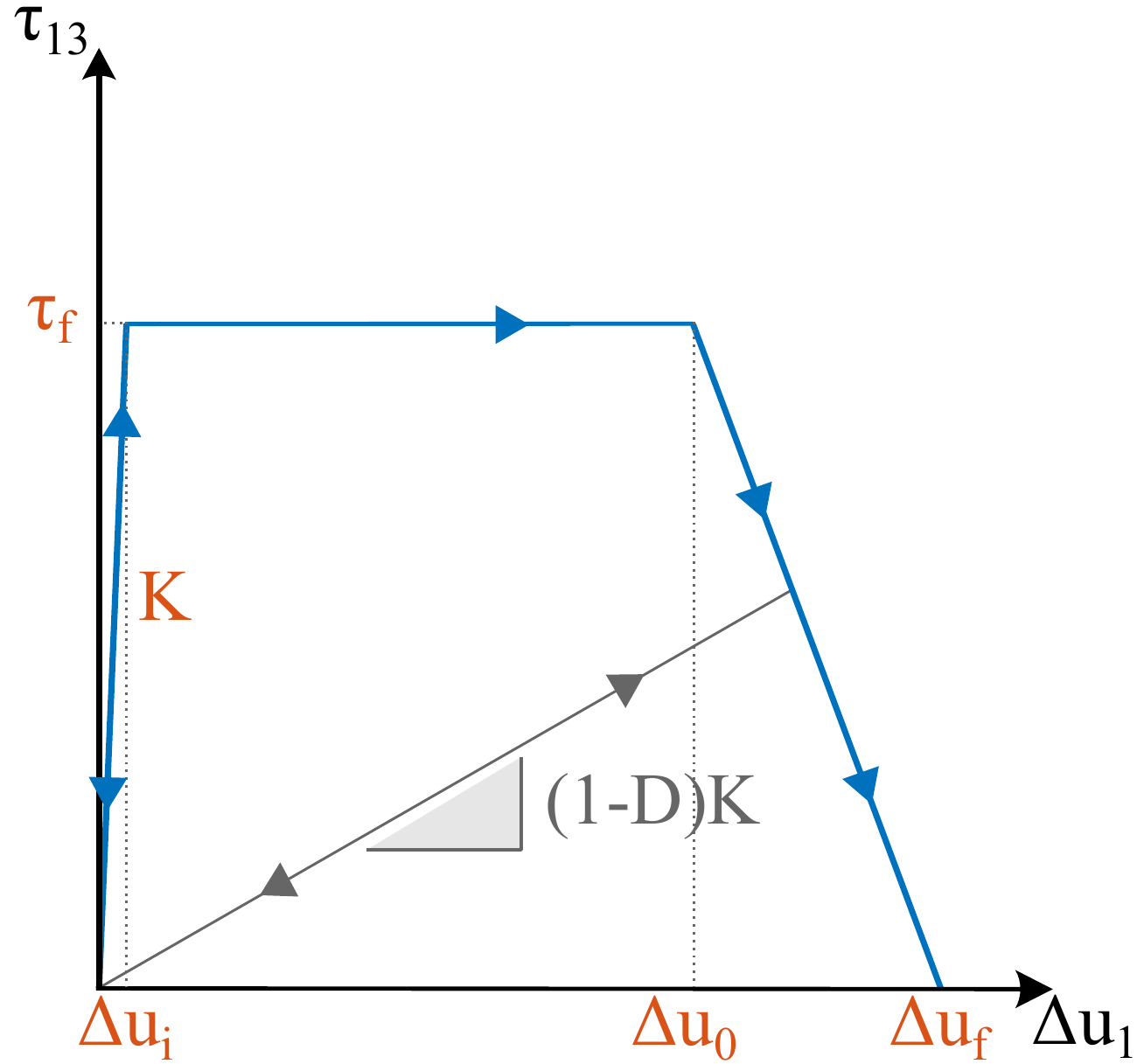}      
    }              
    \caption{Different types of cohesive laws for mode-II interlaminar fracture. 
    (a) CZM-L: Linear softening. 
    (b) CZM-biL: Bilinear softening with a transition at $\Delta u_0$ and $\tau_0$. 
    (c) CZM-pL: Initial plateau with subsequent linear softening at $\Delta u_0$--$\Delta u_\text{f}$.}
    \label{fig_CZM_types}
  \end{figure}

For traction-separation ($\tau_\text{13}$--$\Delta u_\text{1}$) law formulation, many different types of laws were employed and evaluated in this work. 
However, this paper will focus on only three commonly used laws as shown in \cref{fig_CZM_types}, while more complex laws will be described in a separate paper.
Additionally, the potential impact of underlying micromechanics within FPZs on the laws \cite{tran2022cohesive,tran2024numerical} will be investigated in future work.
The initial interface stiffness, damage variable, separation at damage initiation, final (or complete) separation, and maximum (or critical) traction are represented by $K$, $D$, $\Delta u_\text{i}$, $\Delta u_\text{f}$, and $\tau_\text{f}$, respectively. 
The three cohesive laws have different softening laws: linear softening, bilinear softening, and initial plateau with subsequent linear softening.
These laws are also called bilinear, trilinear, and trapezoidal forms, respectively, based on their whole curve shapes \cite{heidari2017}. 
In this work, the laws led to three types of CZMs which were named CZM-L, CZM-biL, and CZM-pL for linear softening, bilinear softening, and plateau-linear softening laws, respectively.
It should be noted that potential plastic deformations or dissipation at the initial crack tip \cite{chandra2003} were not considered in these laws and will be investigated in future work.
                   
All the laws have initial elastic regions with $K$ up to $\tau_\text{f}$.
The initial interface stiffness was assumed to be $K=1\times10^6$ MPa/mm based on the work of Turon et al. \cite{turon2007} and Zhao et al. \cite{zhao2013} for T300/977-2 and T700/QY9511 carbon-fiber/epoxy composites, respectively.
The fracture energy $G_\text{f}$ of the material for mode-II interlaminar fracture is given by \cite{bazant1998-1}
\begin{equation}\label{e_CZM_G}
G_\text{f}=\int_0^{\Delta u_\text{f}} \tau_{13} ~d\Delta u_1,
\end{equation}
which is equivalent to the areas below the cohesive curves.
Additionally, damage variables can be defined from the cohesive curves.
Detailed discussions on the damage variable can be found in Ref. \cite{barbero2013}.
In the initial elastic region, the damage variable is zero (i.e., $D=0$). 
Once $\tau_\text{f}$ is reached, damage progression is initiated at $\Delta u_\text{i}$ and the damage variable grows along the softening curves (i.e., $0<D\leq 1$). 
Under softening, the interface stiffness is reduced to {$(1-D)K$}.
The fracture energy of the material is completely released at $\Delta u_\text{f}$ (i.e., $D=1$).

\subsection{Utilization of the experimental data for modeling}

With the experimental data in \Cref{tab_exp,tab_DIC}, it was relatively easy to set up and calibrate the cohesive parameters to match the individual experimental data. 
The initial simulations were done based on the parameters in \Cref{tab_Abaqus_LEFM}.
The $\Delta u_\text{f}$ values were fixed as $\Delta u_\text{max}$ in \Cref{tab_DIC} assuming complete CZM development at $P_\text{max}$, while $G_\text{f,LEFM}$ was slightly calibrated from the values in \Cref{tab_exp}.
The parameters $\tau_\text{13}$ and $\Delta u_\text{1}$ to model the three representative specimens appeared very different; however, a single set of parameters should exist for the specimen material as its material properties. 
This inconsistency was intended to expose many challenges in achieving a single cohesive law for geometrically scaled specimens of a single quasibrittle material.

\begin{table}[t!]\caption{Cohesive parameters and simulation results}\label{tab_Abaqus_LEFM}
\centering
\begin{tabular}{lcccccc}
 \toprule
\textbf{CZM label} & $\boldsymbol{\tau}_0$ / $\boldsymbol{\tau}_\text{f}$ & $\Delta\boldsymbol{u}_0$ / $\Delta\boldsymbol{u}_\text{f}$ & $\boldsymbol{G}_\text{f}$ & \textbf{FPZ} & $\boldsymbol{P}_\text{max}$ & \textbf{Error}\\
& (MPa) & (\textmu m) & (N/mm) & (mm) & (N)& (\%) \\
\midrule
S1SP4 CZM-L1 & - / 173.4 & - / 6.92 & 0.60 & 0.496 & 431.9 & 0.21 \\
S1SP4 CZM-biL1 & 70 / 220 & 3.32 / 6.92 & 0.60 & 0.430 & 431.3 & 0.07 \\
S1SP4 CZM-pL1 & - / 110 & 4.10 / 6.92 & 0.60 & 0.705 & 435.0 & 0.94 \\
\midrule
S2SP4 CZM-L2 & - / 165.7 & - / 8.75 & 0.725 & 0.695 & 719.7 & 0.09 \\
S2SP4 CZM-biL2 & 50 / 250 & 4.10 / 8.75 & 0.725 & 0.320 & 719.0 & 0.00 \\
S2SP4 CZM-pL2 & - / 100 & 5.85 / 8.75 & 0.725 & 1.057 & 724.7 & 0.79 \\
\midrule
S3SP3 CZM-L3 & - / 87.9 & - / 25.47 & 1.12 & 3.531 & 1,153.4 & $-0.06$ \\
S3SP3 CZM-biL3 & 50 / 170 & 5.74 / 25.47 & 1.12 & 3.137 & 1,154.6 & 0.05 \\
S3SP3 CZM-pL3 & - / 52 & 17.66 / 25.47 & 1.12 & 5.408 & 1,155.2 & 0.10 \\
\bottomrule
\end{tabular}
\end{table}

\begin{figure}[t!]
\centering
    \subfloat[\label{fig_S1SP4_mLEFM_a}]{%
      \includegraphics[width=0.48\textwidth]{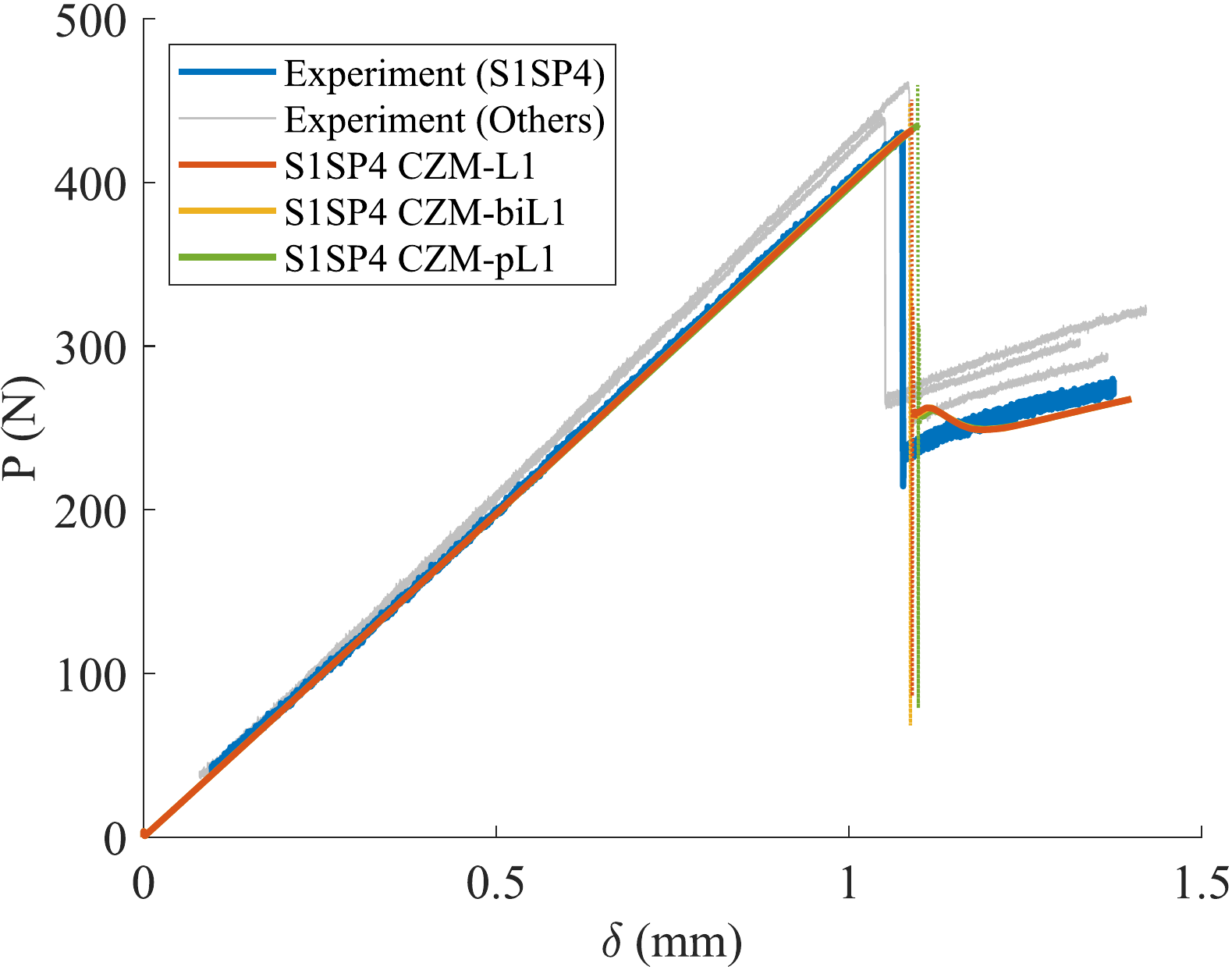}      
    } 
    \hfill
    \subfloat[\label{fig_S1SP4_mLEFM_b}]{%
      \includegraphics[width=0.48\textwidth]{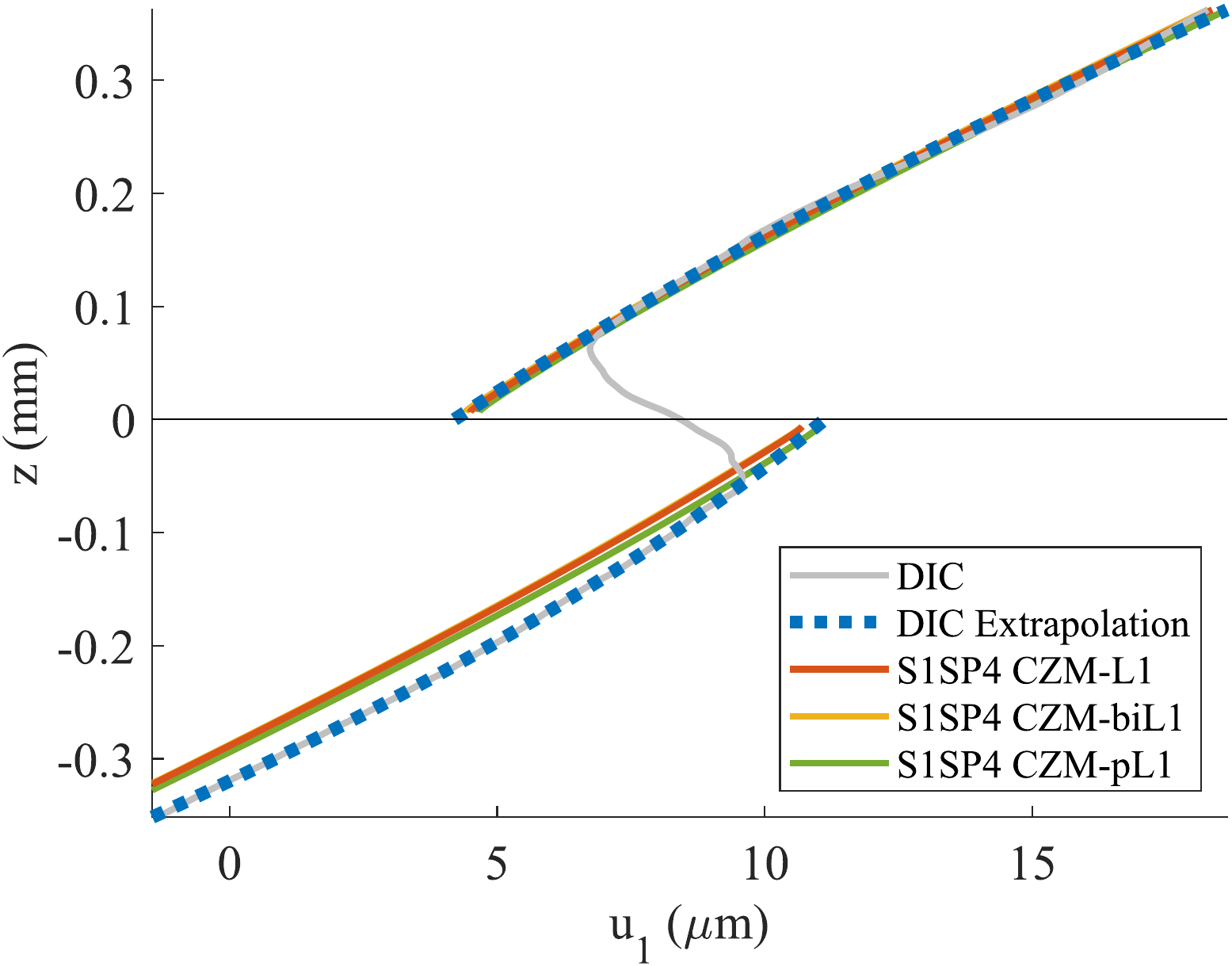}      
    }                   
    \caption{Simulation results of the S1SP4 model based on the parameters in \Cref{tab_Abaqus_LEFM}. 
    The experimental curves in \cref{fig_loading_curves_a,fig_du_size1} were reproduced here for comparison.
    (a) Load-displacement curves.  
	(b) In-plane displacement $u_1$ variation through the thickness at the initial crack tip under $P_\text{max}$. 
}
	\label{fig_S1SP4_mLEFM}
  \end{figure}

\begin{figure}[t!]
\centering
    \subfloat[\label{fig_S2SP4_mLEFM_a}]{%
      \includegraphics[width=0.48\textwidth]{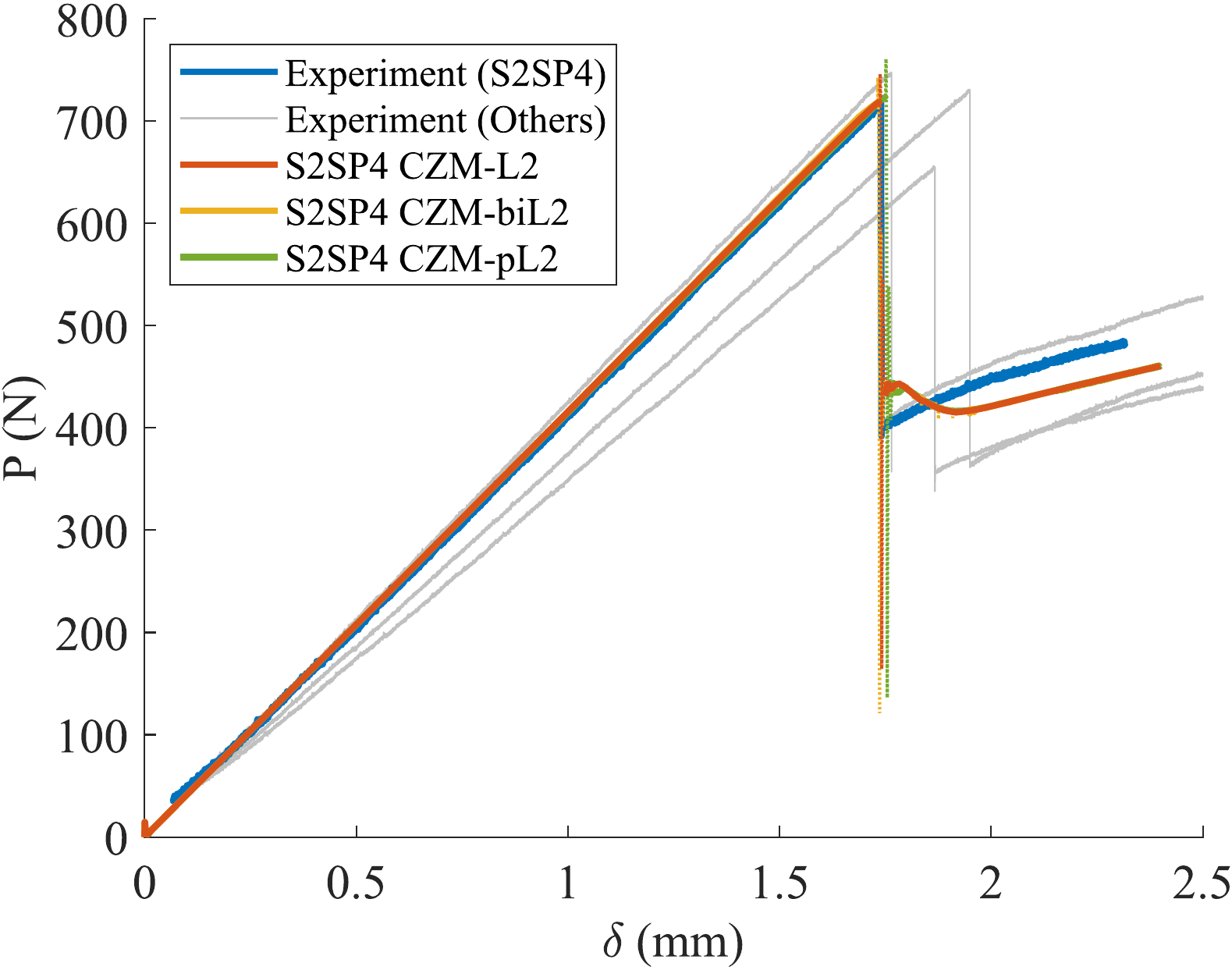}      
    } 
    \hfill
    \subfloat[\label{fig_S2SP4_mLEFM_b}]{%
      \includegraphics[width=0.48\textwidth]{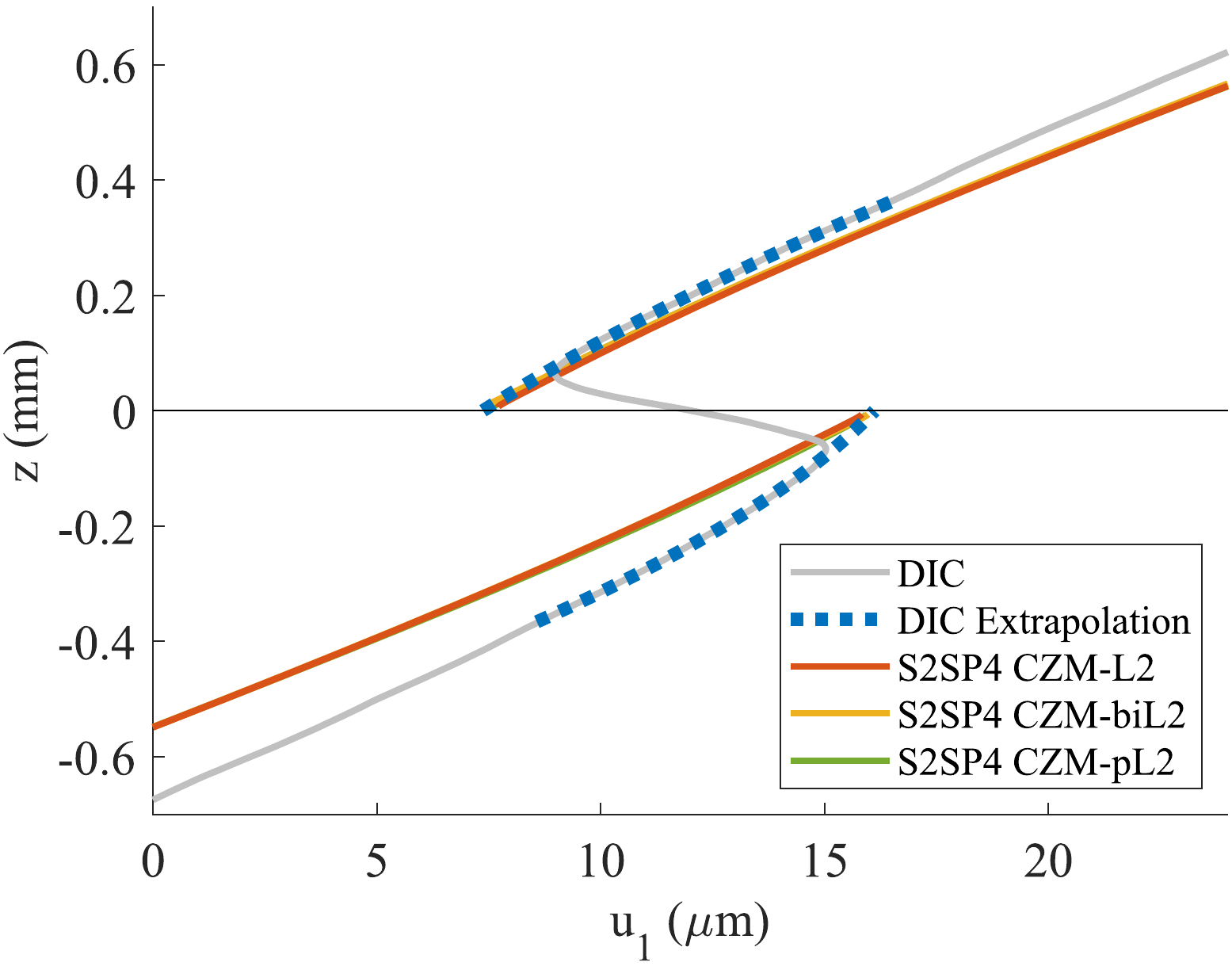}      
    }                   
    \caption{Simulation results of the S2SP4 model based on the parameters in \Cref{tab_Abaqus_LEFM}. 
    The experimental curves in \cref{fig_loading_curves_b,fig_du_size2} were reproduced here for comparison.
    (a) Load-displacement curves.  
	(b) In-plane displacement $u_1$ variation through the thickness at the initial crack tip under $P_\text{max}$. 
}
	\label{fig_S2SP4_mLEFM}
  \end{figure}

\begin{figure}[t!]
\centering
    \subfloat[\label{fig_S3SP3_mLEFM_a}]{%
      \includegraphics[width=0.48\textwidth]{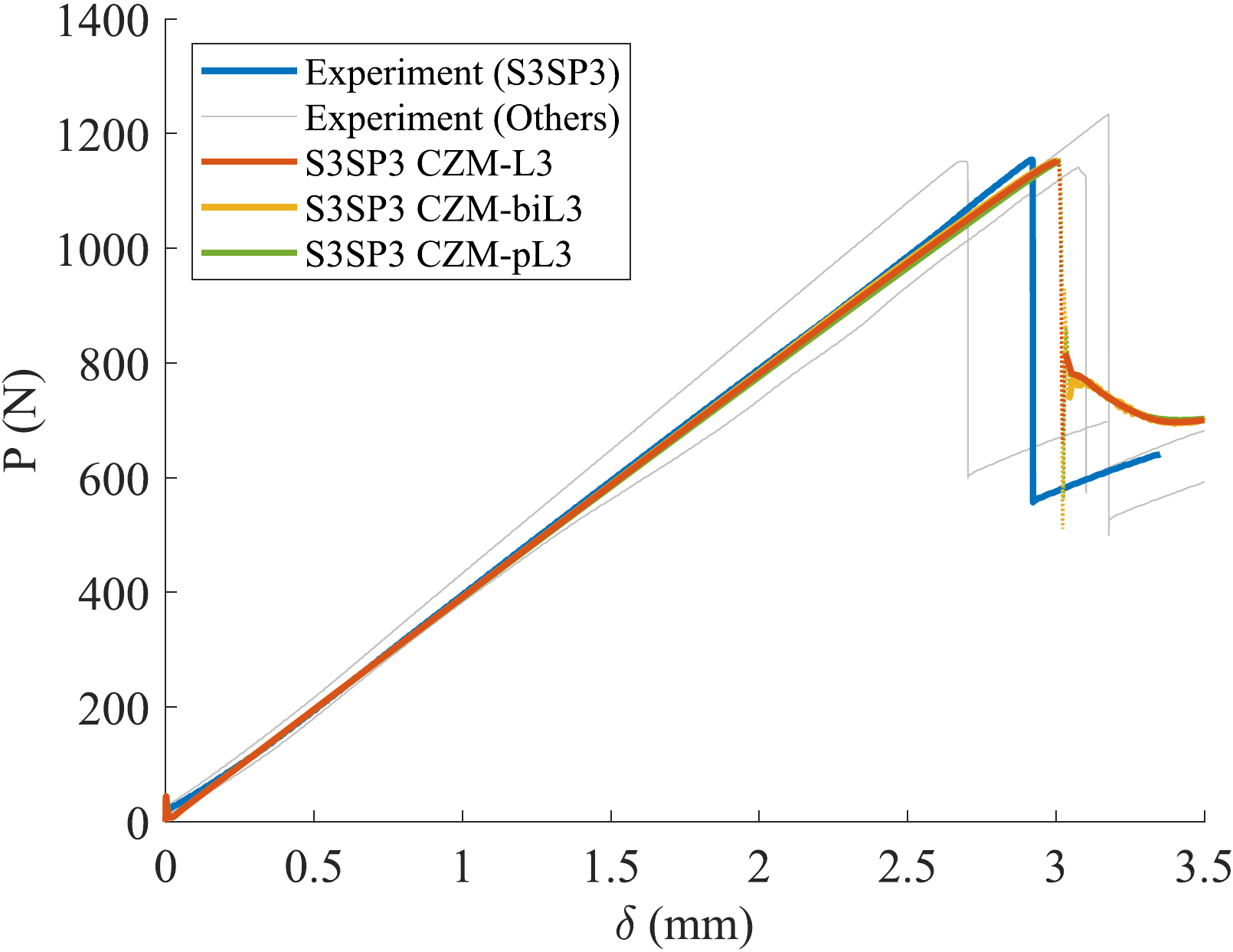}      
    } 
    \hfill
    \subfloat[\label{fig_S3SP3_mLEFM_b}]{%
      \includegraphics[width=0.48\textwidth]{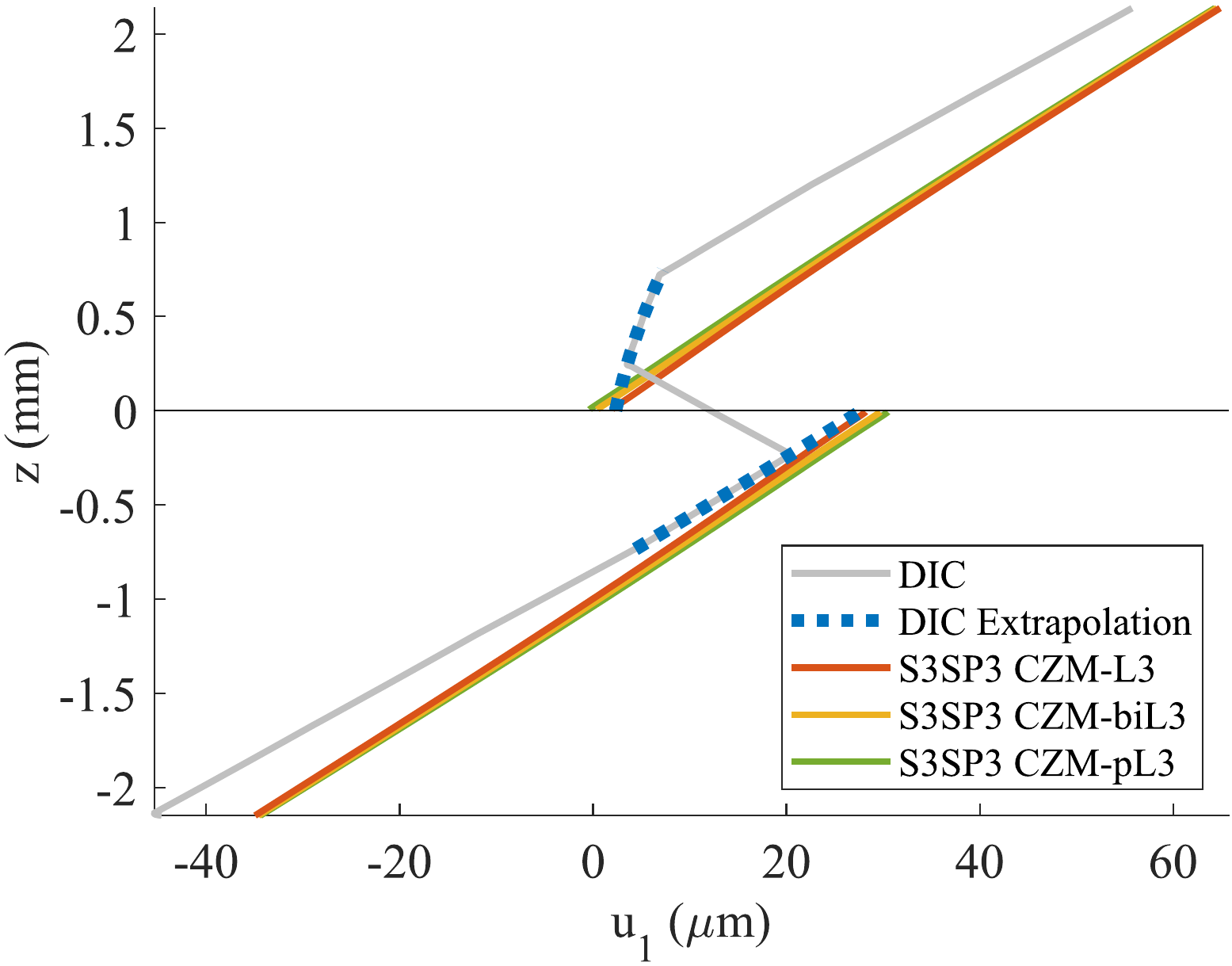}      
    }                   
    \caption{Simulation results of the S3SP3 model based on the parameters in \Cref{tab_Abaqus_LEFM}. 
    The experimental curves in \cref{fig_loading_curves_c,fig_du_size3} were reproduced here for comparison.
    (a) Load-displacement curves.  
	(b) In-plane displacement $u_1$ variation through the thickness at the initial crack tip under $P_\text{max}$.
}
	\label{fig_S3SP3_mLEFM}
  \end{figure}

The simulation results are presented in \cref{fig_S1SP4_mLEFM,fig_S2SP4_mLEFM,fig_S3SP3_mLEFM} for the S1SP4, S2SP4, and S3SP3 models, respectively.
For global behavior validation, the load-displacement curves are presented together with the experimental data (see \cref{fig_S1SP4_mLEFM_a,fig_S2SP4_mLEFM_a,fig_S3SP3_mLEFM_a}).
It needs to be noted that once $P_\text{max}$ was reached, all the curves showed initial sudden rises and subsequent snapdowns.
To differentiate these dynamic behaviors from the quasi-static response, it was considered as a dynamic behavior when the $P$--$\delta$ slope (i.e., $dP/d\delta$) got 10 times higher than $dP/d\delta$ of the linear regime.
The dynamic regimes were drawn with dotted lines in the $P$--$\delta$ curves, and the $P_\text{max}$ values were determined just before the dynamic regimes began. 
More studies will be done in future work to control these dynamic behaviors. 
All cases in \Cref{tab_exp} showed excellent agreement with the experimental data (mostly less than $0.21 \%$ errors).
Additionally, the different types of cohesive laws led to almost identical $P$--$\delta$ curves. 
Lastly, the $P$--$\delta$ curves of the S3SP3 model showed marginal softening near $P_\text{max}$.
This would imply that $\Delta u_\text{f}$ could possibly be overestimated based on $\Delta u_\text{max}$ due to the relatively low resolution of the macroscopic DIC data as discussed in \Cref{sec:exp_through_thickness}. 

Local behavior validation was made by analyzing $u_\text{1}$ at the initial crack tip under $P_\text{max}$ as shown in \cref{fig_S1SP4_mLEFM_b,fig_S2SP4_mLEFM_b,fig_S3SP3_mLEFM_b}.
Similarly to the $P$--$\delta$ curves, the different types of cohesive laws led to almost identical $u_\text{1}$ curves. 
It should be noted that the experimental $u_\text{1}$ curve needed to be slightly shifted for comparison with the simulation results. 
Additionally, as discussed in \Cref{sec:exp_through_thickness}, the $u_\text{1}$ curves of the S3SP3 specimen were not reliable due to the relatively low resolution of the macroscopic DIC data. 
Therefore, the analysis of the local behaviors is focused on the S1SP4 and S2SP4 models in this section.
The slopes of the $u_1$ curves of the S1SP4 model showed excellent agreement with the experimental data, while the S2SP4 model showed reasonable agreement with the experimental data. 
More importantly, the extrapolated parts of the shifted experimental curves for the S1SP4 and S2SP4 specimens were in good agreement with the simulation results, which would substantiate the presence of superficial traction along the FPZ induced by the surface paint.
Therefore, the actual separation values could be estimated by the extrapolated curves; that is, $\Delta u_\text{max}$ would be close to the actual separation values.
Lastly, the FPZ sizes of the models at $P_\text{max}$ were significantly smaller than the experimental measurements.
The plateau-linear softening cases formed the largest FPZs, while the smallest FPZs were shown in the bilinear cases.
This phenomenon would imply that the actual damage initiations occurred at lower traction levels than the $\tau_\text{f}$ values in \Cref{tab_Abaqus_LEFM}.

\subsection{A single cohesive law and partial CZM development}

The cohesive parameters in \Cref{tab_Abaqus_LEFM} showed excellent results for the corresponding size sets but did not work well for the other sizes; that is, the $\tau_\text{13}$--$\Delta u_\text{1}$ parameters were highly sensitive to geometric scaling. 
Additionally, the parameters were based on different fracture energy values for different sizes.
This approach is often called a geometry-dependent cohesive law.
However, this approach would contradict the fact that cohesive parameters are material properties, and thus a single law should exist for a quasibrittle material.
It needs to be noted that the fracture energies for the three sizes in \Cref{tab_Abaqus_LEFM} were all smaller than $G_\text{f}$ obtained from the size effect analysis in \Cref{sec:exp_results_global_size_effect}.
Furthermore, the insensitivity of the global behaviors (i.e., the load-displacement curves) to the traction-separation types was observed from the simulation results. 
This phenomenon made it extremely challenging to characterize the shape of the cohesive laws by merely relying on load-displacement curve validation. 
Lastly, the through-thickness deformation analysis of the experimental data showed an increase in the separation values at $P_\text{max}$ with a size increase (see \Cref{{tab_DIC}}).
This phenomenon could imply that smaller energy was released at $P_\text{max}$ in the smaller (or more scaled-down) specimens due to partial CZM development at a lower $\Delta u_\text{1}$ level.

\begin{figure}[t!]
\centering
    \subfloat[\label{fig_CZM_single_a}]{%
      \includegraphics[width=0.48\textwidth]{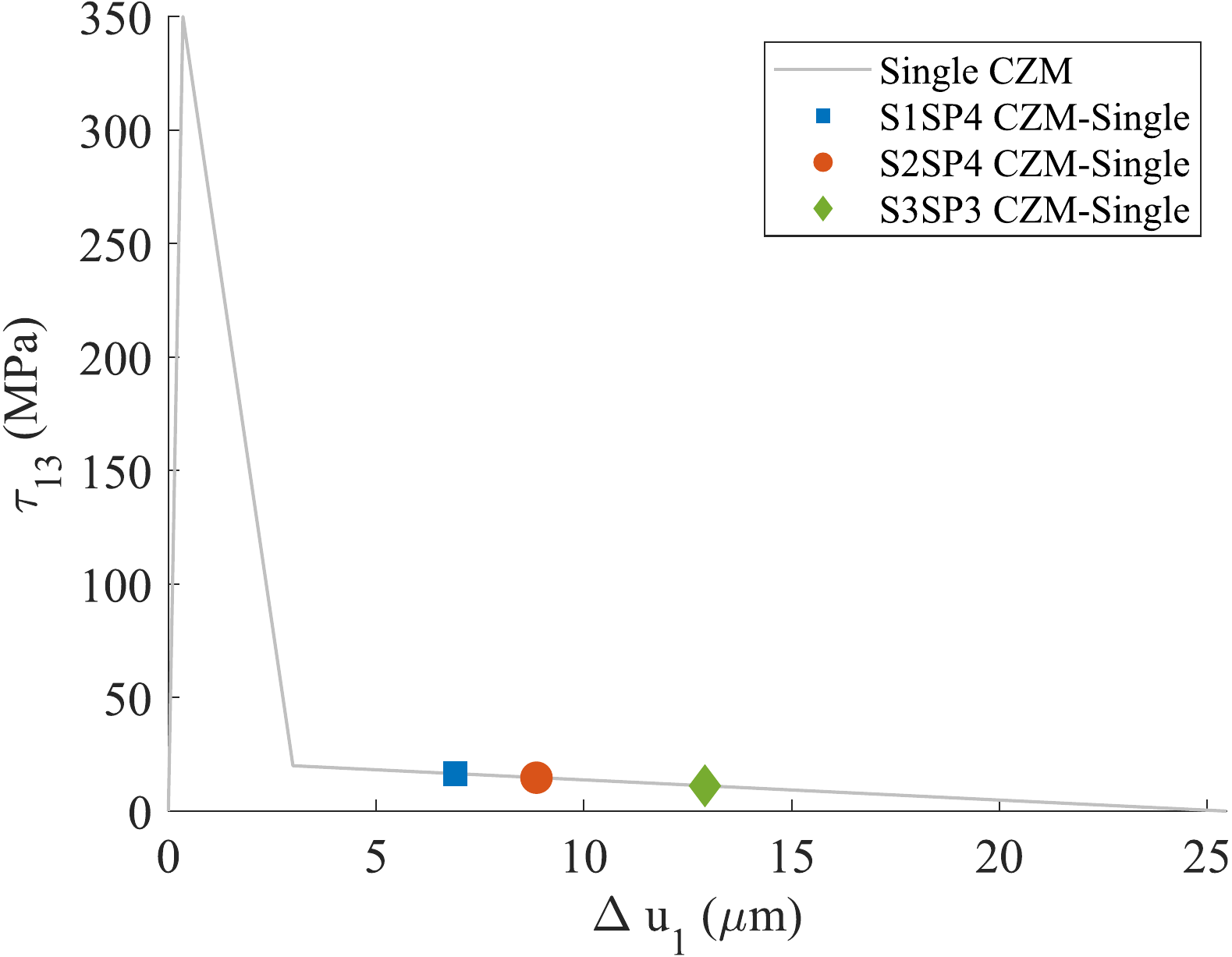}      
    } 
    \hfill
    \subfloat[\label{fig_CZM_single_b}]{%
      \includegraphics[width=0.48\textwidth]{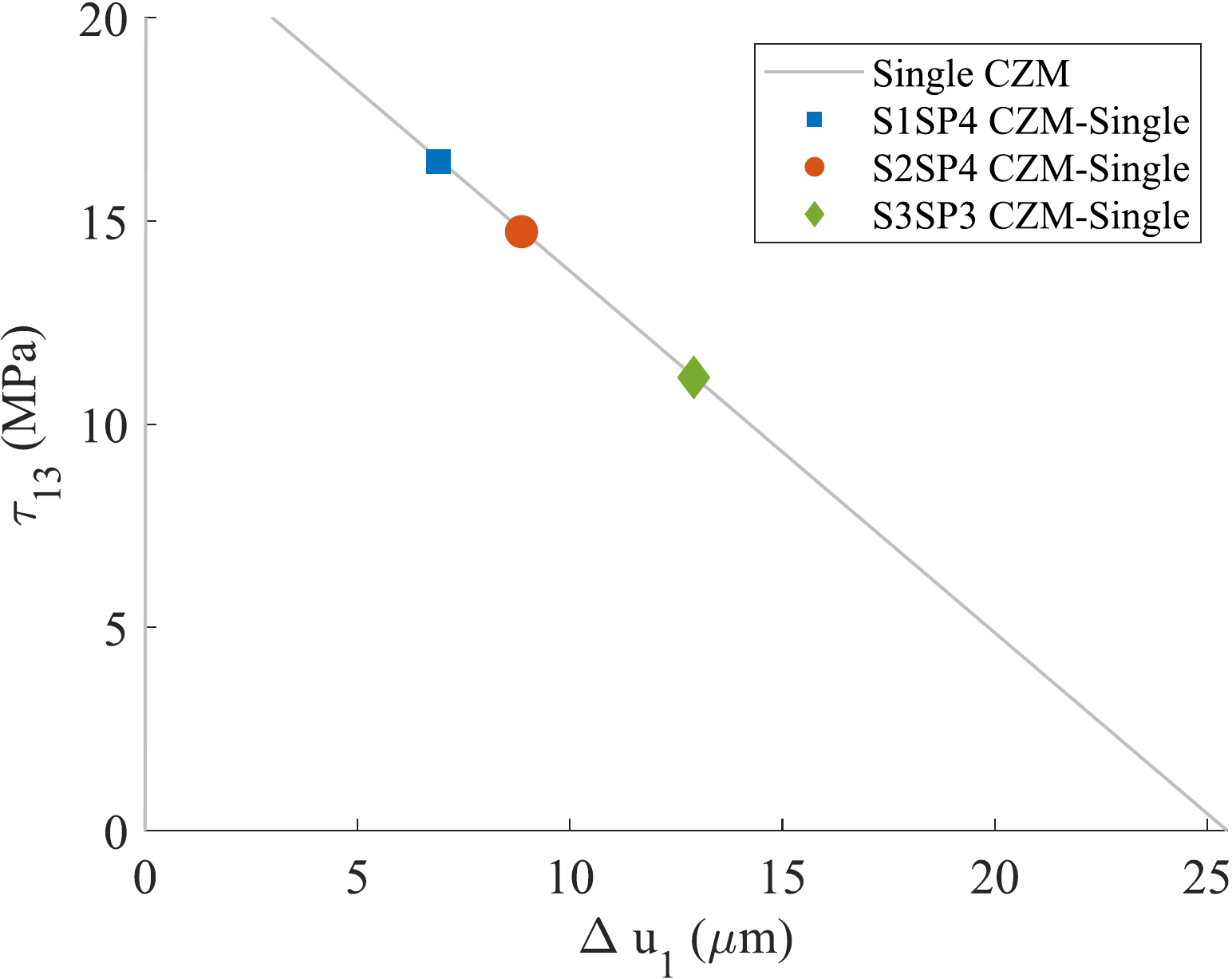}      
    }          
    \caption{A single cohesive law for the specimen material. 
    The markers indicate the levels of $\tau_\text{13}$ and $\Delta u_\text{1}$ at the initial crack tip of the models at $P_\text{max}$.
    (a) On a full scale. 
	(b) On a zoomed-in scale between $\tau_{13}=0$ and $20$ MPa.
}
	\label{fig_CZM_single}
  \end{figure}

To address these issues, a lot of efforts were made in this work to develop a single cohesive law for the specimen materials and match the experimental data from the three different sizes with the single law. 
Details on characterizing and calibrating a single set of $\tau_\text{13}$--$\Delta u_\text{1}$ parameters will be discussed in a separate paper. 
In this paper, only a cohesive law based on bilinear softening is introduced as shown in \cref{fig_CZM_single}.
This law was built with $\tau_\text{f}=350$ MPa, $\tau_\text{0}=20$ MPa, $\Delta u_0=3.00$ $\mu$m, $\Delta u_\text{f}=25.47$ $\mu$m, and $G_\text{f}=0.72$ N/mm (see \cref{fig_CZM_single_a}).
The $\Delta u_\text{f}$ value was set based on $\Delta u_\text{max}$ of the S3SP3 specimen.
It needs to be noted that $G_\text{f}$ used in this law was smaller than $G_\text{f}$ obtained from the size effect analysis.

\begin{figure}[t!]
\centering
    \subfloat[\label{fig_Abaqus_single_a}]{%
      \includegraphics[width=0.48\textwidth]{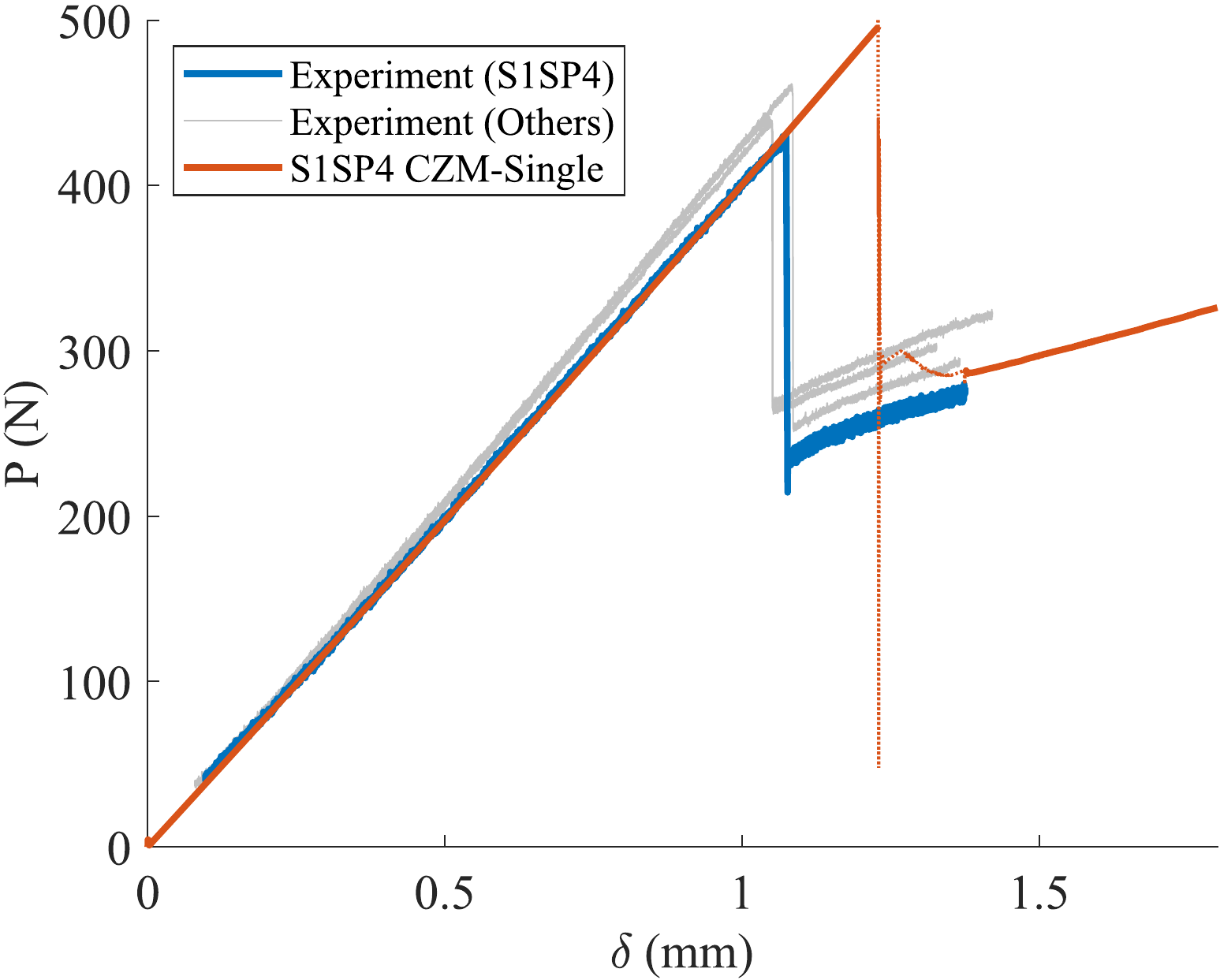}      
    } 
    \hfill
    \subfloat[\label{fig_Abaqus_single_b}]{%
      \includegraphics[width=0.48\textwidth]{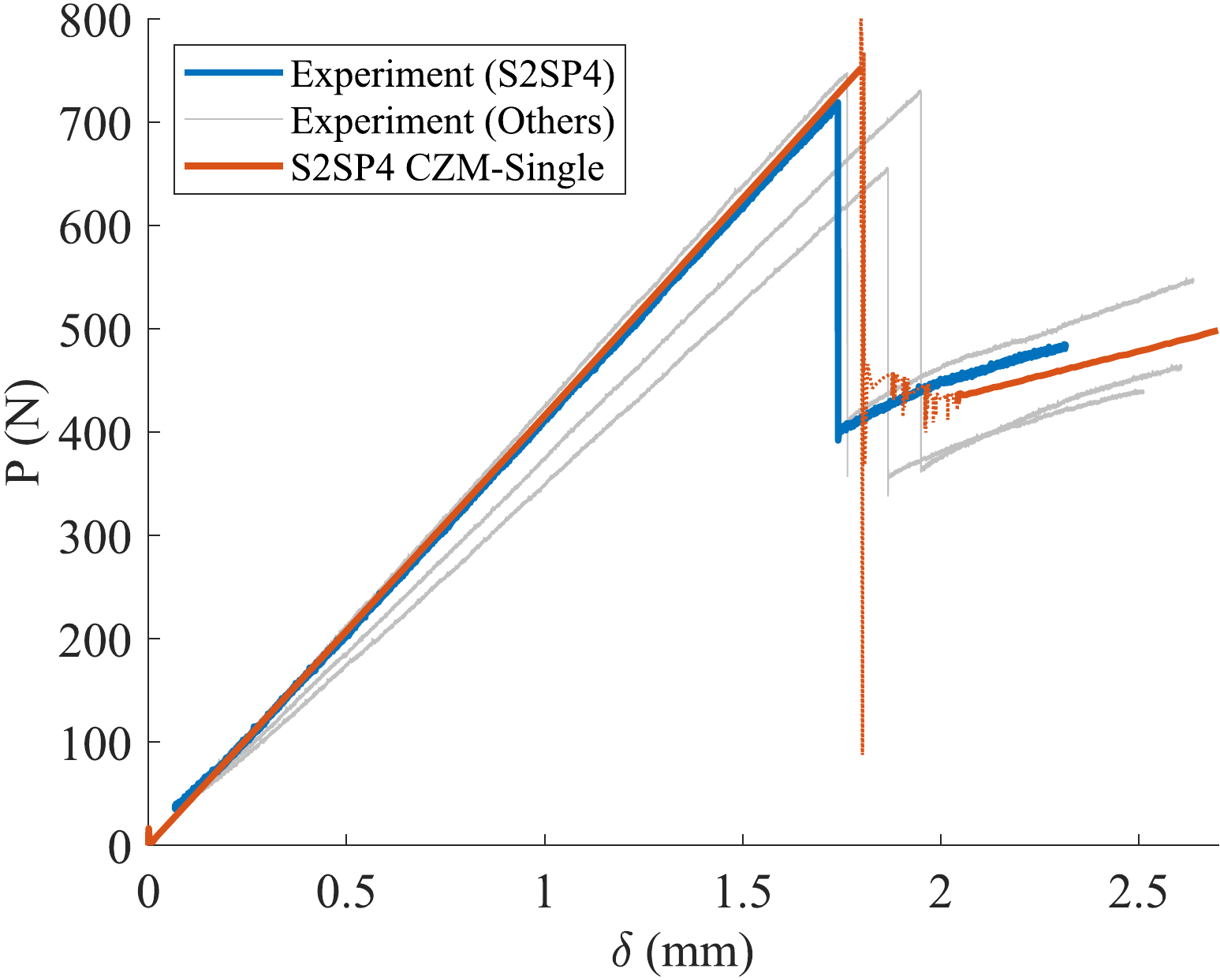}      
    }       
    \hfill
    \subfloat[\label{fig_Abaqus_single_c}]{%
      \includegraphics[width=0.48\textwidth]{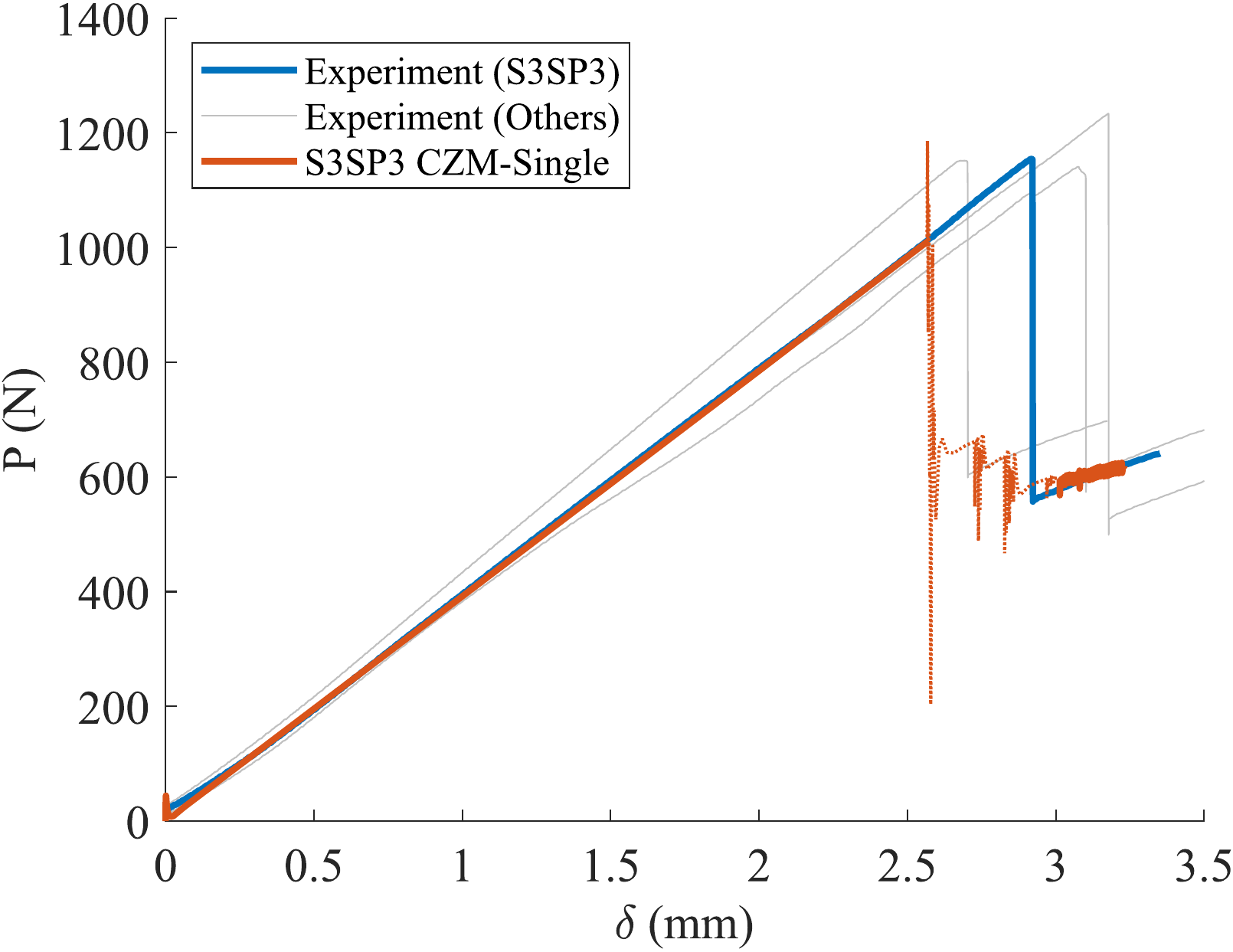}      
    } 
    \hfill
    \subfloat[\label{fig_Abaqus_single_d}]{%
      \includegraphics[width=0.48\textwidth]{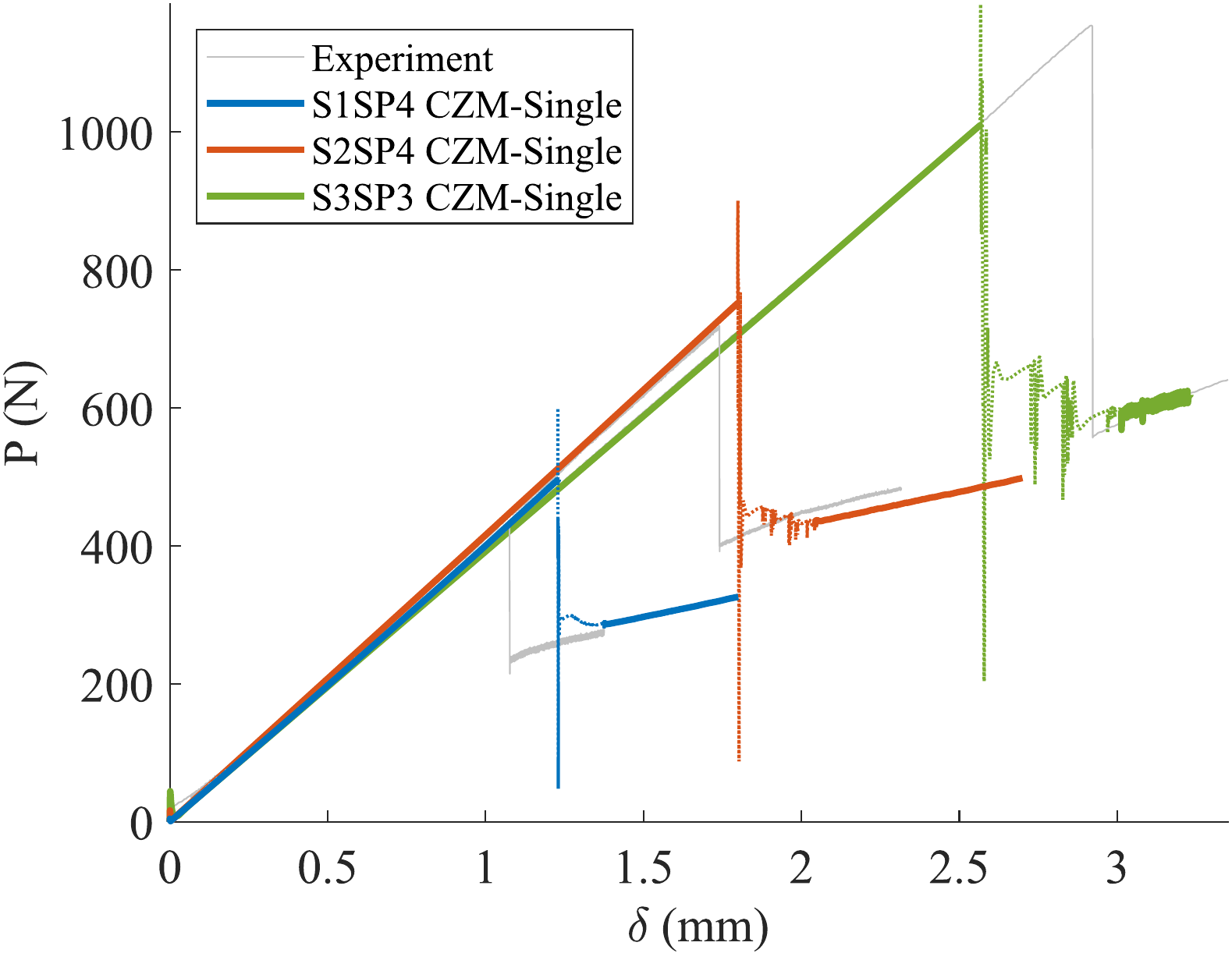}      
    }     
    \caption{Simulation results of the models with a single cohesive law shown in \cref{fig_CZM_single}. 
    The experimental curves in \cref{fig_loading_curves} were reproduced here for comparison. 
    (a) The S1SP4 model. 
	(b) The S2SP4 model. 
    (c) The S3SP3 model. 
	(d) Collection of all results with the corresponding experimental curves.
}
	\label{fig_Abaqus_single}
  \end{figure}

The simulation results are summarized in \Cref{tab_Abaqus_CZM_single} and are illustrated with a label named CZM-Single in \cref{fig_Abaqus_single}.
The single cohesive law made good agreement with $P_\text{max}$ of the S2SP4 specimen (see \cref{fig_Abaqus_single_b}), whereas it overestimated and underestimated $P_\text{max}$ for the S1SP4 and S3SP3 specimens, respectively (see \cref{fig_Abaqus_single_a,fig_Abaqus_single_c}).
Additionally, the FPZ at $P_\text{max}$ was significantly smaller than the experimental measurements. 
However, partial CZM development was well realized as shown in \cref{fig_CZM_single_b} and $\Delta u$ at the initial crack tip of the S1SP4 and S2SP4 models under $P_\text{max}$ showed excellent agreement with the experimental measurements of $\Delta u_\text{max}$ in \Cref{tab_DIC}.
The $\Delta u$ value of the S3SP3 model was significantly smaller than the experimental $\Delta u_\text{max}$ measurement but was closer to $\Delta u_\text{min}$.
As shown in \cref{fig_Abaqus_single_d}, the single cohesive law overall performed reasonably well capturing the post-peak responses of all sizes including snapdrops.

\begin{table}[t!]\caption{Simulation results at $P_\text{max}$ from a single cohesive law in \cref{fig_CZM_single}}\label{tab_Abaqus_CZM_single}
\centering
\begin{tabular}{lccccc}
 \toprule
\textbf{CZM label} & $\boldsymbol{\tau}$ & $\Delta \boldsymbol{u}$ & \textbf{FPZ} & $\boldsymbol{P}_\text{max}$ & \textbf{Error}\\
& (MPa) & (\textmu m) & (mm) & (N)& (\%) \\
\midrule
S1SP4 CZM-Single & 16.47 & 6.91 & 0.384 & 495.8 & 15.03 \\
S2SP4 CZM-Single & 14.74 & 8.86 & 0.583 & 751.9 & 4.58 \\
S3SP3 CZM-Single & 11.15 & 12.92 & 1.093 & 1,154.6 & $-12.46$ \\
\bottomrule
\end{tabular}
\end{table}

Bažant et al. showed the effect of crack-parallel stress at the crack front on the mode-I fracture energies and effective FPZ sizes of quasibrittle materials having finite FPZ sizes through the gap tests of notched concrete beams under three-point bending \cite{bazant2020-1,bazant2020-2}.
From similar tests using carbon-fiber/epoxy composite specimens, Salviato et al. showed that crack-parallel compressive stress could lead to a reduction in the mode-I fracture energy of the material but an increase in the effective FPZ size \cite{salviato2023}.
For mode-II interlaminar fracture, crack-parallel stress could also make a significant impact, particularly when a crack propagates above or below the neutral axis under three-point bending.
As a first step towards developing a single cohesive law for mode-II incorporating size effect, all the specimens in this work had the initial cracks induced along their neutral axes (i.e., their midplanes) to minimize crack-parallel stress at the crack front.
Potential geometric imperfections in these specimens, however, might have led to non-negligible crack-parallel stress, which could have possibly contributed to the discrepancy between the experimental data and the simulation results from the single cohesive law.
To address this issue, more efforts will be made to investigate the potential impact of crack-parallel stress on mode-II interlaminar fracture in future work and thus improve the single cohesive law.

\section{Conclusions}\label{sec:conclusion}

This paper was focused on proposing an experimental framework to characterize a cohesive law for mode-II interlaminar fracture and demonstrating its implementation with modeling and simulations. 
The experimental data of the global fracture behaviors revealed the limitation of a compliance calibration method which is widely used based on linear elastic fracture mechanics.  
The method failed to estimate the fracture energy of the specimen material as a single material property value, showing an increase in the fracture energy with scaling up of the specimen dimensions.  
To address this issue, Bažant's type-II size effect law was applied to the experimental data of the global fracture behaviors in the scaled specimens and a single fracture energy value was obtained. 
For the experimental characterization of local fracture behaviors, the digital image correlation (DIC) data were analyzed using the three steps: coordinate transformation, curve fitting, and through-thickness deformation analysis. 
The proposed method was effective in experimentally characterizing the in-plane separation magnitudes at the initial crack tips in the specimens at their fracture loads from the microscopic DIC data. 
It was also shown that the data resolution of a commonly used macroscopic DIC method (employed only for the largest-size specimens) was not sufficient for the proposed method. 
Finally, it was observed that the separation magnitude and the size of the fracture process zone increased as the specimen size increased, which could be caused by the size effect and imply partial development of cohesive laws at the fracture loads in the smaller specimens.

The specimens were modeled based on different types of cohesive laws. 
Initially, the laws were built with the experimentally measured separation values and slightly calibrated fracture energies from the compliance calibration method assuming complete cohesive law development at the fracture load. 
This modeling approach was intended to validate the proposed experimental framework. 
The simulation results showed excellent agreement with the experimental data of the corresponding sizes but did not work for the other sizes. 
Therefore, the initial modeling and simulation results contradicted the existence of a single cohesive law as a material property of a quasibrittle composite material.
Furthermore, the global behaviors (i.e., the load-displacement curves) of the models were insensitive to the cohesive law types. 
This phenomenon made it extremely challenging to characterize the shape of the cohesive laws by merely relying on load-displacement curve validation. 
To address these issues, a single cohesive law of the specimen material was proposed based on bilinear softening.
For the global fracture behaviors, the simulation results showed good agreement with the experimental data of the mid-size specimen while showing reasonable agreement with the other sizes.
The post-peak responses were particularly well captured for all sizes.
For local fracture behaviors, the single law well captured partial cohesive law development by showing excellent agreement with the microscopic DIC measurement of the separation values at the fracture loads. 
However, the fracture energy of the single law was smaller than the energy obtained from the size effect analysis, while the sizes of the fracture process zones (FPZs) were smaller than the experimental measurements. 

In future work, the experimental setup will be improved to apply the microscopic DIC testing method to all sizes. 
Additionally, more efforts will be made to incorporate the fracture energy obtained from the size effect analysis in the single cohesive law and to match the experimentally measured FPZ sizes using the single law.
Lastly, the impact of potential plastic deformations (or dissipation) and crack-parallel compressive stress at the initial crack tip on global and local fracture behaviors will be investigated.



\section{Acknowledgments}

The authors gratefully acknowledge the support of the Air Force Office of Scientific Research (Grant numbers: FA9550-19-1-0031 and ARCTOS 162643-19-26-C1) and the Air Force Research Laboratory's Structural Sciences Center at Wright-Patterson Air Force Base.





\bibliographystyle{elsarticle-num} 
\bibliography{cas-refs_r1}






\end{document}